\documentclass[a4paper,11pt]{article}
\pdfoutput=1 \usepackage{amssymb}
\usepackage{amsmath}
\usepackage{braket}
\usepackage{mathtools}
\usepackage[usenames,dvipsnames,svgnames,table]{xcolor}
\usepackage[utf8]{inputenc}
\usepackage{color}
\usepackage{cite}
\usepackage{subcaption}
\usepackage{lmodern}
\usepackage{footnote}
\usepackage{glossaries-extra}
\setabbreviationstyle[acronym]{long-short}
\glssetcategoryattribute{acronym}{nohyperfirst}{true}

\usepackage{slashed}
\usepackage[pdftex]{graphicx}
\usepackage{multirow}
\usepackage{jheppub} 
\usepackage{here}
\usepackage{hyperref}
\usepackage{verbatim}
\usepackage[justification=centering, singlelinecheck=false]{caption}
\usepackage{booktabs}
\usepackage{cleveref}

\Crefname{equation}{Eq.}{Eqs.}

\usepackage{longtable}
\usepackage{tabu}
\usepackage{multirow}

\newcommand{\cK}[0]{\mathcal K}

\newcommand{\cM}[0]{\mathcal M}

\newcommand{\df}[0]{\mathrm{df}}
\newcommand{\iso}[0]{{\rm iso}}

\newcommand{\Kisozero}[0]{{\cK_{\df,3}^{\iso,0}}}
\newcommand{\Kisoone}[0]{{\cK_{\df,3}^{\iso,1}}}
\newcommand{\Kdf}[0]{{\cK_{\df,3}}}

\newcommand{\kdf}{\mathcal{K}_{\text{df},3} }

\newcommand{\bm}[0]{\boldsymbol}

\newcommand{\Kisotwo}[0]{{\cK^{\iso,2}_{\df,3}}}

\newcommand{\Luscher}[0]{Luscher:1986n2,Luscher:1991n1}
\newcommand{\RG}[0]{Rummukainen:1995vs}
\newcommand{\KSS}[0]{Kim:2005gf}
\newcommand{\HSQCa}[0]{Hansen:2014eka}
\newcommand{\HSQCb}[0]{Hansen:2015zga}

\newcommand{\BHSQC}[0]{Briceno:2017tce}
\newcommand{\BHSnum}[0]{Briceno:2018mlh}
\newcommand{\BHSK}[0]{Briceno:2018aml}
\newcommand{\dwave}[0]{Blanton:2019igq}
\newcommand{\largera}[0]{Romero-Lopez:2019qrt}

\newcommand{\HHanal}[0]{Blanton:2019vdk}
\newcommand{\isospin}[0]{Hansen:2020zhy}

\newcommand{\BSnondegen}[0]{Blanton:2020gmf}

\newcommand{\Akakia}[0]{Hammer:2017uqm}
\newcommand{\Akakib}[0]{Hammer:2017kms}

\newcommand{\MDpi}[0]{Mai:2018djl}

\newcommand{\MD}[0]{Mai:2017bge}

\newcommand{\Akakinum}[0]{Doring:2018xxx}
\newcommand{\HSrev}[0]{Hansen:2019nir}

\newacronym{CMF}{CMF}{center-of-momentum frame}

\preprint{}

\title{ Interactions of two and three mesons including higher partial waves from lattice QCD}
\author[1]{Tyler D. Blanton}
\affiliation[1]{Physics Department, University of Washington, Seattle, WA 98195-1560, USA}
\author[2,3,4]{, Andrew D. Hanlon}
\affiliation[2]{Helmholtz-Institut Mainz, Johannes Gutenberg-Universit\"at, 55099 Mainz, Germany}
\affiliation[3]{GSI Helmholtzzentrum f\"ur Schwerionenforschung, 64291 Darmstadt, Germany}
\affiliation[4]{Physics Department, Brookhaven National Laboratory, Upton, New York 11973, USA}
\author[5]{, Ben H\"orz}
\affiliation[5]{Nuclear Science Division, Lawrence Berkeley National Laboratory, Berkeley, CA 94720, USA}
\author[6]{, Colin Morningstar}
\affiliation[6]{Department of Physics, Carnegie Mellon University, Pittsburgh, Pennsylvania 15213, USA}
\author[7]{, Fernando Romero-L\'opez}
\affiliation[7]{IFIC, CSIC-Universitat de Val\`encia, 46980 Paterna, Spain}
\author[1]{, and Stephen R. Sharpe}

\emailAdd{blanton1@uw.edu}
\emailAdd{ahanlon@bnl.gov}
\emailAdd{hoerz@lbl.gov}
\emailAdd{cmorning@andrew.cmu.edu}
\emailAdd{fernando.romero@uv.es}
\emailAdd{srsharpe@uw.edu}

\abstract{ 
We study two- and three-meson systems composed either of pions or kaons
at maximal isospin using Monte Carlo simulations of lattice QCD.
Utilizing the stochastic LapH method, we are able to determine hundreds of two- and three-particle energy levels, in nine different momentum frames, with high precision.
We fit these levels using the relativistic finite-volume formalism based on a generic effective field theory
in order to determine the parameters of the two- and three-particle K-matrices.
We find that the statistical precision of our spectra is sufficient to 
probe not only the dominant $s$-wave interactions, but also those in $d$ waves.
In particular, we determine for the first time
a term in the three-particle K-matrix that contains two-particle
$d$ waves. 
We use three $N_f=2+1$ CLS ensembles with pion masses of $200$, $280$, and $340\;$MeV.
This allows us to study the chiral dependence of the scattering observables, 
and compare to the expectations of chiral perturbation theory.
}

\allowdisplaybreaks
\begin{document}

\maketitle
\flushbottom
\clearpage

\section{Introduction}
 \label{sec:intro}



The era of three-particle spectroscopy in lattice QCD (LQCD) has recently begun, 
with several studies of three-meson systems at maximal isospin appearing in the literature~\cite{Beane:2007es,Horz:2019rrn,Blanton:2019vdk,Mai:2019fba,Culver:2019vvu,Fischer:2020jzp,Hansen:2020otl,Alexandru:2020xqf,Beane:2020ycc,Brett:2021wyd}.
These works make use of advances in numerical methods~\cite{Morningstar:2011ka,Peardon:2009gh} for determining the finite-volume spectrum from LQCD, as well as
in the theoretical formalism needed to relate the spectrum to infinite-volume scattering amplitudes~\cite{Polejaeva:2012ut,\HSQCa,\HSQCb,\BHSQC,\BHSnum,\BHSK,\dwave,\largera,Hansen:2020zhy,Blanton:2020gmf,Blanton:2020gha,Hansen:2021ofl,Blanton:2021mih,\Akakia,\Akakib,\Akakinum,Romero-Lopez:2018rcb,Pang:2019dfe,Romero-Lopez:2020rdq,Muller:2020wjo,Muller:2020vtt,\MD,\MDpi,Guo:2017ism,Klos:2018sen,Guo:2018ibd,Pang:2020pkl}
(the latter reviewed in Refs.~\cite{\HSrev,Mai:2021lwb}).
These studies have shown the expected result that the major determinant of the three-particle spectrum is the interaction between pairs of particles. By contrast, the determination of three-particle scattering quantities has been found to be more challenging---their contribution is suppressed by an additional volume factor. 
Indeed, while some work finds evidence for a 
nonzero three-particle interaction~\cite{Blanton:2019vdk,Fischer:2020jzp}, 
other studies find no significant signal~\cite{Brett:2021wyd,Alexandru:2020xqf,Hansen:2020otl}.

In this paper, we study two- and three-particle systems composed either of pions
or kaons at maximal isospin, specifically $2\pi^+$, $3\pi^+$, $2K^+$, and $3K^+$.
We aim to significantly advance the study of multiparticle systems 
by including a much larger number of frames and a range of quark masses, 
and employing state-of-the-art methods, such as the stochastic LapH method~\cite{Morningstar:2011ka},
to obtain excited spectral levels with increased precision.
We use the three-particle formalism developed in the generic relativistic field theory (RFT) approach~\cite{\HSQCa,\HSQCb}, 
which is the only method that has been explicitly worked out including higher partial waves~\cite{\dwave}. 
With hundreds of levels available---an increase of about an order of magnitude over previous work---we are able to determine elements of the three-particle K-matrix
that were previously unexplored.

We use three  $N_f=2+1$ CLS ensembles that follow a chiral trajectory where the trace of the quark mass matrix is kept constant. The lightest pion mass is $M_\pi \simeq 200$~MeV, and the corresponding kaon mass is $M_K\simeq 480$~MeV. 
This allows us to study the dependence on quark (or pion/kaon) masses of the various scattering observables, allowing a comparison with the expectations of chiral perturbation theory (ChPT).  
As will be seen later, our results also indicate that we need to include not only the leading-order $s$-wave interactions, but also $d$-wave two-particle interactions as well as
 three-particle interactions in which pairs are in a relative $d$ wave.\footnote{%
 We stress that the total angular momentum of the three-particle interaction is not required to be nonzero
 when a pair sub-interaction is in a $d$ wave, 
 because the relative angular momentum between the pair and  the third particle can be nonzero.}

This paper is organized as follows. 
In \Cref{sec:lattice}, we describe how we determine the finite-volume spectrum using LQCD.
\Cref{sec:QC} contains a compilation of necessary theoretical background: 
the finite-volume formalism, the strategy for fitting the spectrum with the quantization conditions,
the parametrizations that we use for two- and three-particle K-matrices, 
and the results from ChPT to which we compare.
We present our fits of the quantization conditions to the spectrum in \Cref{sec:results}, 
and interpret the results in \Cref{sec:discussion}. 
We summarize our main conclusions in \Cref{sec:conc}.  
\Cref{app:A} collects some necessary group-theoretical results, \Cref{app:B} summarizes the operators used in this work, and \Cref{app:C} lists the energy levels used in our fits.

\section{ Computation of finite-volume energies}
 \label{sec:lattice}

 The computational methods follow closely those employed in Ref.~\cite{Horz:2019rrn}, and we review only the high-level features in this section.
 \subsection{Interpolating operators}
 \label{ssec:interps}
 In order to extract the finite-volume spectrum reliably from LQCD, it is imperative to include interpolating operators with good overlap onto the states of interest.
 Operators annihilating a two-pion and three-pion state are given by a sum over products of single-pion annihilation operators, each projected to definite momentum $\textbf{p}_i$, 
\begin{align}
  \pi \pi^{(\textbf{P}, \Lambda)} &= c^{(\textbf{P},\Lambda)}_{\textbf{p}_1,\textbf{p}_2} \pi_{\textbf{p}_1} \pi_{\textbf{p}_2}, \label{eqn:twopi} \\
  \pi \pi \pi^{(\textbf{P}, \Lambda)} &= c^{(\textbf{P},\Lambda)}_{\textbf{p}_1,\textbf{p}_2,\textbf{p}_3} \pi_{\textbf{p}_1} \pi_{\textbf{p}_2} \pi_{\textbf{p}_3}.  \label{eqn:threepi}
\end{align}
The corresponding creation operators are obtained by taking the Hermitian conjugate of the annihilation operators.
The momentum combinations encoded in the Clebsch-Gordan coefficients $c^{(\textbf{P}, \Lambda)}$ are chosen such that the resulting operators transform according to the irreducible representation (irrep) $\Lambda$ of the little group of the total momentum $\textbf{P} = \sum_i \textbf{p}_i$.
In this basis, the finite-volume Hamiltonian assumes a block-diagonal form, reflecting the symmetry of the cubic spatial volume or its boosted deformations, thus greatly simplifying the extraction of the spectrum.

In addition to the total momenta $\textbf{d}^2 = L^2 / (2 \pi)^2 \textbf{P}^2 \le 4$ commonly used in previous work, total momenta up to $\textbf{d}^2 \le 9$ are included in the present work (excluding momentum of type $\textbf{d}=[122]$).
In particular, the trivial irrep with total momentum of type $\textbf{d}=[003]$ proves to be useful in the three-particle sector, since it provides an additional data point in the threshold region purely on kinematical grounds.
The group-theoretical projection of meson-meson operators proceeds along the lines of Refs.~\cite{Basak:2005aq,Basak:2005ir,Morningstar:2013bda}.
Our choices for representation matrices of the elements of the little group of the newly included momentum classes $\textbf{d} = [012]$ and $\textbf{d} = [112]$ are given in \Cref{app:irrepconventions}.
Three-pion operators are then obtained by iteratively applying the two-meson coupling procedure~\cite{Horz:2019rrn,Woss:2019hse}, and all interpolators used in this work are tabulated in \Cref{app:interpolators}.

Two-kaon and three-kaon annihilation operators are given by replacing the constituent pion annihilation operators with kaon annihilation operators in~\Cref{eqn:twopi,eqn:threepi}, with the Clebsch-Gordan coefficients unchanged.

\subsection{Correlation functions}
Calculating correlation functions of operators of the form of~\Cref{eqn:twopi,eqn:threepi} requires a method to handle quark propagation from all spatial sites on the source time slice to all spatial sites on the sink time slice in order to be able to perform the momentum projection for each constituent hadron individually.
Such a method is furnished by \emph{distillation} \cite{Peardon:2009gh}, which treats quark propagation in a low-dimensional subspace that encodes the information relevant for hadronic physics.

The distillation subspace is spanned by the $N_\mathrm{ev}$ lowest-lying eigenvectors of the three-dimensional gauge-covariant Laplacian on each time slice of the lattice.
The projection of quark fields into this subspace is equivalent to a smearing of the quark fields with an approximately Gaussian profile having a characteristic smearing radius.
In order to keep the smearing width constant as the physical spatial volume of the lattice is increased, the number of retained eigenvectors needs to be scaled linearly with the volume.
The concomitant increase in computational cost can be ameliorated by employing stochastic estimators in the distillation subspace~\cite{Morningstar:2011ka}.
In this so-called \emph{stochastic LapH method}, a valence quark line is estimated stochastically,
\begin{align}
Q_{a\alpha,b\beta}(x,y) = \lim_{N_r \rightarrow \infty} \frac{1}{N_r} \sum_{r,d}\phi^{(r,d)}_{a\alpha}(x) \rho_{b\beta}(y)^{(r,d)*} \label{e:qprop},
\end{align}
using $N_r$ independent sets of diluted~\cite{Foley:2005ac} stochastic combinations of LapH eigenvectors as sources $\rho(y)$ and corresponding solutions of the Dirac equation $D \phi = \rho$, where $r$ and $d=1,\dots,N_\mathrm{dil}$ denote the noise and dilution indices, respectively, $a, b$ are color indices, and $\alpha, \beta$ are Dirac spin indices.

This method to treat quark propagation affords great flexibility as the computationally-expensive solutions of the Dirac equation can be re-used across several spectroscopy projects.
In a subsequent step, quark sources and solutions are combined into meson functions with support only on the source or sink time slice~\cite{Morningstar:2011ka}.
The meson functions are rank-two tensors with two open dilution indices and a compound label identifying their spin and spatial structure, meson momentum, and pair of quark noises.

The final step in computing correlation functions consists of performing tensor contractions over dilution indices of sets of meson functions according to the Wick contractions of quark fields \cite{Morningstar:2011ka}, and forming the linear combinations of individual momentum assignments governed by the group-theoretical projections discussed in \Cref{ssec:interps}.
This step becomes more computationally expensive as systems of an increasing number of valence quarks are considered, and with a naive implementation the associated cost completely dominates that of the whole calculation already for three-meson systems.
A significant speedup can be achieved by systematically eliminating all redundant computation through \emph{common subexpression elimination}~\cite{10.1007/11758501_39}. The application of this idea in the present LQCD context was described in Ref.~\cite{Horz:2019rrn}.
The  publicly available implementation\footnote{\url{https://github.com/laphnn/contraction_optimizer}} is not restricted to systems of mesons and was used recently to speed up a calculation of two-baryon systems by nearly three orders of magnitude~\cite{Horz:2020zvv}. In the present work, these improvements enable the computation of up to 20,000 distinct correlation functions per gauge configuration and source time, encompassing the evaluation of up to one billion individual diagrams (as defined in Ref.~\cite{Horz:2019rrn}).

 \subsection{Ensemble details}
 \begin{table}
   \centering
   \tabcolsep=0.15cm
   \begin{tabular}{c c c c c c c c c}
     \toprule
     & $(L/a)^3 \times (T/a)\phantom{^3}$ & $M_\pi \, [\mathrm{MeV}]$ & $M_K \, [\mathrm{MeV}]$ & $N_\mathrm{cfg}$ & $t_\mathrm{src}$ & $N_\mathrm{ev}$ & dilution & $N_r$(l/s) \\
     \midrule
     N203 & $48^3 \times 128$ & 340 & 440 & 771 & 32, 52 & 192 & (LI12,SF) & 6/3 \\
     N200 & $48^3 \times 128$ & 280 & 460 & 1712 & 32, 52 & 192 & (LI12,SF) & 6/3 \\
     D200 & $64^3 \times 128$ & 200 & 480 & 2000 & 35, 92 & 448 & (LI16,SF) & 6/3 \\
     \bottomrule
   \end{tabular}
   \caption{Overview of the lattice geometry, approximate pseudoscalar masses, number of gauge configurations $N_\mathrm{cfg}$, fixed source-time positions $t_\mathrm{src}$, number of Laplacian eigenvectors $N_\mathrm{ev}$, employed dilution scheme, and number of independent quark noises $N_r$ used to estimate light and strange quark propagation for ensembles used in this work. The dilution scheme notation is explained in Ref.~\cite{Morningstar:2011ka}. All ensembles share the same lattice spacing $a \approx 0.064 \, \mathrm{fm}$.}
   \label{tab:ensems}
 \end{table}
 Calculations in this study are performed on three ensembles at a fixed lattice spacing generated through the CLS effort \cite{Bruno:2014jqa}.
 The simulations use $N_\mathrm{f}=2+1$ nonperturbatively $O(a)$-improved Wilson fermions and the L\"uscher-Weisz gauge action with tree-level coefficients.
 They are performed along a chiral trajectory keeping the trace of the quark mass matrix fixed, so a heavier-than-physical light quark mass implies a lighter-than-physical strange quark mass.
 An overview of the three ensembles used in this work as well as the computational setup is given in \Cref{tab:ensems}.
 The lattice spacing on these ensembles was determined to be $a = 0.06426(76) \, \mathrm{fm}$ using the linear combination $\frac{2}{3}(F_K + \frac{1}{2} F_\pi)$ of decay constants to set the scale \cite{Bruno:2016plf}.
 In addition, the decay constants used in this work were computed in Ref.~\cite{decay_constants} and are reproduced in~\Cref{tab:decay_constants} for convenience.
\begin{table}
   \centering
   \begin{tabular}{c c c }
     \toprule
     & $ M_\pi / F_\pi$  & $M_K/F_K$ \\
     \midrule
     N203 & 3.4330(89) & 4.1530(72) \\
     N200 & 2.964(10) & 4.348(11) \\
     D200 & 2.2078(67) & 4.5132(93) \\
     \bottomrule
   \end{tabular}
   \caption{Pion and kaon decay constants determined in Ref.~\cite{decay_constants}.}
   \label{tab:decay_constants}
\end{table}
 Open temporal boundary conditions were imposed when generating the ensembles to avoid topological charge freezing at fine lattice spacings \cite{Luscher:2012av}.
 Consequently, translation invariance in the temporal direction is broken, and the position of source operators is fixed to the values given in \Cref{tab:ensems} rather than being randomized on each gauge configuration.
 The effect of the boundary conditions on spectral quantities is expected to decay exponentially with the distance from the boundary, with the decay constant governed by the lightest state with vacuum quantum numbers,  which is expected to be a two-pion state for the quark masses used in this work.
 They are thus expected to be most pronounced on the ensemble with the lightest pion mass, the D200 ensemble in the set used in this work.
 In a previous study on the same ensemble, $t_\mathrm{src} = 32$ was found to have negligible temporal boundary effects \cite{Andersen:2018mau}.
 The sources placed in the bulk of the lattice for this study are thus expected to be sufficiently far away from the boundary.
 Additionally, the correlators are always constructed such that the sink times are toward the center of the lattice with respect to the source position.
 Therefore, the additional source at $t_\mathrm{src} = 92$ for the D200 ensemble required backward-time correlators. The operators used for the backward-time correlators are related to the operators in the forward-time correlators by a parity and charge-conjugation transformation.

 Except for a twofold increase in statistics on the D200 ensemble and the use of an improved dilution scheme on the N200 ensemble, solutions of the Dirac equation for the light quark are re-used from a previous spectroscopic calculation supporting the lattice determination of the hadronic vacuum polarization contribution to the anomalous magnetic moment of the muon \cite{Gerardin:2019rua}.

\subsection{Analysis of correlation functions}

Matrices of correlation functions are computed for a wide range of total momenta and irreps.
In a first step, the data is averaged over equivalent momenta, irrep rows, and source times on each gauge-field configuration.
The subsequent analysis is performed using jackknife resampling and for each total momentum-squared/irrep separately.
We provide the jackknife samples of the resulting spectrum in HDF5 format as ancillary files with the arXiv submission.

\subsubsection{Single-hadron energies}
\begin{table}
   \centering
   \tabcolsep=0.14cm
   \begin{tabular}{c | c c c c | c c c c }
     \toprule
     & $a M_\pi$ & $M_\pi L$ & $t_{\rm fit}$ & $\chi^2_{\rm red}$ & $a M_K$ & $M_K L$ & $t_{\rm fit}$ & $\chi^2_{\rm red}$ \\
     \midrule
     N203 & 0.11261(20) & 5.4053(96) & $[20-40]$ & 3.20 & 0.14392(15) & 6.9082(72) & $[16-40]$ & 3.49 \\
     N200 & 0.09208(22) & 4.420(11) & $[15-36]$ & 1.62 & 0.15052(14) & 7.2250(67) & $[17-40]$ & 1.89 \\
     D200 & 0.06562(19) & 4.200(12) & $[18-40]$ & 1.54 & 0.15616(12) & 9.9942(77) & $[23-40]$ & 1.42 \\
     \bottomrule
   \end{tabular}
   \caption{Single-hadron energies at rest, determined from single-exponential fits.
       The range of time separations included in the fit is given by $t_{\rm fit}$.
       The high $\chi^2_{\rm red}$ for N203 is discussed in the text.}
   \label{tab:single-hadron-energies}
\end{table}
Pion and kaon correlators are computed in each momentum frame, and then used to extract the single-hadron energies for each total momentum-squared.
We employ a single-exponential correlated-$\chi^2$ fit to these correlation functions.
The results for the kaon and pion energies at rest are given in \Cref{tab:single-hadron-energies}.
{ The high $\chi^2_{\rm red} \equiv \chi^2/\text{dof}$ for the single-hadron fits on N203 could be reduced by rebinning the data (as was done on N200 and D200).}
However, due to the smaller number of configurations for N203, rebinning of the data leads to an unstable covariance matrix when fitting the full set of multi-hadron energies to the quantization condition (an issue discussed further below).

\begin{figure}
\begin{subfigure}{.32\textwidth}
  \centering
  \includegraphics[width=\linewidth]{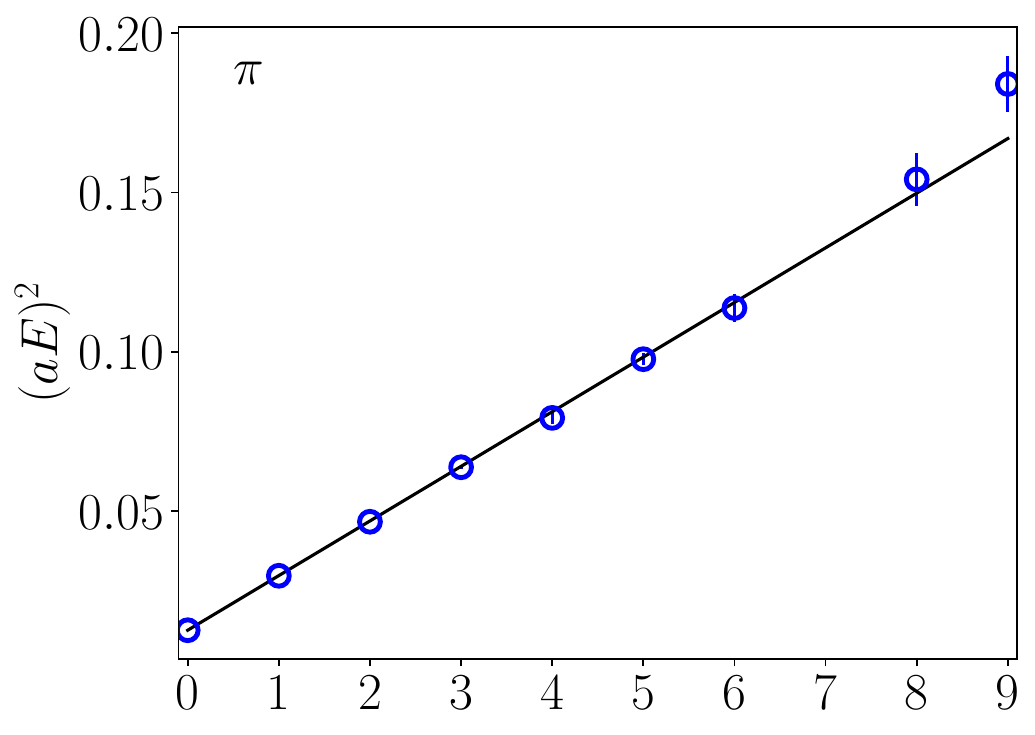}  
  \includegraphics[width=\linewidth]{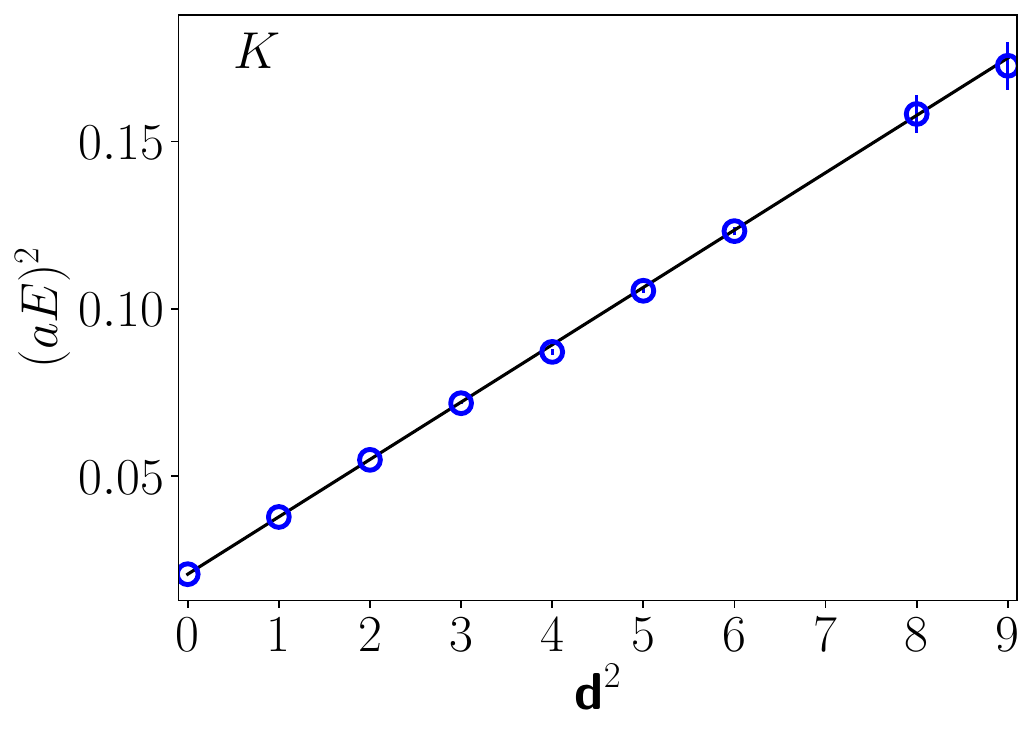}  
  \caption{N203}
  \label{fig:disp-n203}
\end{subfigure}
\begin{subfigure}{.32\textwidth}
  \centering
  \includegraphics[width=\linewidth]{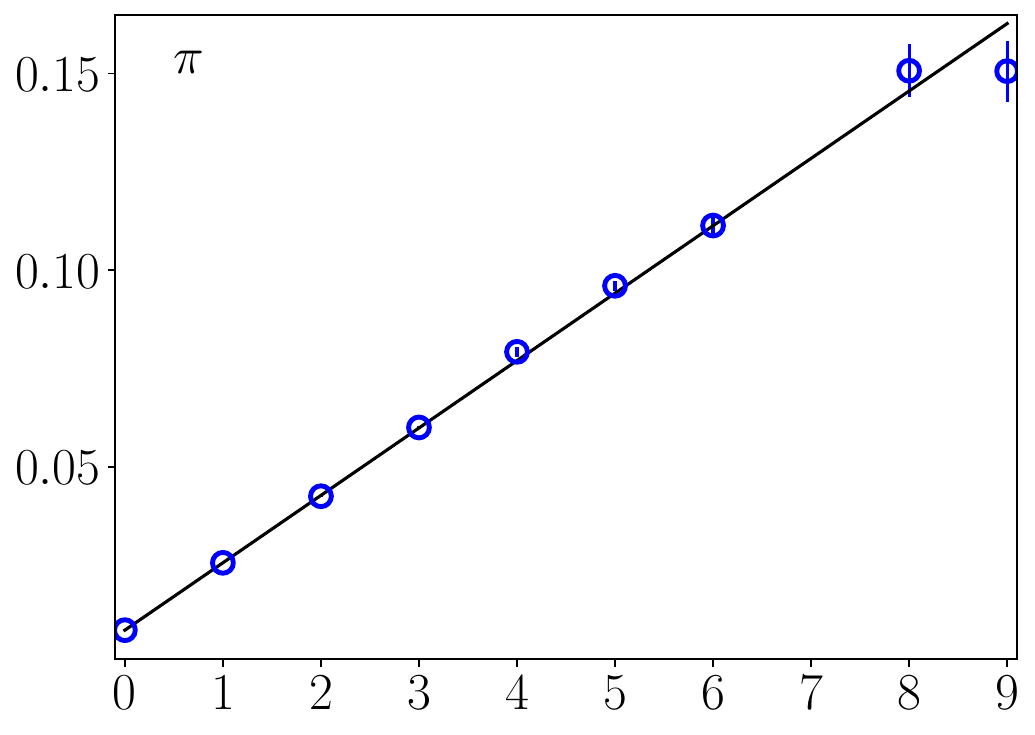}  
  \includegraphics[width=\linewidth]{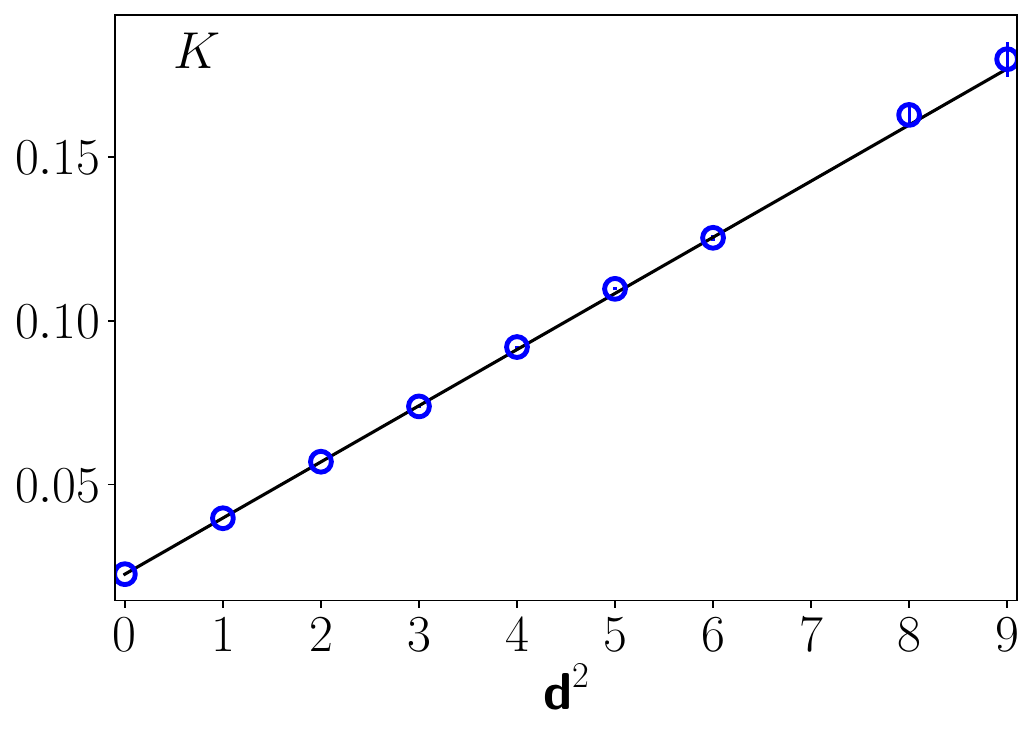}  
  \caption{N200}
  \label{fig:disp-n200}
\end{subfigure}
\begin{subfigure}{.32\textwidth}
  \centering
  \includegraphics[width=\linewidth]{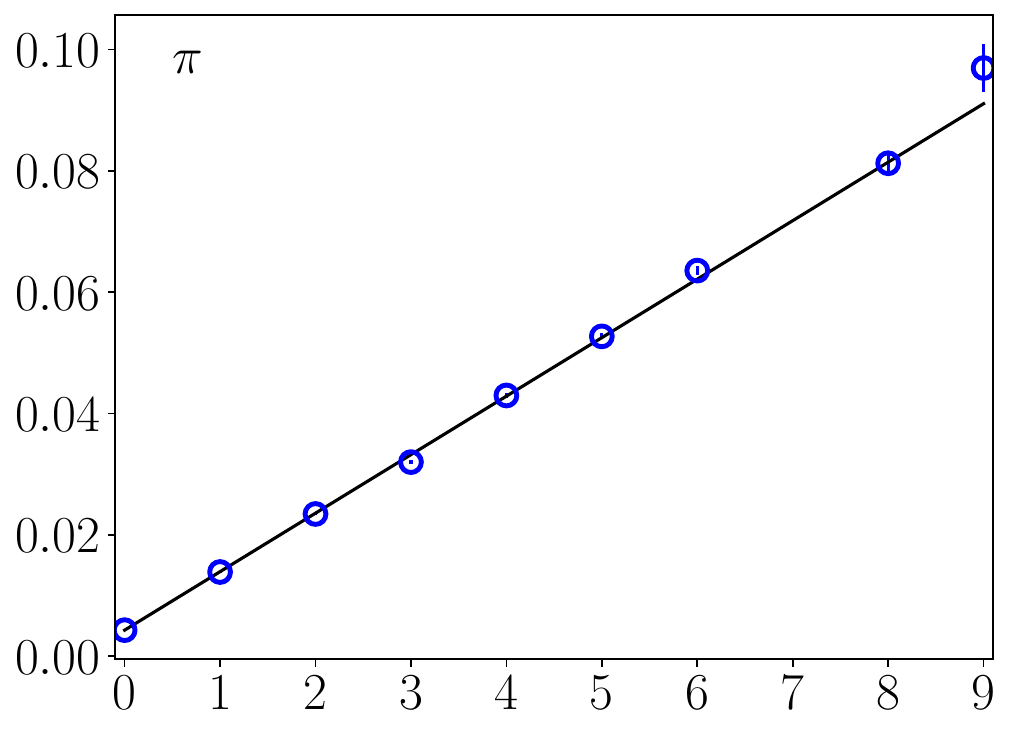}  
  \includegraphics[width=\linewidth]{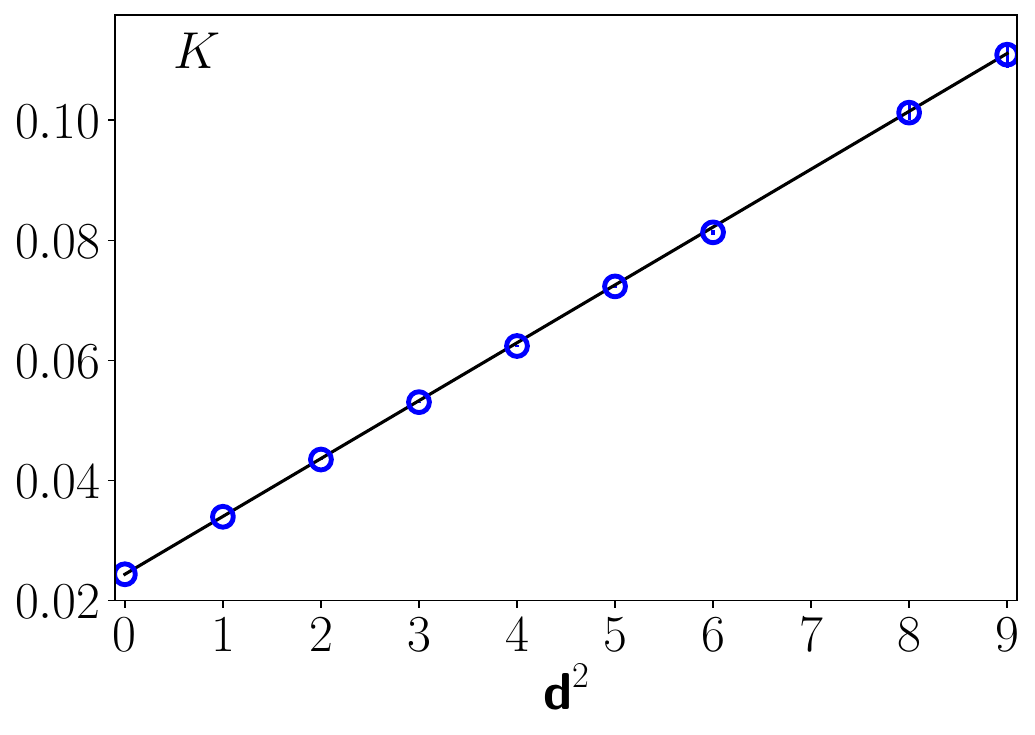}  
  \caption{D200}
  \label{fig:disp-d200}
\end{subfigure}
\caption{Single-pion (top row) and single-kaon (bottom row) energies vs. $\textbf{d}^2$ for the frames we use.
The continuum dispersion relations from the energies at rest are shown as black lines.}
\label{fig:disp}
\end{figure}

When fitting the multi-hadron spectrum to the quantization condition, the continuum dispersion relation is assumed to be valid for the pion and kaon, and therefore only the energies for the single hadrons at rest are actually needed.
We can test this assumption by studying the dispersion relation of our single-hadron states.
The results, shown in \Cref{fig:disp}, show no appreciable disagreement with the continuum dispersion, suggesting that discretization effects are small.

\subsubsection{Multi-hadron spectrum extraction}
In order to extract not only ground states but also a tower of excited states in each irrep, the eigenvectors ${v_n}$ determined from the generalized eigenvalue problem (GEVP) \cite{Michael:1985ne,Luscher:1990ck}
\begin{align}
 C(t_d) v_n = \lambda_n C(t_0) v_n ,
 \label{eqn:fixedgevp}
\end{align}
are used to form the correlation functions of rotated operators with `optimal' overlap onto the $n$th state in the spectrum,
\begin{align}
 \hat C_n(t) = (v_n, C(t) v_n) ,
\end{align}
where the parentheses denote an inner product over the operator indices.
The diagonalization is performed for a single $(t_0, t_d)$ only, keeping $t_0 \gtrsim t_d/2$ \cite{Blossier:2009kd}, but we checked for a range of sensible values that the resulting spectrum is independent of that choice.

Two- and three-pion finite-volume energies are extracted from single-exponential corre-lated-$\chi^2$ fits to the ratios
\begin{align}
  R_n(t) = \frac{\hat C_n(t)}{\prod_i C_\pi(\textbf{p}_i^2, t)},
  \label{eqn:fitratio}
\end{align}
%
where the product in the denominator is over two or three single-pion correlation functions, respectively, with momenta chosen to match the closest noninteracting energy to the $n$th state.
The ratio \Cref{eqn:fitratio} gives access to the energy splitting between the interacting and noninteracting state at sufficiently large time separation, and the lattice energy is reconstructed using the single-pion mass and dispersion relation.
Fits are performed to the ratio data in the time separation range $[t_\mathrm{min}-t_\mathrm{max}]$ with $t_\mathrm{max} = 40$ fixed on all ensembles (with the exception of the pion spectra on N200, for which $t_\mathrm{max} = 36$ led to a more consistent analysis).
The lower bound of the fit window is chosen such that the residual excited-state contamination is subdominant compared to the statistical uncertainty.
The analysis of the two-kaon and three-kaon data proceeds analogously with the product in the denominator of \Cref{eqn:fitratio} replaced with single-kaon correlation functions.

\begin{figure}
\begin{subfigure}{.32\textwidth}
  \centering
  \includegraphics[width=1.00\linewidth]{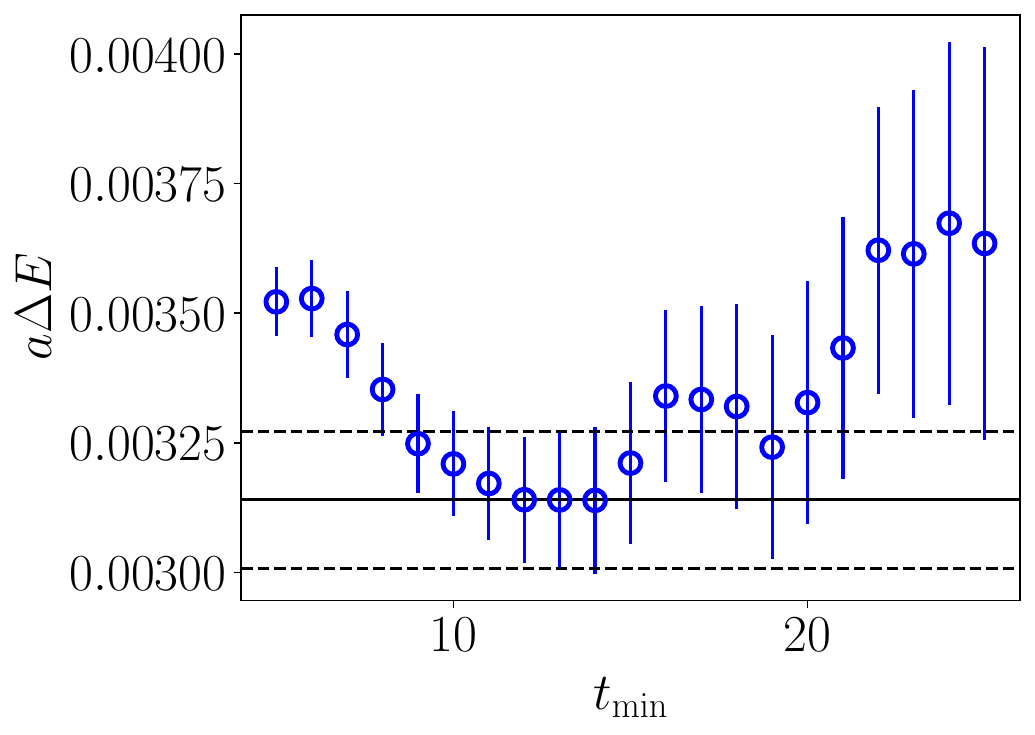}  
  \caption{N203, $t_{\rm min} = 13$}
  \label{fig:tmin_N203}
\end{subfigure}
\begin{subfigure}{.32\textwidth}
  \centering
  \includegraphics[width=1.00\linewidth]{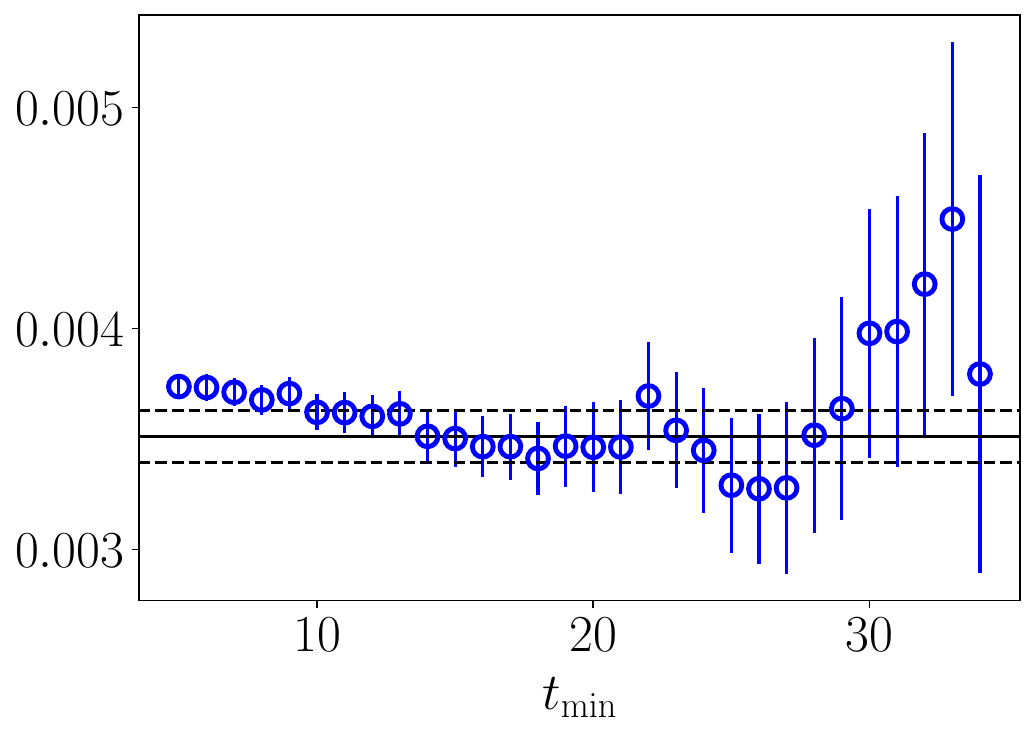}  
  \caption{N200, $t_{\rm min} = 14$}
  \label{fig:tmin_N200}
\end{subfigure}
\begin{subfigure}{.32\textwidth}
  \centering
  \includegraphics[width=1.00\linewidth]{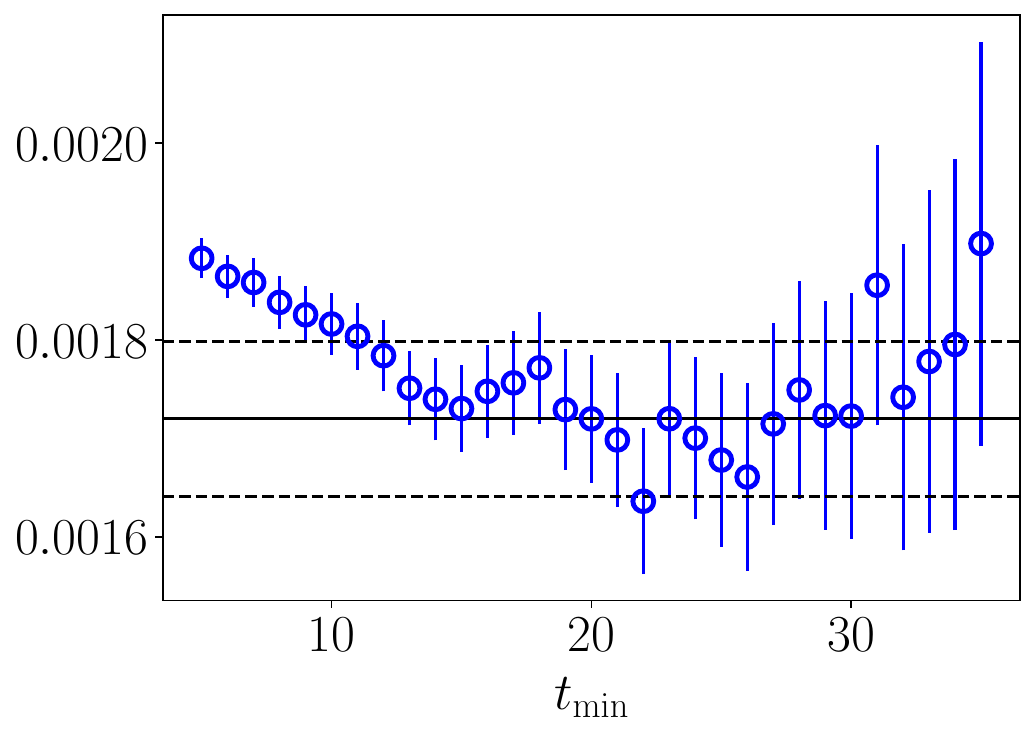}  
  \caption{D200, $t_{\rm min} = 23$}
  \label{fig:tmin_D200}
\end{subfigure}
\caption{The extracted energy shift in the lab frame as a function of the smallest time separation included in the fit, $t_{\rm min}$, for the three-kaon $A_2 (9)$ ground state energy on each ensemble.
  The fits are to the correlator ratio, and a single-exponential is used for the model.
  The central value and error of the energy shift for the chosen fit is indicated with the horizontal black solid and dashed lines, respectively.}
\label{fig:tmin}
\end{figure}

Fitting to the ratio of correlators, rather than directly to the rotated correlator, generally allows for an earlier choice of $t_{\rm min}$ and thus a more precise energy extraction.
{ The dependence of these fits on the choice of $t_{\rm min}$ is shown in \Cref{fig:tmin} for the three-kaon $A_2(9)$ ground-state energy on each ensemble.}
The earlier plateau in the correlator ratio can be understood as coming from a partial cancellation of correlations and excited states between the numerator and denominator.
However, one disadvantage is the loss of a monotonic decrease in the effective energy of the correlator ratio.
Therefore, we verify consistency between the results from the ratio fit and direct fits to the rotated correlators.

\begin{figure}
  \centering
  \includegraphics{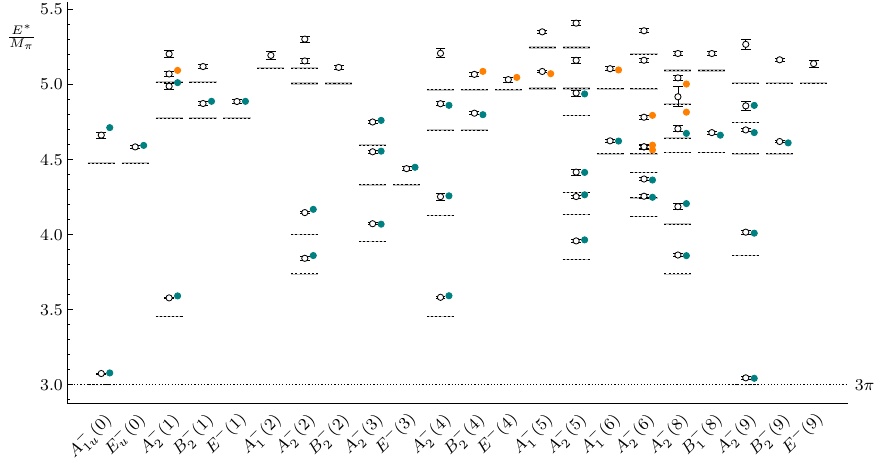}
  \includegraphics{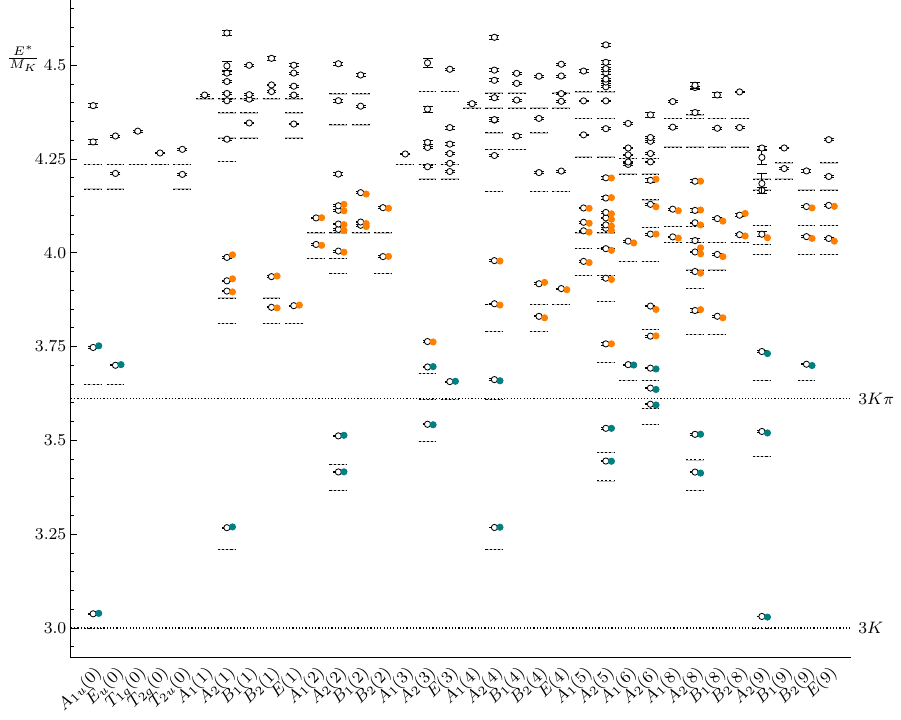}
  \caption{Center-of-mass energies of three pions~(top) and  three kaons~(bottom) on N200.
  The various irreps and momentum-squared are listed at the bottom.
  The dashed lines specify the free energy levels, and the open circles denote the interacting energies.
  Free energy levels may be degenerate---see~\Cref{app:B}.
  Several thresholds are indicated with dotted lines.
  The colored circles indicate the central values of the resulting energies from a global fit to the two- and three-particle quantization condition;
  we use the parameters from fit 3 of Table~\ref{tab:fitpiN200} and fit 4 of Table~\ref{tab:fitN200K} for the pion and kaon spectra, respectively.
  The teal and orange circles correspond to energies included and not included, respectively, in the fit.}
  \label{fig:n200_three_pion_spectrum}
\end{figure}

The resulting energy splitting $a \Delta E$ from the fits to the correlator ratio are converted to absolute energies in the center-of-momentum frame $a E^\ast$ by using the extracted single-hadron energies at rest with the continuum dispersion relation.
All of the extracted three-kaon and three-pion energies on N200 are shown in \Cref{fig:n200_three_pion_spectrum}, along with the energies resulting from a global fit using the two- and three-particle quantization condition, to be described below.

The global fits account for the correlations between all levels, i.e., those with both two and three particles of a given type.
Since there are many levels in a given fit (up to 77) one might be concerned about the reliability of our determination of the covariance matrix.
To assess this, we compared fits
to the spectrum with both bootstrap and jackknife resampling and found
them to be consistent. On N200 and D200, we also found consistency for
the central values of our fits when rebinning the data by 2 and 3,
with the errors increasing due to autocorrelations. On N203, we found
that rebinning the data leads to inconsistent fits, which can be
explained by the much smaller number of configurations as compared to N200
and D200.
In fact, using a smaller subset of energies on N203, such that the covariance matrix is stable with the rebinned data, still gives consistent results with the same subset of energies without rebinning,
suggesting the rebinning is not necessary in the final analysis on N203.
With our final choices of rebinning (3 for both N200 and
D200, and none for N203), the number of bins is roughly consistent
across all ensembles. This is approximately an order of magnitude
larger than the number of energies going into our largest fits. All of
our consistency checks suggest this is sufficient for a reliable
estimate of the covariance matrix.



\section{Formalism and fitting strategy}
\label{sec:QC}

In this section we summarize the formalism that we use to fit the spectrum, 
describe our overall fitting strategy, and present the
parametrizations of the K-matrices that are used to describe the two- and three-particle
interactions. In addition we collect various results from ChPT
that are needed to analyze the mass dependence of our results.

\subsection{Quantization conditions}

To relate the finite-volume spectrum to infinite-volume scattering parameters, we 
need both two- and three-particle quantization conditions. 
The former is standard---the original L\"uscher quantization condition~\cite{\Luscher}
 generalized to moving frames in a relativistic formalism~\cite{\RG,\KSS}.
For the latter, we use the results of the relativistic field theory (RFT) approach.
This formalism, developed in Ref.~\cite{\HSQCa}, holds for
three identical, spinless particles within the kinematic range for which
only a single three-particle channel can go on shell.
It applies for relativistic particles, and leads to Lorentz-invariant
scattering amplitudes
as long as one uses the relativistic form of the kinematic functions (a point discussed
extensively in Ref.~\cite{\BSnondegen}).
Practical implementation of the quantization condition requires truncating the angular
momentum of the interactions between pairs of particles. We follow Refs.~\cite{\dwave,\largera}
and include both $s$ and $d$ waves ($p$ waves being forbidden for identical particles).
Where we break new ground is the inclusion of $s$ and $d$ waves in moving frames;
previous work in the RFT approach
with moving frames was restricted to $s$ waves~\cite{\HHanal}.

We present here only a summary of the formalism and its implementation,
since most of the details are the same as in Refs.~\cite{\dwave,\largera,\HHanal}.
The formalism is derived for a
 generic
continuum effective field theory 
restricted to a cubic spatial box  of length $L$.
Thus the allowed total momenta are drawn from the finite volume set:
$\bm P = (2\pi/L) \bm d$, where $\bm d \in \mathbb Z^3$.
For a given choice of $\bm P$ and $L$,
the two-particle spectrum is given by the energies $E_2$ that solve
\begin{equation}
\det \left[F(E_2,\bm P,L)^{-1} + \cK_2(E_2^*) \right] = 0\,,
\label{eq:QC2}
\end{equation}
where $E_2^*=\sqrt{E_2^2-\bm P^2}$ is the center-of-mass frame (CMF) energy,
$F$ is a kinematical function to be discussed shortly,
and $\cK_2$ is the two-particle K-matrix given in \Cref{eq:K2def} below.
The three-particle spectrum is given by the energies $E$ that solve
\begin{equation}
\det \left[F_3(E,\bm P,L)^{-1} + \cK_{\df,3}(E^*) \right] = 0\,,
\label{eq:QC3}
\end{equation}
where $E^*=\sqrt{E^2-\bm P^2}$,
$F_3$ will be discussed shortly,
and $\cK_{\df,3}$ is a three-particle K-matrix discussed around
\Cref{eq:Kdfthreshold}.
The quantization conditions hold up to exponentially suppressed corrections,
which should be small for our values of $M_\pi L$ and $M_K L$.

Although the two quantization conditions have a similar form, they differ substantially in the details.
The determinant in \Cref{eq:QC2} runs over indices $\{\ell,m\}$, which denote the
angular momentum of the two particles in their CMF,
while that in \Cref{eq:QC3} runs over $\{k\ell m\}$, where $k$ 
represents the three-momentum of one of the particles, $\bm k$, which is drawn from
the finite-volume set, while $\{\ell,m\}$ are the
relative angular momenta of the other two particles in their CMF.
The formalism has a built-in cutoff, such that only a finite number of values of $\bm k$ contribute,
but the sum over $\{\ell,m\}$ in both quantization conditions must be truncated by hand.
As noted above, here we consider $\ell_{\rm max}=2$.

Another difference between \Cref{eq:QC2,eq:QC3} concerns the first entry
in the determinants. For two particles, the matrix $F$ is a purely kinematic function
(a ``L\"uscher zeta function''), which encodes the effect of working in finite volume,
and, for a general moving frame, couples $s$ and $d$ waves.
We use the form given in Appendix A of Ref.~\cite{\dwave}, extended to moving frames.
For three particles, $F_3$ contains not only kinematical functions 
($F$, together with an additional function, $G$), but also the
two-particle K-matrix. Again, the details are given in Ref.~\cite{\dwave}, except
here extended to moving frames. The extension of the implementation of the
$s$-wave-only formalism to moving frames is described in the supplementary material
to Ref.~\cite{\HHanal}. The generalization to include $d$ waves is straightforward.

The final difference between the two quantization conditions concerns the
second term in the determinants. For two particles, the $\cK_2$ that appears
is a version of the two-particle K-matrix, differing from the standard choice by some
additional cutoff terms (discussed below). It is an infinite-volume quantity that is 
algebraically related to the physical two-particle scattering amplitude $\cM_2$. 
For the three-particle quantization condition, what appears instead is
a three-particle K-matrix, $\cK_{\df,3}$, which is an infinite-volume but cutoff-dependent
amplitude, related to the physical three-particle amplitude $\cM_3$ through integral
equations~\cite{\HSQCb}. What matters here is that $\cK_{\df,3}$ is a real, analytic
function of Lorentz invariants, and thus can be simply parametrized, as discussed below.

Both quantization conditions can be block-diagonalized by projecting onto 
 irreps of
the appropriate symmetry group for the given choice of $\bm P$.
This is the little group LG($\bm P$), composed of elements of the cubic group $O_h$ 
that leave $\bm P$ invariant. We implement this projection using the
formalism developed in Ref.~\cite{\dwave}, extended in Ref.~\cite{\HHanal}, and here
generalized to allow for the inclusion of two-particle $d$ waves in moving frames.
For given choices of $\cK_2$ and $\cK_{\df,3}$,
the solutions to the quantization condition are obtained by tracking the small eigenvalues of
the matrices in the determinants, irrep by irrep,
and numerically determining where they cross zero. 
This procedure has been independently implemented in two Python codes, and in
a (much slower) Mathematica code, so that all numerical results 
presented below have been cross-checked.

As noted in previous implementations of the RFT approach, there are choices of the
functions $\cK_2$ and $\cK_{\df,3}$, as well as of $M_\pi L$ and $M_K L$,
for which there are unphysical solutions to the
three-particle quantization condition~\cite{\BHSnum,\dwave}.
Their source is not fully understood, but is likely due to a combination of
numerically enhanced exponentially suppressed errors (which are not controlled in the
derivation) and the use of overly restrictive choices for $\cK_2$ and $\cK_{\df,3}$.
 We have checked that, for the fits presented in the next section,
there are no unphysical solutions in the energy range considered.

Another source of error that we do not systematically control is due to discretization effects.
Strictly speaking, the quantization conditions hold only in the continuum limit, since we
are assuming relativistic dispersion relations for intermediate particles, and restrict
the two- and three-particle interactions using Lorentz invariance.
The potential size of discretization errors is unclear.
On the one hand,
the dispersion relations shown in \Cref{fig:disp} show no indication of large discretization effects,
consistent with the findings of
an investigation of resonant $I=1$ pion scattering on ensembles with the same action~\cite{Andersen:2018mau}.
On the other hand,
a recent study of two-baryon scattering amplitudes from
LQCD serves as a cautionary tale about potentially large discretization effects in these types of
calculations~\cite{Green:2021qol}.
What is clear is that more work, involving multiple lattice spacings, is needed to
demonstrate control over discretization effects.

\subsection{Fitting strategy}

We now describe the overall strategy used for fitting. 
An important consideration is that there is not a one-to-one relationship between the
energy levels and the K-matrices---this holds only for the two-particle spectrum in the approximation
of a single partial wave. Thus one must parametrize the K-matrices and perform a global fit
to the whole set of levels.
We perform such fits to the two-particle spectrum alone, 
and to the combined two- and three-particle spectra.
We note that, on a given ensemble, these two spectra are correlated, as are the individual levels,
and we perform a fully correlated fit. 
More precisely, our fits to the spectrum are carried out by  minimizing the following 
$\chi^2$ function with respect to the parameters in the K-matrices, $\{ p_n \}$,
\begin{equation}
\chi^2(\{ p_n \}) = \sum_{ij} \left(E_i - E_i^\text{QC}(\{ p_n \}) \right)C_{ij}^{-1} \left(E_j - E_j^\text{QC}(\{ p_n \})\right)\,,
\end{equation}
where $E_j$ are the measured energy levels, with covariance matrix $C$,
and $E_i^\text{QC}(\{ p_n \})$ are the energy levels predicted by the quantization conditions
for a given choice of $\{ p_n\}$. The model functions yielding the $E_i^\text{QC}(\{ p_n \})$ 
depend on the data through the particle masses and box length $L$, so the covariance of the residuals 
$r_i=E_i-E_i^\text{QC}(\{ p_n \})$ does not equal ${\rm cov}(E_i,E_j)$ \cite{Morningstar:2017spu}.
However, we use $C_{ij}={\rm cov}(E_i,E_j)$ instead of ${\rm cov}(r_i,r_j)$ since it makes
little difference to the final fit parameters but significantly simplifies the analysis.
We estimate $C$ by means of jackknife samples,
ignoring the sample to sample variations in $M_\pi$ and $M_K$, which are very small effects
compared to the variations in the $E_i$.
We do not apply the correction discussed in Ref.~\cite{Toussaint:2008ke} since any resulting change to the final fit parameters are expected to be insignificant.

In order to estimate the uncertainties of the best fit parameters, we use the derivative method. This uses the matrix of covariances between the parameters $p_n$ and $p_m$,
given by 
\begin{equation}
V_{nm} = \left( \frac{\partial E^\text{QC}_i}{\partial p_n} C^{-1}_{ij}  
\frac{\partial E^\text{QC}_j}{\partial p_m} \right)^{-1}\,,
\end{equation}
where the partial derivatives are evaluated numerically at the minimum of $\chi^2$. 
The main advantage of this approach is that one avoids refitting the spectrum in each sample,
 which is computationally very expensive for the three-particle case. 
 To ensure the validity of this procedure, 
 we have checked for two-particle fits that it yields almost identical uncertainty estimates
as obtained with the standard jackknife method. 
We have also checked that using bootstrap instead of jackknife samples does not
lead to significant changes in the results.

One might have thought that, instead of a global fit,
a better procedure would be to fix $\cK_2$ from fits to the
two-particle spectrum, and then use the result as input into the three-particle spectrum, so as
to determine $\cK_{\df,3}$. Such a procedure would, however, fail to make use of the
constraints on two-particle interactions arising from the interactions of pairs in the three-particle
system. Indeed, the three-particle spectrum provides significant additional constraints
on two-particle interactions, since there are three interacting pairs.

As a final comment on fitting, we consider the appropriate range of $E^*$ to use.
Since we are considering isosymmetric QCD, G parity is an exact symmetry, and
there are no transitions between sectors with even and odd numbers of pions.
Thus, for two pions, the range of validity of the quantization condition is
$0 < E^* < 4 M_\pi$, while for three pions it is $M_\pi < E^* < 5 M_\pi$.
For kaons, by contrast, there is no constraint from G parity, and
the process $K^+K^+ \to K^+ K^+ \pi^0$ is allowed, for example if the kaons
are in a relative $d$ wave. Also allowed is $K^+ K^+ \to K^+ K^0 \pi^+$.
Because of this, the upper bounds on the validity of the quantization conditions for
two and three kaons are, strictly speaking, $2 M_K + M_\pi$ and $3 M_K+M_\pi$, respectively.
We expect, however, the coupling to the channels with an additional pion to be very small.
This is because these transitions are induced by the chiral anomaly, 
and thus, in ChPT, by the Wess-Zumino-Witten (WZW) term. 
However, the WZW term itself does not lead to a $K^+ K^+ \to K^+ K^0 \pi^+$ transition,
due to its antisymmetry; the closest transition that is induced is
$K^+ K^0 \to K^+ K^0 \pi^0$. To obtain the desired transition requires an additional loop
(which attaches a  $K^0 K^+ \to K^0 K^+$ vertex).
This implies that the process is at least of next-to-next-to-leading order (NNLO) in
ChPT, since the WZW term is itself of next-to-leading order (NLO).
Thus we expect the coupling to an additional pion to turn on only far above threshold, 
where the $p^4$ suppression is less significant.
On the other hand, we do not include mixed-flavor operators in our calculations
  to capture these extra levels above threshold. Although this could result in unreliable energy extractions in this region,  we expect that the overlap of our operators with these mixed-flavor states is small enough to still obtain an accurate determination of the desired energies.
Because of this we have included in the fits to $2K^+$ and $3K^+$ some levels
that lie slightly above the strict threshold. Details are given in the next section.

\subsection{Parametrizations of K-matrices}
\label{subsec:Kmat}

We now summarize the parametrizations that we use for $\cK_2$ and $\cK_{\df,3}$.
The former is given by
\begin{equation}
\cK_2(E_2^*)_{\ell' m'; \ell m} = \delta_{\ell' \ell} \delta_{m'm} \cK_2^{(\ell)}(E_2^*)\,,
\label{eq:K2def}
\end{equation}
where the subscript on $E_2^*$ emphasizes that this is the CMF energy
of a {\em two}-particle system,
and
\begin{equation}
\cK_2^{(\ell)}(E_2^*) = \frac{16\pi E_2^*}{q \cot \delta_\ell(q) + |q| [1 - H(q^2) ]}\,.
\end{equation}
Here $q$ is the magnitude of the three-momentum of each of the two particles in their CMF,
\begin{equation}
4(q^2 + M^2) = E_2^{*2}\,,
\end{equation}
and $H(q^2)$ is a smooth cutoff function that equals unity for physical scattering ($q^2\ge 0$).
We follow all previous work implementing the RFT formalism and use the form of $H(q^2)$ from Ref.~\cite{\HSQCa}. We stress that the term proportional to $[1- H(q^2)]$ vanishes identically for physical scattering. It is not necessary for the two-particle quantization condition,
although it can be included ($H$ dependence for subthreshold two-particle
finite-volume states is cancelled by the corresponding dependence of $F$),
but it is essential for the three-particle case, 
where subthreshold momenta are included down to $q^2=-M^2$.
We also note that the additional scheme dependence in $\cK_2$ introduced in
Ref.~\cite{\largera} is not needed here as there are no resonances.

To complete the parametrization of the two-particle phase shifts we need to specify the
forms we use for $\cot\delta_\ell$. For $s$-wave scattering we consider two choices. The first is motivated by the ChPT expressions for the scattering of identical pseudo-Goldstone bosons (e.g.,
$\pi^+\pi^+$ and $K^+K^+$), which includes the Adler zero below threshold:
\begin{equation}
\frac{q}{M} \cot \delta_0(q) = \frac{M E_2^*}{E_2^{*2} - 2 z^2 M^2} \left( B_0 + B_1 \frac{q^2}{M^2} 
+ B_2 \frac{q^4}{M^4}\right)\,.
\label{eq:Adler0}
\end{equation}
This contains the dimensionless parameters $z^2$, $B_0$, $B_1$, and $B_2$.
We stress that, here and below, the full expressions contain an infinite set of higher order terms
in $q^2$ that we are setting to zero.
We use this expression for both pions and kaons, with $M=M_\pi$ and $M_K$, respectively.
At lowest order in ChPT, $z^2=1$, and we perform both fits with $z^2$ fixed to this value,
and others allowing it to vary. We also do fits with and without the parameter $B_2$.

The second form is the effective-range expansion (ERE)
\begin{equation}
\frac{q}{M} \cot \delta_0(q) = -\frac1{M a_0} + \frac{r_0 q^2}{2M} + \frac{P_0 r_0^3 q^4}{M} \,,
\label{eq:ERE0}
\end{equation}
given in terms of the scattering length $a_0$, effective range $r_0$, and quadratic parameter $P_0$.
We use the sign convention in which the scattering length is positive for repulsive interactions.
In principle, the radius of convergence of the ERE is given by the position of the Adler zero, 
although this form has been used beyond this radius
in many previous LQCD-based analyses of above-threshold scattering.
The parameters in \Cref{eq:Adler0} are related to those of the ERE through
\begin{equation}
M a_0 = - \frac{2-z^2}{B_0}\,,\quad
M^2 a_0 r_0 =  \frac{2+z^2}{2-z^2} - \frac{2 B_1}{B_0} \,.
\label{eq:conversion}
\end{equation}
We use the combination $M^2 a_0 r_0$ rather than $M r_0$ alone, since  the former does not diverge
in the chiral limit,  whereas the latter can [see \Cref{eq:r0pi_SU2} below].

For $d$-wave scattering we use a simpler, one-parameter form,
\begin{equation}
\frac{q^5}{M^5} \cot \delta_2 = \frac{E_2^*}{2M} D_0 - 1 \,,
\label{eq:simpledelta2}
\end{equation}
as we find that this provides an adequate description of our data.
The overall factor of $E_2^*=\sqrt{s}$ is adopted from standard continuum analyses
(see, e.g., Refs.~\cite{Yndurain:2007qm,Kaminski:2006qe})
and implies that higher-order terms in the ERE for $\delta_2$ are present,
although with fixed coefficients. 
The factor of $-1$ is chosen to avoid unphysical poles in the subthreshold 
scattering amplitude, which is given by 
\begin{equation}
\cM_2^{(\ell)} = \frac{16 \pi E_2^*}{q \cot\delta_\ell + |q|} \qquad (q^2 < 0)\,.
\end{equation}
Such poles arise for $|q|\approx M$ from the $D_0$ term alone
if the interactions are repulsive (as they are here).
The use of this ad hoc factor is adapted from that used in continuum analyses of $s$-wave
scattering~\cite{Yndurain:2007qm}. 
The $d$-wave scattering length is then given by $M^5 a_2=-1/(D_0 - 1)$,
using the conventional definition in which $a_2^{1/5}$ has dimensions of length,
and the same sign convention as for $a_0$.

For the three-particle K-matrix we use the threshold expansion worked out in Ref.~\cite{\dwave}.
This expands $\Kdf(\bm p'_1,\bm p'_2,\bm p'_3;\bm p_1,\bm p_2,\bm p_3)$
in powers of $\Delta=(E^{*2}-9M^2)/(9 M^2)$ and related quantities, where $\bm p_i$ 
($\bm p'_i$) are the initial (final) momenta for on-shell three-to-three scattering.
We use the expansion through quadratic order,
\begin{equation}
\Kdf = \cK^{\text{iso,0}}_{\df,3} + \cK^{\text{iso,1}}_{\df,3} \Delta +  \cK^{\text{iso,2}}_{\text{df,3}} \Delta^2  + \cK_A \Delta_A  + \cK_B \Delta_B\,,
\label{eq:Kdfthreshold}
\end{equation}
where and $\cK^{\text{iso,0}}_{\df,3}$,  $ \cK^{\text{iso,1}}_{\df,3}$, $ \cK^{\text{iso,2}}_{\df,3}$,
$\cK_A$, and $\cK_B$ are real constants.
The first three terms depend only on the overall CMF energy, but not otherwise on the
momenta of the particles, and are referred to as ``isotropic'' contributions.
The final two terms do depend on momenta through the 
dimensionless quantities $\Delta_A$ and
$\Delta_B$, which are defined in Ref.~\cite{\dwave}, and are of quadratic order in $\Delta$.
\Cref{eq:Kdfthreshold} is the most general form consistent
with Lorentz symmetry, time-reversal, parity, and particle exchange symmetry.
To use this result in the quantization condition, \Cref{eq:QC3}, one must decompose $\Kdf$ into
the $\{k\ell m\}$ basis discussed above. How to do so for the rest frame ($\bm P=0$) is explained
in Ref.~\cite{\dwave}, and the generalization to moving frames is straightforward.
We also note that, when decomposed into irreps of the little group of the various frames,
the $\cK_{\df,3}^{\rm iso}$ and $\cK_A$ terms contribute only to trivial irreps,
while the $\cK_B$ term can also contribute to nontrivial irreps,
and does so to all the nontrivial irreps included in our fits.
It is for this reason that $\cK_B$ can be determined
more easily than $\cK_A$. In fact, in our minimal fits we use only the three parameters
$\cK^{\text{iso,0}}_{\df,3}$,  $ \cK^{\text{iso,1}}_{\df,3}$, and $\cK_B$.

In the two-particle context, it is standard to consider the {\em total} angular momentum $J$ of the
system, and we have done that above when distinguishing $s$- and $d$-wave choices for $\cK_2$,
which have $J=0$ and $2$, respectively.
The finite-volume irreps do not uniquely pick out values of $J$, but do restrict them.
For example, the two-particle $E^+_g$ irrep in the rest frame contains $J=2, 4, \dots$ but not $J=0$,
while trivial irreps in all frames contain $J=0$ as well as higher values.
The relation between irreps and $J$ or helicity components is given, for example,
in Table II of Ref.~\cite{Dudek:2012gj}.

In infinite volume, one can also classify the three-particle interactions by their total angular momentum.
Although this is not discussed in Ref.~\cite{\dwave}, it is straightforward to see from the explicit
forms of $\Delta_A$ and $\Delta_B$ that the $\cK_A$ term contains only $J=0$, 
as do the isotropic terms, while $\Delta_B$ contains $J=0$, $1$ and $2$ components. 
The presence of these higher angular momenta provides an alternative way of understanding 
why only the $\cK_B$ term contributes to nontrivial reps, since these require $J>0$.
We note that a total $J=2$, say, can arise both from an $\ell=2$ contribution to the pair sub-interaction
combined with a relative $s$-wave between the pair and the remaining particle, and
from an $\ell=0$ contribution combined with a  pair-spectator relative $d$-wave, as well as
from other options.

\subsection{Results from chiral perturbation theory}
\label{subsec:ChPT}

We collect in this subsection the results from ChPT that we need when
fitting the dependence of the quantities derived from fits versus $M_\pi^2$.
We use both SU(2) and SU(3) ChPT.
For SU(2) ChPT
to be valid, we need both $M_\pi^2/(4\pi F_\pi)^2 \ll 1$ and
$M_\pi^2/M_K^2 \ll 1$. The former is indeed small, ranging up to $0.075$ on our ensembles.
The latter is significantly larger, with the values $\{0.18, 0.38,0.61\}$ on the 
D200, N200, and N203 ensembles, respectively. Thus SU(2) ChPT is of borderline
applicability for the N203 ensemble.
The rapid rise in this ratio is due to the path along which we take the chiral limit,
namely with $m_{\rm av}=(2 m_q+m_s)/3$ fixed
(with $m_q$ the common up- and down-quark mass and $m_s$ the strange-quark mass).
At leading order (LO) in ChPT, this implies that $M_0^2=(2 M_K^2 + M_\pi^2)/3$ is constant,
so that $M_K$ decreases as $M_\pi$ increases.
The use of this trajectory does, however, improve the range of validity of SU(3) ChPT,
compared to the traditional choice of fixing $m_s=m_s^{\rm phys}$.
The relevant quantity is $M_K^2/(4\pi F_\pi)^2$, which takes the values
$\{0.18, 0.15, 0.12\}$ on our ensembles.

In the following, we use the abbreviations
\begin{equation}
x_\pi = \frac{M_\pi^2}{F_\pi^2} \ \ {\rm and} \ \
x_K = \frac{M_K^2}{F_K^2}\,,
\label{eq:abbrevs}
\end{equation}
where $F_\pi$ and $F_K$ are the physical pion and kaon decays constants,
in the $F_\pi \simeq 92$ MeV convention.
We work mostly at NLO, 
and do not indicate the higher-order terms that are dropped.

We begin with the expression for the pion scattering length in SU(2) 
ChPT~\cite{Gasser:1983yg} (choosing the form commonly used in the lattice community
and given in Ref.~\cite{Aoki:2019cca})
\begin{equation}
M_\pi a_0^{\pi\pi} = \frac{x_\pi}{16\pi} 
\left[ 1 -
\frac{x_\pi}{32 \pi^2} \left( \ell_{\pi\pi} + 1 - 3 \log \frac{x_\pi}2 \right)
\right]\,.
\label{eq:a0pi_SU2}
\end{equation}
Here $\ell_{\pi\pi}$ is a low-energy coefficient (LEC) evaluated at the scale $\sqrt2 F_\pi$.
A subtlety here concerns the chiral trajectory that we follow.
In SU(2) ChPT, $\ell_{\pi\pi}$ depends implicitly on $m_s$, and so has, in principle,
different values for our three ensembles, none of which equal the standard value,
which has $m_s=m_s^{\rm phys}$.
Since $\ell_{\pi\pi}$ is a smooth function of $m_s$, 
the difference between the values on our ensembles and the standard  value is proportional to 
$m_s-m_s^{\rm phys} = 2 (m_q^{\rm phys}-m_q)$,
and thus proportional to $M_\pi^2$. 
As the $\ell_{\pi\pi}$ term is itself of NLO, 
the part proportional to $M_\pi^2$ is effectively of NNLO (or higher order), 
and can be consistently dropped.

The expression for the effective range in SU(2) ChPT is~\cite{Beane:2011sc}
\begin{equation}
M_\pi^2 r_0^{\pi\pi} a_0^{\pi\pi} = 3 + x_\pi \frac{C_3}{2} + \frac{11}{24 \pi^2} x_\pi \log \frac{x_\pi}2 \,,
\label{eq:r0pi_SU2}
\end{equation}
where $C_3$ is an LEC, and is evaluated at the scale $\sqrt2 F_\pi$. Similarly to $\ell_{\pi\pi}$,
$C_3$ can be treated as a constant in an NLO expression.

One can also obtain the NLO expression for the position of the Adler zero in pion
scattering. Rewriting the result of Ref.~\cite{Yndurain:2002ud}, we have
\begin{equation}
z^2 = 1 - \frac{x_\pi}{32 \pi^2} \left[ \ell_z + \frac16 - \frac{11}3 \log \frac{x_\pi}2 \right]\,,
\label{eq:SU2_Adler}
\end{equation}
where $\ell_z$ is an LEC evaluated at $\sqrt2 F_\pi$, which is related to the
standard SU(2) ChPT LECs (see, e.g., Ref.~\cite{Aoki:2019cca}) by
\begin{equation}
3 \overline{\ell}_3 + 8 \overline{\ell}_2 = 3 \ell_z - 11 \log (x_\pi/2)\,.
\end{equation}

The $d$-wave $I=2$ pion amplitude vanishes at LO in ChPT. Although
a NNLO expression exists in the literature, our results are insufficient to attempt a fit to this form. 
Furthermore, there is a subtlety related to a change in sign of the phase shift at physical
pion masses just above threshold, an issue we discuss below when we analyze our results.

Expressions in SU(3) ChPT for the pion and kaon scattering lengths can be obtained from
Refs.~\cite{Chen:2005ab,Chen:2006wf}. The NLO terms depend not only on the pion
and kaon masses, but also on that of the $\eta$ meson. Within these terms, it is consistent
to rewrite the $\eta$ mass using the LO expression $3 M_\eta^2= 4 M_K^2-M_\pi^2$,
and also to treat $F_\pi$ and $F_K$ as interchangeable. In this way, we can write the
results as functions of $x_\pi$ and $x_K$ alone, finding
\begin{align}
M_\pi a_0^{\pi\pi} &= \frac{x_\pi}{16\pi}
\left[1 + \frac{x_\pi}{16 \pi^2} \left( 
\frac32 \log \frac{x_\pi}{16\pi^2}
+\frac1{18} \log \frac{4 x_K - x_\pi}{48 \pi^2} 
- \frac49
- 256 \pi^2 L_{\pi\pi}
\right) \right]\,,
\label{eq:a0pi_SU3}
\\
\begin{split}
M_K a_0^{KK} &= \frac{x_K}{16\pi}
\bigg[1+ \frac{x_K}{16\pi^2} \bigg(
\log \frac{x_K}{16\pi^2}
-  \frac{x_\pi}{4 (x_K-x_\pi)}  \log \frac{x_\pi}{16\pi^2}
\\
&\qquad \qquad+ \frac{20 x_K - 11 x_\pi}{36 (x_K - x_\pi)}
 \log \frac{4 x_K - x_\pi}{48 \pi^2} 
- \frac79 -256 \pi^2 L_{\pi\pi}
\bigg)\bigg]\,.
\end{split}
\label{eq:a0K_SU3}
\end{align}
These two expressions contain only a single LEC, $L_{\pi\pi}$, which is here evaluated
at the scale $4\pi F_\pi$. We stress that, in SU(3) ChPT, there is no subtlety due to our
choice of chiral trajectory, since all dependence on $m_s$ is explicit. In other words,
the trajectory is encoded into the values of $x_\pi$ and $x_K$.

Expressions in SU(3) ChPT for the kaon effective range and Adler zero position could,
in principle, be extracted from the results given in Ref.~\cite{Alexandru:2020xqf} for
the $K^-K^-$ scattering amplitude. We have not done so, however, as our simulation results for
$M_K^2 a_0^{KK} r_0^{KK}$ presented below lie close to unity, and thus very far from
the LO chiral prediction of $3$. This indicates very large NLO corrections, and a breakdown
in convergence. 

An alternative would be to use SU(2) ChPT, treating the kaon as a heavy source for pions,
following Ref.~\cite{Roessl:1999iu}. However, the analysis for kaon scattering has not been
carried out (and would require methods similar to that used to study $NN$ scattering in EFT).
Thus we use simple analytic parametrizations of $M_K a_0^{KK}$,
$M_K^2 a_0^{KK} r_0^{KK}$, and $1/D_0^{KK}$, fitting them to linear functions of $x_\pi$.

Finally, we turn to the chiral predictions for the three-particle K-matrix.
At tree level (LO) $\Kdf$ is purely isotropic, with only $\cK_{\df,3}^{\iso,0}$ and $\cK_{\df,3}^{\iso,1}$ 
being nonzero~\cite{Blanton:2019vdk}
\begin{equation}
M^2 \cK_{\df,3}^{\iso,0} = 18 \frac{M^4}{F^4} = 18 (16 \pi M a_0)^2 \,,
\qquad
M^2\cK_{\df,3}^{\iso,1} = 27 \frac{M^4}{F^4} = 27 (16 \pi M a_0)^2 
\,.
\label{eq:Kdf_LO}
\end{equation}
These expressions hold both for three-$\pi^+$ and three-$K^+$ systems, 
using the corresponding masses and scattering lengths.
The other constants in \Cref{eq:Kdfthreshold}  ($\Kisotwo, \mathcal{K}_A$, and $\mathcal{K}_B$) can appear first at NLO,
and thus are suppressed by at least one additional power of $1/F^2 \propto M a_0$.



\clearpage

\section{Fitting the spectrum}
 \label{sec:results}

In this section, we present the results of fits to the two- and three-particle spectra,
describing in turn the results for pions and kaons.
The parametrizations explained in \Cref{subsec:Kmat} are used.
We comment on the main features of the fit, 
such as goodness of fit and which parameters are needed, 
but leave the interpretation of the results for
the fit parameters themselves to the following section.
An example of these fits on the N200 ensemble was already shown in \Cref{fig:n200_three_pion_spectrum}. 
We also note that the precise set of levels used in the fits of this section is given in \Cref{app:C}.

\subsection{Fits to the spectrum of two and three pions}
\label{subsec:pionfits}
  
We start with fits to the energy levels in the $2\pi^+$ and $3\pi^+$ sectors. 
For the two-pion $s$-wave interaction, we use the Adler-zero form, given in \Cref{eq:Adler0},
rather than the ERE form.
The former has been found previously to provide a better description for light pions~\cite{Blanton:2019igq,Fischer:2020jzp}, and we confirm this result below.
When $d$ waves are included, we use the expression in \Cref{eq:simpledelta2}.
For $\Kdf$, we consider only three of the terms in \Cref{eq:Kdfthreshold}:
the leading two isotropic ones, $\cK_{\df,3}^{\iso,0},  \cK^{\text{iso,1}}_{\df,3}$
(referred to below as $s$-wave terms), and $\cK_B$ (referred to as a $d$-wave term). 
We find that this choice provides a good description of the levels, whereas the
other parameters in $\Kdf$ are poorly determined if included.

We present results for the following set of representative fits:
\begin{enumerate}
\item 
A fit solely to two-particle energies, including both $s$- and $d$-wave interactions. 
This fit shows the information that can
be extracted from the two-particle spectrum alone, and thus is a useful point of comparison 
for three-particle fits.
\item 
A combined fit to two- and three-pion energies  
including only levels in the trivial irreps in each frame, and with only $s$-wave contributions
for both two- and three-particle interactions. Furthermore, the position of the Adler zero is 
fixed to its LO ChPT value ($z^2=1$). 
This is the type of fit used in previous work~\cite{Blanton:2019igq,Fischer:2020jzp},
albeit to fewer frames than available here.
\item 
A combined fit to all two- and three-pion levels, including all irreps, and
including both $s$- and $d$-wave interactions. 
We again fix $z^2=1$, and do not include the quadratic parameter $B_2$ from
\Cref{eq:Adler0}.
We call this the ``minimal'' complete fit.
\item 
An extension of the minimal fit in which we do not fix the position of the Adler zero.
The goal is to check the validity of the $z^2=1$ hypothesis.
\item 
An extension of the minimal fit in which we keep $z^2=1$, but allow $B_2$ to vary
in order to test whether this can improve the fit. This fit also gives information on the convergence
of the $q^2$ expansion in the Adler-zero form.
\end{enumerate}
The fits can, in principle, extend in the CMF energy up to the inelastic thresholds, which are
$4 M_\pi$ and $5 M_\pi$, respectively, for two- and three-pion spectra. 
 In practice, we quote results from fits to a smaller energy range on two
of the three ensembles to avoid possible issues with the estimate of the covariance matrix.
Fits to levels within varying energy ranges yield compatible results, even
though $\chi_\text{red}^2$ increases as more levels are included.
 This $\chi_\text{red}^2$ behavior is to be expected; we parameterize the K-matrices by expanding about the appropriate thresholds, so it is no surprise that our models become less accurate if too many high-energy levels are included in fits.
This may indicate that
the threshold expansions used in our parametrizations are breaking down for higher values
of the CMF energy.

The results of the fits are shown in \Cref{tab:fitpiN203,tab:fitpiN200,tab:fitpiD200},
in each case ordered from left to right as in the list above. 
 The tables give the number of degrees of freedom, from which the number of levels included
in the fits can be seen. For example, on the N203 ensemble, we fit to 38 two-particle
levels and 35 three-particle levels.
 In addition to quoting results for the fit parameters themselves, we also quote,
in the last two rows, the results for the $s$-wave scattering length and effective range,
obtained using \Cref{eq:conversion}, to facilitate a more direct comparison of the results of the fits.

We draw several global conclusions from the results in the tables.
\begin{enumerate}
\item[a.]
The values of $\chi^2_{\rm red}$ for the best fits are generally reasonable, 
although always larger than unity.
Such values are typical in the analysis of lattice 
spectra~\cite{Fischer:2020jzp,Hansen:2020otl,Brett:2021wyd}, 
and may indicate the relevance of neglected systematic uncertainties, 
e.g., discretization effects, or exponentially suppressed effects in the quantization condition.
Goodness of fit becomes poorer for lighter pion masses,
possibly indicating a reduction in the range of validity of our truncated threshold expansions.
\item[b.]
The inclusion of two- and three-particle
$d$-wave interactions yields a better description of the data, as shown
by the smaller $\chi^2_{\rm red}$ compared to fits of type 2. This result is particularly striking as the
levels included in this fit type are those in the trivial irreps, which are the least sensitive to
$d$-wave interactions. Were we to attempt a fit to all irreps without including the $d$-wave terms,
a significantly higher $\chi^2_{\rm red}$ would result, as shown by the significance of the
parameter $D_0$ in the fits in which it is included. 
We will further elaborate on these points when discussing fits to the kaon spectra.
\item[c.]
The results for $M_\pi a_0^{\pi\pi}$ and $M_\pi^2 a_0^{\pi\pi} r_0^{\pi\pi}$ are consistent across all five fits on all ensembles
(with the fit 1 result for $M_\pi^2 a_0^{\pi\pi} r_0^{\pi\pi}$ on the N203 ensemble being the only outlier), 
indicating that we are obtaining a consistent description of the two-particle interactions from
two- and three-pion levels.
\item[d.]
A comparison of the results of fits 1 and 4 suggests that the addition of the three-particle levels
improves the precision with which we can extract two-particle interactions, most notably
for the $d$-wave parameter $D_0^{\pi\pi}$.
\item[e.]
The minimal fits (fit 3) have essentially the same $\chi^2_{\rm red}$ as those of
fits 4 and 5, and the results for $z^2$ (in fit 4) and $B_2$ (in fit 5) are consistent with
unity and zero, respectively. Thus we conclude that the minimal description is adequate for
our dataset, and use the parameters from this fit in our investigation of quark-mass dependence
in the following subsection.
\end{enumerate}

We close this section by showing the results of two additional fits to the N203 ensemble
that motivate the choice of fits presented above. 
First, we test whether the $s$-wave parameters in $\kdf$ that we previously set to zero,
$\cK_{\df,3}^{\rm iso,2}$ and $\cK_A$,
are relevant for a description of our data. 
We use the ensemble with the heaviest pion mass for this test, since these parameters
are of higher order in the chiral expansion than those we keep in our standard fits.
We show in \Cref{tab:fitfullpiN203} the result of a fit in which
all ten parameters in both the two- and three-particle interactions
are turned on. We observe
that the additional parameters do not lead to a reduction in $\chi^2_{\rm red}$,
but do lead to much larger uncertainties in the fit parameters associated with $s$-wave interactions.
Since we have not added additional $d$-wave parameters, we expect the
results for these parameters to be unchanged, which is confirmed by the fit.
Not visible from the table is the fact that the correlation between the additional fit parameters is substantial.
We conclude that there is no need to include the additional parameters  in order to represent our data.

Our second goal is to check whether the use of the Adler-zero form remains appropriate 
at heavier pion masses. 
To study this, we perform a minimal fit to the N203 data
using the ERE form for the two-particle $s$-wave interaction,  \Cref{eq:ERE0}.
We find, as shown in \Cref{tab:fitEREpiN203}, that this fit is significantly disfavored compared
to the Adler-zero form.

\begin{table}[h!]
\centering
\begin{tabular}{c||c|c|c|c|c}
\multicolumn{6}{c}{ \bf N203($\mathbf{\pi}$) }\\ \hline  \hline
 &   \multicolumn{1}{c|}{($2\pi, \ell\leq 2$) } &   {($2\pi/3\pi, \ell=0$) } &  \multicolumn{3}{c}{($2\pi/3\pi, \ell\leq 2$) } \\ \cline{2-6}
 & Fit 1  & Fit 2  & Fit 3  & Fit 4  & Fit 5  \\ \hline  \hline
$z^2$&0.78(13) & 1.0(fixed)&1.0(fixed)& 0.90(10) &  1.0(fixed)   \\ \hline
$B_0$& -5.85(56) & -4.88(9) & -4.86(8) & -5.24(39) & -4.77(10) \\ \hline
$B_1$&-1.70(23)  &-2.27(12) &-2.06(10)& -1.92(17)  & -2.41(24) \\ \hline
$B_2$&--- & &---   & ---  & 0.20(13)  \\ \hline
$D_0^{\pi\pi}$&-154(23) & --- &-137(16) & -137(16) & -136(16)  \\ \hline
$M_\pi^2\Kisozero$&--- & 240(210) &650(210)   &  540(240)  &  520(230)\\ \hline
$M_\pi^2\Kisoone$& --- &-1690(330) &-2100(470) & -1950(490)   &  -2000(470) \\ \hline
$M_\pi^2\mathcal K_B$& ---  & ---&-2400(800)  & -2500(800)  & -2500(800) \\ \hline
\hline
dof  &38-4 & 54-4 & 73-6& 73-7& 73-7   \\ \hline
$\chi^2_{\rm red}$ & 1.02 & 1.87& 1.39 & 1.40 & {1.37} \\ \hline
\hline
$M_\pi a_0^{\pi\pi}$ & 0.2082(51) & 0.2050(38) &0.2059(34)   & 0.2092(47) &0.2097(42)\\ \hline
$M_\pi^2 a_0^{\pi\pi} r_0^{\pi\pi}$ &  1.70(24) & 2.06(6) & 2.15(5)  &  1.92(22) &1.99(12) 
\end{tabular}
\caption{Fit results for pions on the N203 ensemble with $M_\pi L=5.405365$, using the Adler-zero form for the $s$-wave phase shift. Each fit contains all $2\pi^+$ and $3\pi^+$ energy levels (in the appropriate irreps) below $E^*_{2,\rm max}=3.46M_\pi$ and $E^*_{\rm max}=4.46M_\pi$, respectively. 
} \label{tab:fitpiN203}
\end{table}

 \begin{table}[h!]
\centering
\begin{tabular}{c||c|c|c|c|c}
\multicolumn{6}{c}{ \bf N200($\mathbf{\pi}$) }\\ \hline  \hline

 &   \multicolumn{1}{c|}{($2\pi, \ell\leq 2$) } &   {($2\pi/3\pi, \ell=0$) } &  \multicolumn{3}{c}{($2\pi/3\pi, \ell\leq 2$) } \\ \cline{2-6}
 & Fit 1  & Fit 2  & Fit 3  & Fit 4  & Fit 5  \\ \hline  \hline
$z^2$&0.96(13)   &  1(fixed)  & 1(fixed)  & 0.96(11)  & 1(fixed)    \\ \hline
$B_0$& -7.00(70) & -6.54(18) & -6.45(16) &-6.67(58) & -6.33(19)\\ \hline
$B_1$&  -1.70(27) & -2.06(19) & -1.90(15) & -1.83(23) & -2.26(36)\\ \hline
$B_2$& ---  &  ---- & --- &---  & 0.15(14)   \\ \hline
$D_0^{\pi\pi}$& -280(80) & --- &   -241(52) &  -240(52) & -237(50) \\ \hline
$M_\pi^2\Kisozero$& ---  & 460(160) & 500(160)&470(170) & 480(160)  \\ \hline
$M_\pi^2\Kisoone$& --- &   -1000(230) & -1040(330)&-1030(330) &-1060(330) \\ \hline
$M_\pi^2\mathcal K_B$&  ---  & --- & -840(550)   & -890(560) & -930(560)  \\ \hline
\hline
dof  & 39-4   &  53-4& 72-6 & 72-7 & 72-7 \\ \hline
$\chi_\text{red}^2$& 0.83 & 1.38 & {1.17} & 1.19 & {1.17}  \\ \hline
\hline
$M_\pi a_0^{\pi\pi}$ & 0.1481(62) & 0.1528(43) & 0.1550(40)&  0.1565(56) & 0.1579(48) \\ \hline 
$M_\pi^2 a_0^{\pi\pi} r_0^{\pi\pi}$ & 2.37(37) & 2.37(7) & 2.41(6) & 2.28(32) & 2.29(13)
\end{tabular} 
\caption{Fit results for pions on the N200 ensemble using $M_\pi L=4.419849$, with notation as in \Cref{tab:fitpiN203}.
 Each fit contains all $2\pi^+$ and $3\pi^+$ energy levels (in the appropriate irreps) below $E^*_{2,\rm max}=4M_\pi$ and $E^*_{\rm max}=5M_\pi$, respectively.
}
\label{tab:fitpiN200}
\end{table}

\begin{table}[h!]
\centering
\begin{tabular}{c||c|c|c|c|c}
\multicolumn{6}{c}{ \bf D200($\mathbf{\pi}$) }\\ \hline  \hline
 &   \multicolumn{1}{c|}{($2\pi, \ell\leq 2$) } &   {($2\pi/3\pi, \ell=0$) } &  \multicolumn{3}{c}{($2\pi/3\pi, \ell\leq 2$) } \\ \cline{2-6}
 & Fit 1  & Fit 2  & Fit 3  & Fit 4  & Fit 5  \\\hline  \hline
$z^2$& 0.83(21) & 1.0(fixed)& 1.0(fixed) & 0.81(18) & 1.0(fixed)  \\ \hline
$B_0$& -13.0(1.7)& -11.56(62) & -11.64(56)  & -13.0(1.5) & -11.1(7)\\ \hline
$B_1$& -1.7(6)&  -2.49(41) & -2.19(38)  &-1.8(6)   & -3.2(1.0)\\ \hline
$B_2$& ---&  --- & --- & --- & 0.34(33) \\ \hline
$D_0^{\pi\pi}$& -690(380)   & ---  & -640(280)& -640(290) & -620(270) \\ \hline
$M_\pi^2\Kisozero$& ---    & -150(190)  & -100(190) & -150(190) & -130(190) \\ \hline
$M_\pi^2\Kisoone$& --- & 20(180)  & 10(210) & 40(210) & 40(210) \\ \hline
$M_\pi^2\mathcal K_B$&  ---    &---  & -180(240)& -190(240)  & -150(240) \\ \hline
\hline
dof  & 28-4   &41-4 & 52-6 & 52-7 &52-7  \\ \hline
$\chi_\text{red}^2$& 2.29 & 1.99 & {1.67} & 1.68  & 1.68 \\ \hline
\hline
$M_\pi a_0^{\pi\pi}$ & 0.0899(71)  & 0.0866(47) & 0.0859(41) & 0.0913(63)& 0.0898(56)  \\ \hline
$M_\pi^2 a_0^{\pi\pi} r_0^{\pi\pi}$ &   2.16(51) & 2.57(9) & 2.62(8) & 2.10(44) & 2.44(20)
\end{tabular}
\caption{Fit results for pions on the D200 ensemble using $M_\pi L = 4.199492$, with notation as in \Cref{tab:fitpiN203}.
 Each fit contains all $2\pi^+$ energy levels (in the appropriate irreps) below $E^*_{2,\rm max}=3.74M_\pi$ and most $3\pi^+$ levels (in the appropriate irreps) below $E^*_{\rm max}=4.74M_\pi$; we have discarded some high-energy $3\pi^+$ levels with unusually large errors, only including the lowest two ${\bm d}^2=6$ levels and lowest three ${\bm d}^2=8$ levels in the $A_2$ irrep.}
\label{tab:fitpiD200}
\end{table}

\begin{table}[h!]
\begin{minipage}{.45\linewidth}
    \centering
\begin{tabular}{c||c}
\multicolumn{2}{c}{ \bf N203($\mathbf{\pi}$) }\\ \hline  \hline
 & \multicolumn{1}{c}{($2\pi/3\pi, \ell\leq 2$) } \\ \cline{2-2}
 & Full fit  \\ \hline  \hline
$z^2$& 0.96(21) \\ \hline
$B_0$&  -5.0(1.0)  \\ \hline
$B_1$&  -2.2(8) \\ \hline
$B_2$& 0.1(3)  \\ \hline
$D_0^{\pi\pi}$& -140(17)  \\ \hline
$M_\pi^2\Kisozero$& 80(390)  \\ \hline
$M_\pi^2\Kisoone$& 100(1500)  \\ \hline
$M_\pi^2\Kisotwo$&  -1700(1200) \\ \hline
$M_\pi^2\mathcal K_A$&  1300(1300) \\ \hline
$M_\pi^2\mathcal K_B$&  -2900(800) \\ \hline
\hline
dof  &  73-10   \\ \hline
$\chi_\text{red}^2$& 1.37   \\ \hline
\hline
$M_\pi a_0^{\pi\pi}$ & 0.2086(48) \\ \hline
$M_\pi^2 a_0^{\pi\pi} r_0^{\pi\pi}$ & 1.95(32)
\end{tabular}
\caption{Fit for pions on the N203 ensemble including all possible parameters at the order we are working,
 with $M_\pi L$ and levels as in fits 3--5 of \Cref{tab:fitpiN203}.
}
\label{tab:fitfullpiN203}
 \medskip
\end{minipage}\hfill
\begin{minipage}{.45\linewidth}
    \centering
\begin{tabular}{c||c}
\multicolumn{2}{c}{ \bf N203($\mathbf{\pi}$) }\\ \hline  \hline
 & \multicolumn{1}{c}{($2\pi/3\pi, \ell\leq 2$) } \\ \cline{2-2}
 & ERE fit  \\ \hline  \hline
$-(M_\pi a_0^{\pi\pi})^{-1}$& -4.47(8)   \\ \hline
$M_\pi r_0^{\pi\pi}$&  0.89(8)  \\ \hline
$M_\pi^3 (r_0^{\pi\pi})^3 P_0^{\pi\pi}$& -0.57(7) \\ \hline
$D_0^{\pi\pi}$& -144(18)  \\ \hline
$M_\pi^2 \Kisozero$& -160(230) \\ \hline
$M_\pi^2\Kisoone$& -840(470)  \\ \hline
$M_\pi^2 \mathcal K_B$& -3400(800)  \\ \hline
\hline
dof  &  73-7   \\ \hline
$\chi_\text{red}^2$&   1.88 \\ \hline
\hline
$M_\pi a_0^{\pi\pi}$ & 0.2238(39) \\ \hline
$M_\pi^2 a_0^{\pi\pi} r_0^{\pi\pi}$ &  0.80(6)
\end{tabular}
\caption{Fit for pions on the N203 ensemble using the ERE form of the $s$-wave phase shift,
 with $M_\pi L$ and levels as in fits 3--5 of \Cref{tab:fitpiN203}.
}
\label{tab:fitEREpiN203}
 \medskip
\end{minipage}
\end{table}

\subsection{Fits to two and three kaons} 
\label{subsec:kaonfits}
 
We now turn to the fits of the $2K^+$ and $3K^+$ levels.
We first use the same five fits as for pions, listed in the previous section,
with results given in \Cref{tab:fitN203K,tab:fitN200K,tab:fitD200K}. 
Note that we choose  $E^*_{\rm max}$ to be slightly above the relevant inelastic thresholds, 
which are $2M_K + M_\pi$ and $3 M_K+M_\pi$, respectively, for two and three kaons.
The values of $M_\pi/M_K$ for our ensembles can be determined from \Cref{tab:single-hadron-energies}.

Our observations concerning the fits are similar to those described above for the pion fits.
In particular, the inclusion of $d$-wave interactions remains crucial to obtain reasonable fits,
even if considering only trivial irreps. 
To illustrate this, we do a simple exercise using the N203 ensemble of \Cref{tab:fitN203K}. Taking the best fit parameters of fit 3 in the set of energy levels of fit 2, we get $\chi_\text{red}=1.46$. This is much lower than that 
of fit 2, and thus a significantly better description of the levels in trivial irreps is achieved when $D_0$ and $\cK_B$ are included.

The trend of $\chi^2_{\rm red}$ with the pion mass is opposite to that for the pion fits,
but this can be perhaps understood because the kaon mass increases as the pion mass decreases.
Another small difference is that the significance of the difference $z^2-1$, while still
less than $2\sigma$, is greater on the N203 and D200 ensembles than for pions.
For this reason we choose our canonical fit in the next section to be fit 4, i.e.~with $z^2$ left free.

\begin{figure}[b!]
\centering
\includegraphics[width=1.0\linewidth]{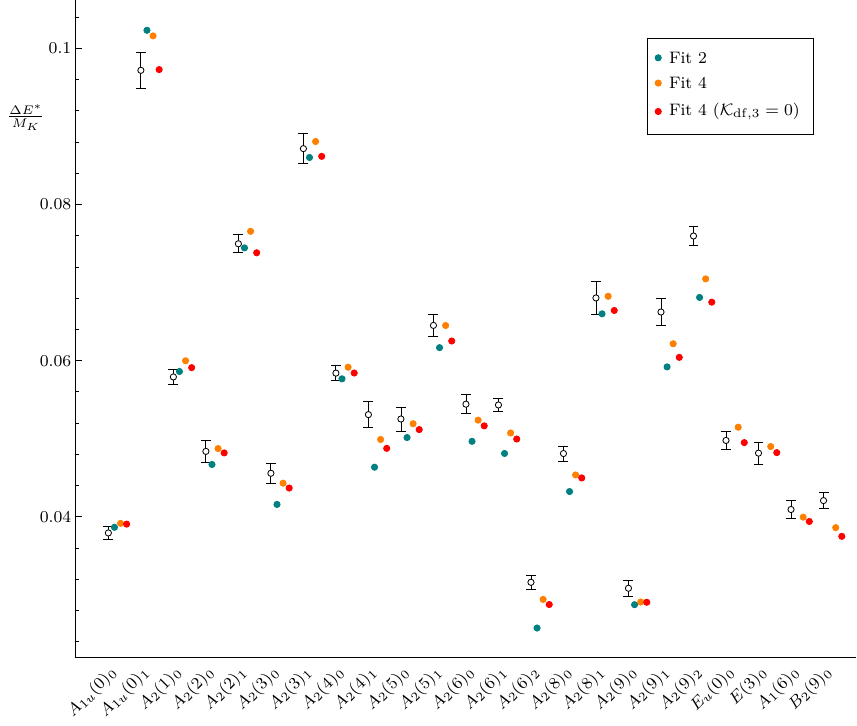}
\caption{Shifts from the noninteracting levels for three-kaon states included
in our fits on the N200 ensemble.
Each energy is labeled at the bottom by the irrep, total momentum-squared, and the energy level
(with $0$ indicating the lowest level in the channel, etc.).
For each level, the shifts determined from the lattice simulation are given by the open circles
with error bars, followed to the right
by the predictions of fit 2 (teal), fit 4 (orange), and fit 4 with $\Kdf=0$ (red), respectively.
The predictions for fit 2 are absent for the four nontrivial irreps on the right, 
as these levels were not included in this fit.
See text for further discussion. \\}
\label{fig:energyshifts}
\end{figure}

As noted in the introduction, 
the dominant contribution to the shift in energies from their noninteracting values
is due to two-particle interactions.
Although our fits indicate that including the terms in $\Kdf$, in particular $\cK_B$, leads to
improved fits, it is interesting to have a visual representation of the contribution of $\Kdf$ to
the shifts. To show this, we have used the quantization conditions to determine the
spectrum on the N200 ensemble taking all the parameters from fit 4 except that we set
$\Kdf$ to zero. 
The resulting energy shifts for the 23 three-kaon levels included in the fits are shown
in Fig.~\ref{fig:energyshifts}, along with those for fits 2 and 4, as well as the results from
the simulations.
Note that the four levels in nontrivial irreps are not included in fit 2.
We draw several conclusions. First, the shifts due to $\Kdf$ are comparable
to our present statistical errors. Second, one can see, particularly from the levels on the right half
of the plot, that including $\Kdf$ improves the fit more often than not. Third, for the trivial irreps
(the first 19 in the plot), the difference between fits 2 and 4 is again comparable to our errors.
Finally, although one cannot determine the goodness of fit from this figure, 
since only diagonal errors are shown,
we can obtain quantitative comparisons by calculating $\chi^2$ for the complete
two- and three-kaon fits for various choices of level.
We find that, if we keep only the 41 levels of fit 2, then fit 4 yields $\chi^2 = 52.5$,
while fit 4 with $\Kdf=0$ gives $\chi^2=180$, to be compared to $101$ for fit 2.
We also note that fit 4 with $\Kdf=0$ yields $\chi^2=192$ on the complete set of
levels, to be compared to $\chi^2=64$ from the full fit 4.

 \begin{table}[b!]
\centering
\begin{tabular}{c||c|c|c|c|c}
\multicolumn{6}{c}{ \bf N203($\mathbf{K}$) }\\ \hline  \hline
 &   \multicolumn{1}{c|}{($2K, \ell\leq 2$) } &   {($2K/3K, \ell=0$) } &  \multicolumn{3}{c}{($2K/3K, \ell\leq 2$) } \\ \cline{2-6}
 & Fit 1  & Fit 2  & Fit 3  & Fit 4  & Fit 5  \\ \hline  \hline
$z^2$& 1.01(10)  &  1(fixed)  & 1(fixed)  &  1.13(7)   & 1(fixed)    \\ \hline
$B_0$& -3.29(30) & -3.37(4) &  -3.28(4)  & -2.88(21) & -3.30(4) \\ \hline
$B_1$& -2.30(17) & -2.29(9) & -2.23(8)&  -2.43(13)  & -2.09(16)  \\ \hline
$B_2$& ---  &  ---- & --- &---  & -0.16(16)  \\ \hline
$D_0^{KK}$& -54(6)  & --- &  -45(4) & -45(4)   & -45(4)   \\ \hline
$\Kisozero$& ---  & -570(190)  & -250(190)   & -10(220)  & -130(220)  \\ \hline
$\Kisoone$& --- & -3700(500)  &-3800(800)   & -4500(900)   &  -4000(800)   \\ \hline
$\mathcal K_B$&  ---  & --- &  -5300(2700)  &  -5600(2700) & -5100(2700)  \\ \hline
\hline
dof  & $33 - 4$ & $46-4$& $ 62 - 6$  &   $62 - 7$ & $62 - 7$  \\ \hline
$\chi_\text{red}^2$& 1.29 & 3.03 &1.57 & {1.55}  & 1.58  \\ \hline
\hline
$M_\pi a_0^{KK}$ & 0.3009(49) & 0.2972(38) & 0.3047(38) & 0.3012(44) & 0.3032(41) \\ \hline
$M_\pi^2 a_0^{KK} r_0^{KK}$ &  1.64(20) & 1.64(6) & 1.64(6) & 1.92(19) & 1.73(11)
\end{tabular} 
\caption{Fit results for kaons on the N203 ensemble with $M_K L=6.908149$, using the Adler-zero form for the $s$-wave phase shift.
 Each fit contains all $2K^+$ and $3K^+$ energy levels (in the appropriate irreps) below $E^*_{2,\rm max}=2.9M_K$ and $E^*_{\rm max}=3.9M_K$, respectively,
except that we only include the lowest $2K^+$ level in the $A_1$ irrep
 in the ${\bm d}^2=9$ frame,  as the two higher-energy levels have unusually large errors. \\
}
\label{tab:fitN203K}
\end{table}

\begin{table}[tbh!]
\centering
\begin{tabular}{c||c|c|c|c|c}
\multicolumn{6}{c}{ \bf N200($\mathbf{K}$) }\\ \hline  \hline
 &   \multicolumn{1}{c|}{($2K, \ell\leq 2$) } &   {($2K/3K, \ell=0$) } &  \multicolumn{3}{c}{($2K/3K, \ell\leq 2$) } \\ \cline{2-6}
 & Fit 1  & Fit 2  & Fit 3  & Fit 4  & Fit 5  \\ \hline  \hline
$z^2$&  1.18(12)   & 1(fixed)  & 1(fixed)  & 1.04(13) & 1(fixed)  \\ \hline
$B_0$&  -2.47(34)   &  -3.00(4)& -2.97(4) &  -2.87(37) & -2.98(4)  \\ \hline
$B_1$&  -2.67(19)   & -2.70(12) &-2.52(10)& -2.57(20) & -2.44(18)   \\ \hline
$B_2$& ---   & --- &---  &  --- & -0.10(22)  \\ \hline
$D_0^{KK}$& -44(7)   & --- & -50(7)&-50(7)&-51(7) \\ \hline
$\Kisozero$& ---   &   -210(350)&110(350)&170(400)&220(410)  \\ \hline
$\Kisoone$& --- &-9200(900) & -9300(1100)&-9500(1200)&-9600(1200)\\ \hline
$\mathcal K_B$&  ---  & ---  & -$24(5)\cdot 10^3$ & -$24(5)\cdot 10^3$ & -$24(5)\cdot 10^3$  \\ \hline
\hline
dof  & 28-4  & 41-4 & 51-6 &51-7  &51-7    \\ \hline
$\chi_\text{red}^2$& 1.79    & 2.72 & {1.43} & 1.46  & 1.46    \\ \hline
\hline
$M_K a_0^{KK}$ & 0.3322(55)     &  0.3327(43) & 0.3366(42)  & 0.3358(50) &0.3355(47)  \\ \hline
$M_K^2 a_0^{KK} r_0^{KK}$ & 1.72(31) & 1.21(9) & 1.30(8)  & 1.36(23)& 1.36(14) 
\end{tabular}
\caption{Fit results for kaons on the N200 ensemble using $M_K L=7.225143$, with notation as in \Cref{tab:fitN203K}.
 Each fit contains all $2K^+$ energy levels (in the appropriate irreps) below $E^*_{2,\rm max}=2.75M_K$ and all $3K^+$ levels (in the appropriate irreps) below $E^*_{\rm max}=3.75M_K$. \\
 }
\label{tab:fitN200K}
\end{table}

\begin{table}[tbh!]
\centering
\begin{tabular}{c||c|c|c|c|c}
\multicolumn{6}{c}{ \bf D200($\mathbf{K}$) }\\ \hline  \hline
 &   {($2K, \ell\leq 2$) } &   {($2K/3K, \ell=0$) } &  \multicolumn{3}{c}{($2K/3K, \ell\leq 2$) } \\ \cline{2-6}
 & Fit 1  & Fit 2  & Fit 3  & Fit 4  & Fit 5  \\ \hline  \hline
$z^2$& 0.73(35)  & 1.0(fixed)& 1.0(fixed) & 0.47(42)  & 1.0(fixed)  \\ \hline
$B_0$& -3.50(94) & -2.89(4) & -2.78(4) & -4.2(1.1) & -2.75(4) \\ \hline
$B_1$& -2.12(49) & -2.58(13)& -2.50(10) & -1.8(6) & -2.92(26)\\ \hline
$B_2$& ---& --- & --- & --- &  0.7(4)  \\ \hline
$D_0^{KK}$&  -20(3)  & ---  & -20(2) & -20(2) &-20(2) \\ \hline
$\Kisozero$& ---&  -880(900)  & -340(900) &-900(1000)  &-1000(1000)  \\ \hline
$\Kisoone$& ---& -10000(3500)  & -6100(4900) & -3900(4900) & -4300(4600) \\ \hline
$\mathcal K_B$&  ---  &---   &  -$6(24) \cdot 10^3$ & -$10(24) \cdot 10^3$  & -$10(24) \cdot 10^3$ \\ \hline
\hline
dof  &  40-4 & 54-4& 77-6 & 77-7 & 77-7  \\ \hline
$\chi_\text{red}^2$& 1.32 & 1.84 &1.34 & {1.31} &1.31\\ \hline
\hline
$M_K a_0^{KK}$ &   0.3612(66) & 0.3459(51) & 0.3591(50)  & 0.3648(59)& 0.3641(58) \\ \hline 
$M_K^2 a_0^{KK} r_0^{KK}$ & 0.95(30) & 1.22(11) & 1.20(9)  & 0.77(24)& 0.88(21)
\end{tabular}
\caption{Fit results for kaons on the D200 ensemble using $M_K L=9.994083$, with notation as in \Cref{tab:fitN203K}.
 Each fit contains all $2K^+$ and $3K^+$ energy levels (in the appropriate irreps) below $E^*_{2,\rm max}=2.53M_K$ and $E^*_{\rm max}=3.53M_K$, respectively. \\
}
\label{tab:fitD200K}
\end{table}

In ChPT, kaon interactions are generally stronger than those of pions due to their heavier mass.
Thus the higher-order parameters in $\kdf$ might have more impact here.
We have tested this with a full 10-parameter fit on the N200 ensemble, chosen as it
has the greatest statistical significance for the lower-order terms in $\kdf$. 
The resulting fit parameters are given in \Cref{tab:fitfullN200}.  
Comparing to the standard fits in \Cref{tab:fitN200K},
we see that adding more parameters does lead to a better fit (unlike for pions),
with the result for common parameters being consistent with those from the standard fits.
However, the large diagonal errors, and the large correlations (not shown),
indicate that there are redundant directions in parameter space, so it is difficult to extract conclusions.
The only exception is that there is an indication of a nonzero $\cK_A$.

\begin{table}[tbh!]
\centering
\begin{tabular}{c||c}
\multicolumn{2}{c}{ \bf N200($\mathbf{K}$) }\\ \hline  \hline
 & \multicolumn{1}{c}{($2K/3K, \ell\leq 2$) } \\ \cline{2-2}
 & Full fit  \\ \hline  \hline
$z^2$& 0.95(50) \\ \hline
$B_0$&  -3.1(1.5)  \\ \hline
$B_1$&  -2.2(1.3) \\ \hline
$B_2$& -0.3(8)  \\ \hline
$D_0^{KK}$& -45(6)  \\ \hline
$\Kisozero$&1200(600)  \\ \hline
$\Kisoone$& -$10(5) \cdot 10^3$  \\ \hline
$\Kisotwo$&  $6(10) \cdot 10^3$\\ \hline
$\mathcal K_A$&   -$12(4) \cdot 10^3$ \\ \hline
$\mathcal K_B$&  -$11(8) \cdot 10^3$ \\ \hline
\hline
dof  &  51-10   \\ \hline
$\chi_\text{red}^2$& 1.34    \\ \hline
\hline
$M_K a_0^{KK}$ & 0.3355(52) \\ \hline
$M_K^2 a_0^{KK} r_0^{KK}$ & 1.41(37)
\end{tabular}
\caption{Fit results for kaons on the N200 ensemble including 
all possible parameters at the order we are working,
 with $M_K L$ and levels as in fits 3--5 of \Cref{tab:fitN200K}. \\  \vspace{1cm}
}
\label{tab:fitfullN200}
\end{table}

\begin{table}[tbh!]
\centering
\begin{tabular}{c||c||c||c}
&\multicolumn{1}{c||}{ \bf D200($\mathbf{K}$) } &\multicolumn{1}{c||}{ \bf N200($\mathbf{K}$) }  &\multicolumn{1}{c}{ \bf N203($\mathbf{K}$) }  \\ \hline  \hline
 &   {($2K, \ell\leq 2$) }  &  {($2K, \ell\leq 2$) } &  {($2K, \ell\leq 2$) }  \\ \hline  \hline
$-(M_K a_0^{KK})^{-1}$& -2.76(5) &  -2.96(4)   & -3.27(5)  \\ \hline
$M_K r_0^{KK}$& 0.53(14) & 0.71(9)  & 0.84(8)    \\ \hline
$M_K^3 (r_0^{KK})^3 P_0^{KK}$& -0.97(42)  & -1.17(18) & -1.14(14)   \\ \hline
$D_0^{KK}$& -20(3)   & -49(9)  &  -58(7)  \\ \hline
\hline
$M_K L$  & 9.994083 & 7.225143 & 6.908149   \\  \hline
$E^*_{2,\rm max}$ &  $2.53 M_K$  & $2.75M_K$ & $2.90 M_K$   \\  \hline
dof  &  40-4 & 28-4 &   33-4    \\ \hline
$\chi_\text{red}^2$& 1.32  & 1.88 &  1.41  \\ \hline
\hline
$M_K a_0^{KK}$ &   0.3623(62)& 0.3377(49)   & 0.3057(45)  \\ \hline
$M_K^2 a_0^{KK} r_0^{KK}$ & 0.76(19)  & 0.95(11)  &  1.03(9) 
\end{tabular}
\caption{Fit results for two kaons using the ERE form of the $s$-wave phase shift.
We stress that here $M_KL$ is a fixed input to the quantization condition, 
chosen to be consistent with the lattice results in Table~\ref{tab:single-hadron-energies}. \\ }
\label{tab:fit2KERE}
\end{table}

\begin{table}[tbh!]
\centering
\begin{tabular}{c||c||c||c}
&\multicolumn{1}{c||}{ \bf D200($\mathbf{K}$) } &\multicolumn{1}{c||}{ \bf N200($\mathbf{K}$) }  &\multicolumn{1}{c}{ \bf N203($\mathbf{K}$) }  \\ \hline  \hline
  &  ($2K/3K, \ell\leq 2$)  &  ($2K/3K, \ell\leq 2$) &  ($2K/3K, \ell\leq 2$)  \\ \hline  \hline
$-(M_K a_0^{KK})^{-1}$& -2.74(4)   & -2.94(4)  & -3.21(4)  \\ \hline
$M_K r_0^{KK}$ & 0.46(12)  & 0.61(7) & 0.80(6)   \\ \hline
$M_K^3 (r_0^{KK})^3 P_0^{KK}$& -0.73(34)  & -1.00(15) &  -1.07(11)   \\ \hline
$D_0^{KK}$& -20(2) &  -55(9)  & -48(4)  \\ \hline
$\Kisozero$& -1000(1000)   &  -220(410) & -750(210)  \\ \hline
$\Kisoone$&  -3200(4900)  & -8500(1200)  & -2100(900) \\ \hline
$\mathcal K_B$& -$10(24) \cdot 10^3$ & -$27(5)\cdot 10^3$ & -3400(2800)     \\ \hline
\hline
{ $M_K L$ } & 9.994083 & 7.225143 & 6.908149   \\  \hline
{ $E^*_{2,\rm max}$, $E^*_{\rm max}$} &  $2.53 M_K, 3.53 M_K$  & $2.75M_K, 3.75 M_K$  & $2.90M_K, 3.90 M_K$   \\  \hline
dof   &  77-7 & 51-7   &   62-7  \\ \hline
$\chi_\text{red}^2$ & {1.30} & 1.47  & 2.05 \\ \hline
\hline
$M_K a_0^{KK}$ & 0.3654(55)    & 0.3400(45)  &  0.3116(39)  \\ \hline
$M_K^2 a_0^{KK} r_0^{KK}$  & 0.67(16)   &   0.82(9) & 1.00(7)
\end{tabular}
\caption{Fit results for kaons using the ERE form of the $s$-wave phase shift.
Here $M_KL$ is a fixed input to the quantization condition, as in Table~\ref{tab:fit2KERE}. \\ }
\label{tab:fit3KERE}
\end{table}

Finally,  
a natural question to ask is whether is it justified 
to include the Adler zero in the parametrization of two-kaon interactions. 
This is the case at the SU(3)-flavor symmetric point, since the kaon interactions there are identical
to those of pions. 
However, the situation when $M_K \gg M_\pi$ is unclear. 
To study this, we have performed fits on all ensembles using
the standard ERE parametrization of the $s$-wave phase shift,
both for the two-kaon spectrum alone and for the combined two- and three-kaon spectra. 
The results are given, respectively, in \Cref{tab:fit2KERE,tab:fit3KERE}.

The results indicate that the ensemble N203, closest to the SU(3) point, 
seems to be better fit by an Adler-zero parametrization. 
By contrast, results from N200 and D200 are equally well described by the ERE form,
with the resulting fit parameters being similar.
This differs from what we found with pions, where the ERE form was disfavored.
One reason for this difference might be that the kaon fits are over a smaller energy range,
and thus lie within the range of convergence of the ERE.




 
\section{Discussion of results} 
\label{sec:discussion}

We now turn to an analysis of the K-matrix parameters obtained from the fits
presented in the preceding section.
We compare their dependence on $M_\pi$ and $M_K$ to the predictions from
ChPT described in \Cref{subsec:ChPT}.
We discuss the parameters for pion and kaon scattering in two separate subsections.

We begin with some general comments.
First, in all fits, we consider only the uncertainty in the scattering parameters (the $y$ coordinates),
treating the $x$ coordinates (which are related to $M_\pi/F_\pi$ and $M_K/F_K$) as error-free. 
This is justified because the relative uncertainty in $x$ coordinates is about an order of magnitude smaller than those in the $y$ coordinates (see \Cref{tab:decay_constants}).
Second, in our fit results and chiral extrapolations, we quote only the statistical uncertainty and
the systematic uncertainty due to the choice of fit.
We do not account for discretization errors, which are of $\mathcal O(a^2)$ and thus
could be $\sim 5\%$, as is the case for the decay constants~\cite{Bruno:2016plf}. We also neglect volume dependence of the form $\exp(-M_\pi L)$ (likely to be at or below the percent level).
Thus our results have systematic errors that are not fully controlled, but break new ground by providing the first lattice results for some quantities.
Finally, when we present extrapolations to the physical point, we use the physical charged
pion and kaon masses ($139.57$ and $493.677\;$MeV, respectively~\cite{Zyla:2020zbs})
and the corresponding decay constants ($130.2$ and $155.7\;$MeV, respectively~\cite{Zyla:2020zbs}).

\subsection{Results for multi-pion systems}
\label{subsec:piondiscussion}

We start with the quantities that describe two- and three-pion scattering amplitudes. 
In the two-pion sector, we will discuss the threshold parameters,  the mass dependence of the position of the Adler zero, and $d$-wave interactions. We compare our results for three pions with previous work that included only $s$-wave interactions, and discuss the chiral behavior of the components of $\kdf$.

\subsubsection{Two-pion threshold parameters}
\label{subsubsec:pionthreshold}

Here we analyze the threshold parameters of two pions at maximal isospin,
the scattering length, $a_0^{\pi\pi}$, and the effective range, $r_0^{\pi\pi}$. 
This system has been studied using LQCD in many earlier works~\cite{Feng:2009ij,Helmes:2015gla,Culver:2019qtx,Mai:2019pqr, Bulava:2016mks,Yamazaki:2004qb,Beane:2005rj, Beane:2007xs,Beane:2011sc,Yagi:2011jn,Fu:2013ffa, Sasaki:2013vxa,Fischer:2020jzp,Blanton:2019vdk,Blum:2021fcp}. 
Threshold parameters have been obtained using two approaches: 
(i) applying the threshold expansion to order $L^{-5}$ to extract the scattering length;
and (ii) using the full spectrum to determine the phase shift as a function of $q^2$.
We use the latter approach, which is required to obtain the effective range,
and uses the input from many more spectral levels.

As explained in the previous section,
we take the results from the minimal fit (fit 3).
We add a systematic uncertainty associated with the dependence on choice of fit,
taken to be the standard deviation of the results of fits 3-5.
The inputs to the chiral fits are summarized in \Cref{tab:th3pi}, 
where we also include the leading-order ChPT predictions for comparison.

\begin{table}[h!]
\centering
\begin{tabular}{c||c|c|c|c}
Ensemble & $M_\pi a^{\pi\pi}_0$  & $\left(M_\pi a^{\pi\pi}_0\right)^\text{LO}$  & $M_\pi^2 r^{\pi\pi}_0 a^{\pi\pi}_0$  &  $\left(M_\pi^2 r^{\pi\pi}_0 a^{\pi\pi}_0\right)^\text{LO}$  \\ \hline \hline
N203($\pi$) &$0.2059(34)_\text{st}(21)_\text{fit}$ & 0.2345(12)  & $2.15(5)_\text{st}(12)_\text{fit}$  &3  \\ \hline
N200($\pi$)   &$0.1550(40)_\text{st}(15)_\text{fit}$ & 0.1747(12)  & $2.41(6)_\text{st}(7)_\text{fit}$  &3\\ \hline
D200($\pi$)  &$0.0859(41)_\text{st}(28)_\text{fit}$ & 0.0970(6) & $2.62(8)_\text{st}(26)_\text{fit}$ & 3 \\ 
\end{tabular}
\caption{Two-pion threshold parameters used for chiral fits, obtained as described in the text, along with the leading-order predictions of ChPT.}
\label{tab:th3pi}
\end{table}

\begin{figure}[h!]
\begin{subfigure}{.5\textwidth}
  \centering
  \includegraphics[width=\linewidth]{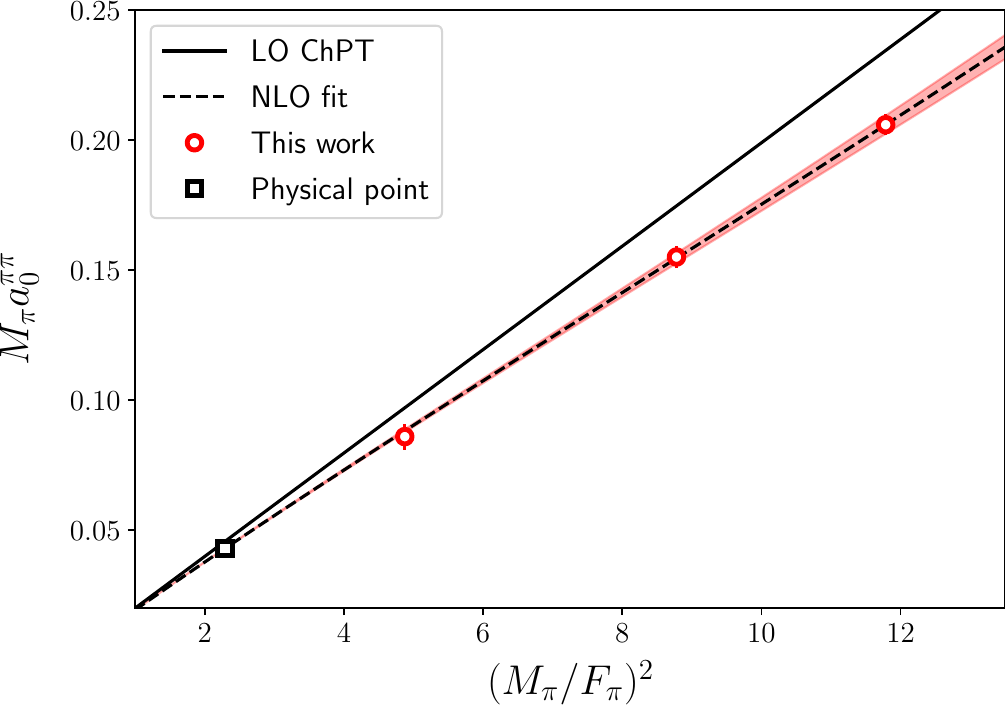}  
  \caption{$M_\pi a_0^{\pi\pi}$.}
  \label{fig:2pia0}
\end{subfigure}
\begin{subfigure}{.5\textwidth}
  \centering
  \includegraphics[width=\linewidth]{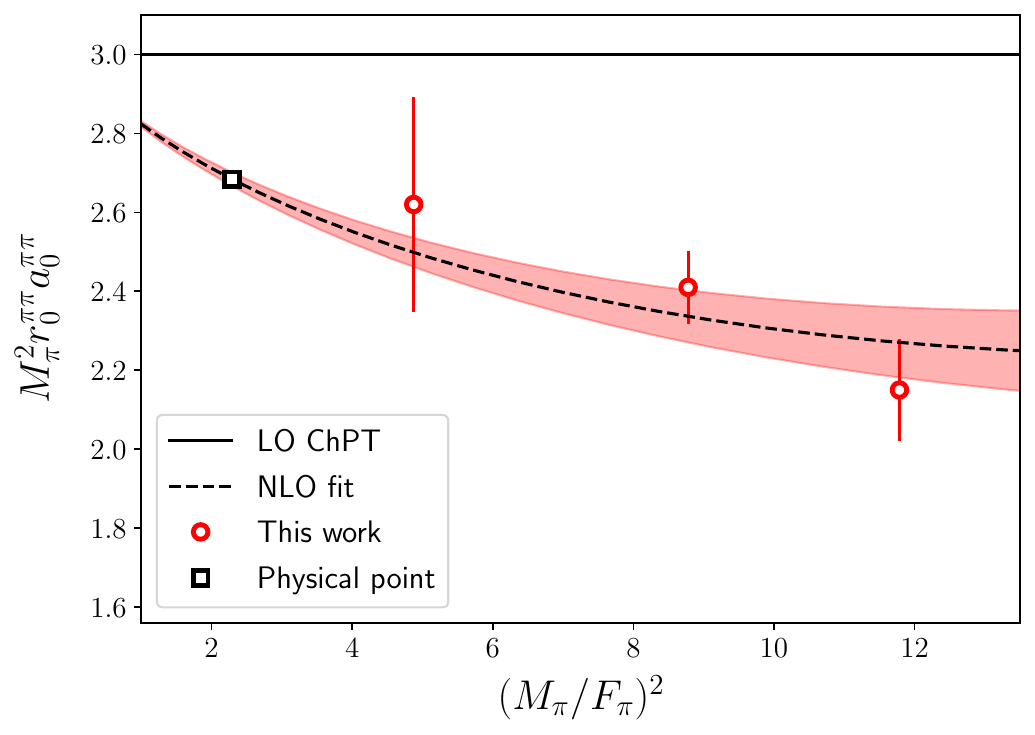}  
  \caption{$M_\pi^2 r_0^{\pi\pi} a_0^{\pi\pi}$.}
  \label{fig:2pir0}
\end{subfigure}
\caption{Chiral fits of the threshold parameters of the isospin-2 $\pi\pi$ system. Solid lines correspond to the LO ChPT predictions, which have no free parameters. The dashed line is a NLO fit to \Cref{eq:a0pi_SU2,eq:r0pi_SU2}. We also include the $1\sigma$ region as a shaded area. 
Circles mark our data points, while squares denote the extrapolated values at the physical point.}
\label{fig:th2pi}
\end{figure}

We fit these results to the SU(2) ChPT expressions, given in \Cref{eq:a0pi_SU2,eq:r0pi_SU2}. 
The fit to the scattering length yields
\begin{equation}
\ell_{\pi\pi} = 7.6(4)\,, \quad \chi^2/\text{dof} = 0.22/(3-1)\,,
\end{equation}
while for the effective range we obtain
\begin{equation}
C_3 = -0.29(2)\,, \quad \chi^2/\text{dof} = 1.7/(3-1)\,.
\end{equation}
We plot these fits in \Cref{fig:th2pi}, which show the complete consistency with
the ChPT expressions.
The extrapolation to the physical point yields
\begin{equation}
(M_\pi a^{\pi\pi}_0)_\text{phys} = 0.0429(1) \,, \quad (M_\pi^2 r_0^{\pi\pi} a^{\pi\pi}_0)_\text{phys} = 2.68(2)\,, \label{eq:a0physSU2}
\end{equation}
also shown in the figures.
We stress again that the errors do not include all sources of systematic uncertainty,
so that these numbers cannot be quantitatively compared to those from other works.
However, we do note that the result for the scattering length agrees within $\sim 3\%$ with the
only fully controlled result (according the FLAG review~\cite{Aoki:2019cca}), 
\begin{equation}
  (M_\pi a^{\pi\pi}_0)_\text{phys} = 0.0442(2)({}^4_0) \,, \qquad [\text{Ref.~\cite{Helmes:2015gla}}]\,.
\end{equation}

\subsubsection{Position of the Adler zero}

In the context of LQCD determinations of the isospin-2 $\pi\pi$ $s$-wave phase shift,
it was proposed in Ref.~\cite{Blanton:2019vdk} to use the 
Adler-zero parametrization, \Cref{eq:Adler0}, 
in place of the ERE form used previously, \Cref{eq:ERE0}.
Explicit inclusion of the Adler zero extends the range of validity in $q^2$,
and is particularly important when including subthreshold amplitudes ($q^2 < 0$),
which is unavoidable when fitting the three-particle spectrum~\cite{Polejaeva:2012ut}.
Indeed, it was found in Ref.~\cite{Blanton:2019vdk}, 
and subsequently supported by the results of Ref.~\cite{Fischer:2020jzp}, 
that the Adler-zero form led to better fits than those using the ERE.
However, in most fits done to date,
the position of the Adler zero has been fixed to its leading-order value in ChPT, $z^2=1$.  
Since our fits are more precise than those done previously, 
we can attempt  to study the chiral behavior of the Adler zero, 
and compare to the NLO prediction from ChPT, \Cref{eq:SU2_Adler}.

\begin{figure}[h!]
  \centering
  \includegraphics[width=0.5\linewidth]{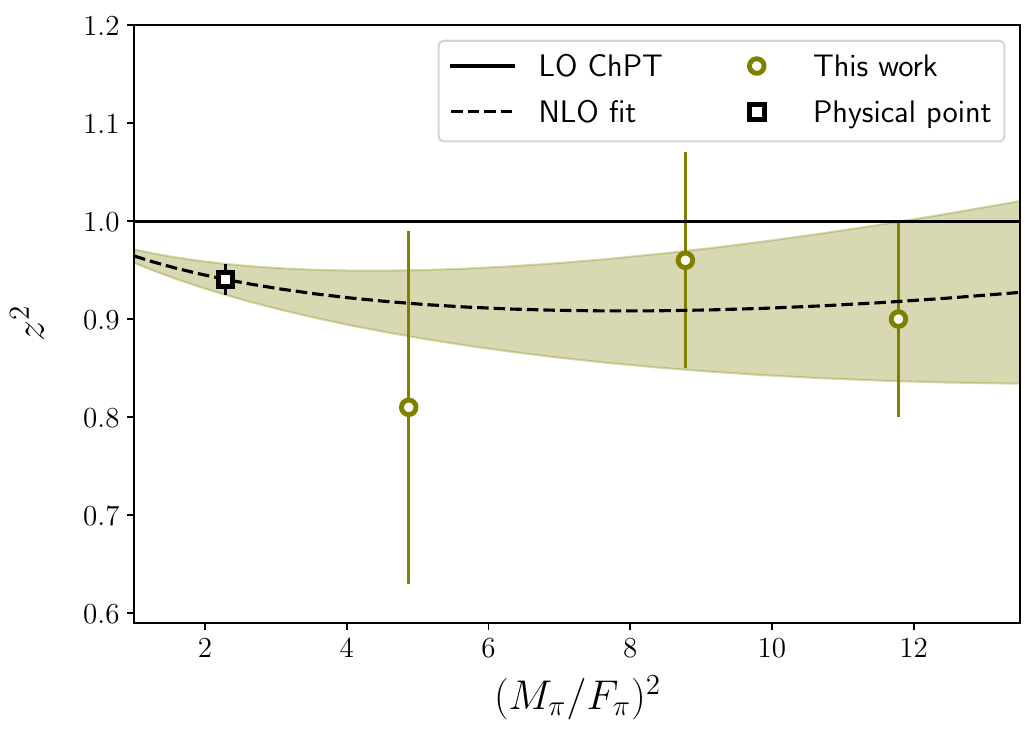}  
  \caption{NLO chiral fit to the position of the Adler zero. Notation as in \Cref{fig:th2pi}. }
  \label{fig:adler}
\end{figure}

We use the results from our fits in which $z^2$ is a free parameter 
(fit 4 in  \Cref{tab:fitpiN203,tab:fitpiN200,tab:fitpiD200} above).
We find 
\begin{equation}
\ell_z = 8.5(2.2), \quad \chi^2/\text{dof} = 0.6/(3-1)\,,
\end{equation}
and show the fit in \Cref{fig:adler}. 
The extrapolation to the physical point yields
\begin{equation}
(z^2)_\text{phys} = 0.94(2)\,.
\end{equation}
Nevertheless, at the level of precision achieved here, it is reasonable to set $z^2=1$,
as the magnitude of the NLO correction is smaller 
than the statistical uncertainty in our determination of $z^2$.

\subsubsection{Two-pion $d$-wave scattering length}

We next analyze our results for $d$-wave interactions.  While the $s$-wave interaction in the $I=2$ channel has been studied extensively on the lattice,
much less is known for the $d$-wave interaction, which to our knowledge has only been
extracted in Refs.~\cite{Dudek:2010ew,Dudek:2012gj,Fischer:2020jzp}.

At physical quark masses, the isospin-2 $d$-wave phase shift exhibits an interesting feature: dispersion  relations, Roy equations, and chiral perturbation theory predict a change of sign near threshold. It starts out positive (attractive) for very small $q^2$
and then passes through zero and becomes increasingly negative 
(repulsive)~\cite{Colangelo:2001df, Pelaez:2004vs, Kaminski:2006yv, Kaminski:2006qe, GarciaMartin:2011cn}.
In terms of $k^5 \cot \delta_2$, this behavior implies a pole slightly above threshold.
It is unclear, however, whether this phenomenon persists for heavier pion masses.
Indeed, Ref.~\cite{Nebreda:2011di} studied $\delta_2$ using ChPT,
and found poor convergence for the region of  pion masses of relevance here.

 In practice, probing the near-threshold dependence of $\delta_2$ using the L\"uscher method
is difficult for two reasons. First, $\delta_2$ is small close to threshold due to 
its scaling as $q^5$,
resulting in tiny shifts of the energy spectrum.
Second, the energy levels most sensitive to the $d$-wave interaction lie
well above threshold for typical values of $M_\pi L$ currently used,
and are thus sensitive to $\delta_2$ only in the region where the phase shift is expected to be negative. 
For this reason we opt to use the simple one-parameter form \Cref{eq:simpledelta2},
which implies a uniformly negative phase shift in all our fits, with $D_0<0$.
As noted in \Cref{subsec:ChPT}, $\delta_2$ vanishes at LO in ChPT.
We therefore expect the $d$-wave scattering length to behave as 
\begin{equation}
M_\pi^5 a_2^{\pi\pi} = -\frac1{D^{\pi\pi}_0-1} \propto \left(\frac{M_\pi^2}{F_\pi^2}\right)^2 \,,
\end{equation}
up to logarithmic corrections, and choose to fit to this simple power-law dependence.

\begin{figure}[h!]
  \centering
  \includegraphics[width=0.5\linewidth]{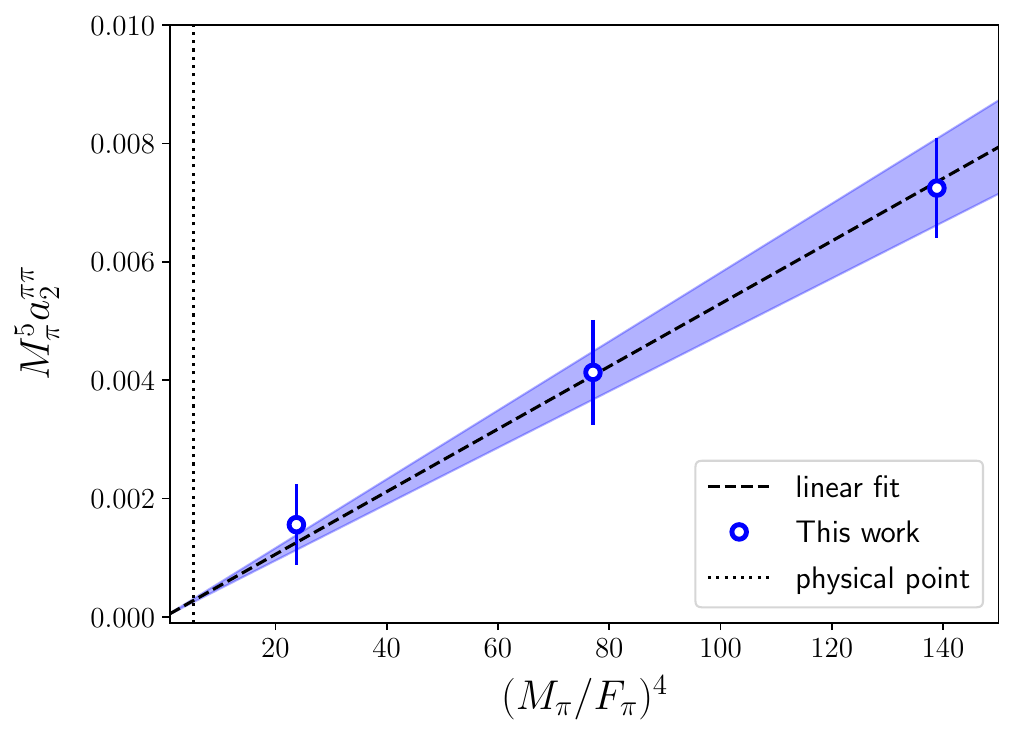}  
  \caption{Dependence of the $d$-wave scattering length on $(M_\pi/F_\pi)^4$. 
  We include a linear fit based on chiral expectations, 
  and indicate the position of the physical point.
  }
\label{fig:dwavedpi}
\end{figure}

 As shown in \Cref{fig:dwavedpi}, our results are in excellent agreement with this behavior. 
Given the possibility discussed above of rapid changes in $\delta_2$ near threshold,
we refrain from quoting a value extrapolated to the physical point.

We can make a qualitative comparison to the results of Ref.~\cite{Fischer:2020jzp}.
The values of the scattering lengths in the two works are of the same order of magnitude, but those
from Ref.~\cite{Fischer:2020jzp} do not show the same monotonic chiral behavior.
The largest tension is at our heaviest pion mass, which is comparable to the heaviest
pion mass used in Ref.~\cite{Fischer:2020jzp}.
We note that our results are based on a global fit including many two- and three-pion
levels in irreps that are sensitive to the $d$-wave amplitude, whereas those from
Ref.~\cite{Fischer:2020jzp} are from just a few two-pion levels in the nontrivial irreps. 
Another difference is that Ref.~\cite{Fischer:2020jzp} employs an $N_f=2$ simulation,
compared to ours that uses $N_f=2+1$ sea quarks. Relative cutoff effects may also be significant, given that our lattice spacing is significantly finer. Thus we view our results as more reliable, although further work will be needed to understand the differences.

\subsubsection{Comparison to previous $s$-wave three-pion fits}

The three-pion coupling at maximal isospin has been extracted in the RFT approach~\cite{Blanton:2019vdk,Fischer:2020jzp,Hansen:2020otl}, the FVU approach~\cite{Mai:2018djl,Brett:2021wyd}, and via the threshold expansion~\cite{Beane:2007es,Beane:2020ycc}. All these studies have included only 
$s$-wave contributions to the two- and three-pion interactions.
In this section, we compare results that we have obtained from fits that closely match those 
used in these previous studies, i.e.,~restricted to $s$-wave interactions and using only
a subset of the moving frames and levels.
Specifically, we compare to the RFT results of Refs.~\cite{Blanton:2019vdk,Fischer:2020jzp}, 
which use the same fit model as this work, allowing for a direct comparison.
We stress, however, that, as seen in \Cref{sec:results}, fitting without including
$d$-wave terms leads to poorer fits (even if only trivial irreps are included).
Thus, the present comparison must be understood as a consistency check.

\begin{table}[h!]
\centering
\begin{tabular}{cccccccc}
Fit                         & $B_0$ & $ B_1$ & $M_\pi^2\Kisozero$ & $M_\pi^2\Kisoone$ & $\chi^2_\text{red}$& $M_\pi a^{\pi\pi}_0$ & $M_\pi^2 r_0^{\pi\pi} a^{\pi\pi}_0$ \\ \hline\hline
This work & -12.8(8)  & -2.2(4)    & -380(190)  & 220(160)     & 0.85    & 0.078(5) &   2.66(8)
\\ \hline
 Ref.~\cite{Blanton:2019vdk}  &   -11.1(7) &  -2.4(3)    & 550(330) & -280(290)   & 1.45 & 0.090(5) & 2.57(8)
\end{tabular}
\caption{Global fits to the two- and three-pion spectrum of D200 including only $s$-wave interactions. We use the 22 energy levels as in Ref.~\cite{Blanton:2019vdk}: eleven $2\pi^+$ levels (all in trivial irreps), and eleven $3\pi^+$ levels, including three in nontrivial irreps.
}
\label{tab:compareD200}
\end{table}

We start with a direct comparison to Ref.~\cite{Blanton:2019vdk}. 
That work used the two- and three-pion spectra of Ref.~\cite{Horz:2019rrn}, 
which were calculated on the D200 ensemble also used here. 
The present determination of the spectrum differs in several ways:
(i) we increased statistics by including measurements on more gauge configurations as
well as adding a second source time per configuration;
(ii) we rebin data in blocks of three configurations to ameliorate autocorrelation;
(iii) following these changes, we re-assessed the excited-state systematics of the spectrum extraction.
The fit results are shown in \Cref{tab:compareD200}, 
along with those from Ref.~\cite{Blanton:2019vdk}. 
Note that this is a new fit, different from those presented in \Cref{subsec:pionfits}.
We stress that we are fitting to the same set of levels,
so differences in fit parameters result solely from updates in the energies of the levels.
We observe a substantial difference between the two fits, which, including correlations,
is about $3\sigma$. 
In particular, the determination of $M_\pi^2 \Kisozero$,
which was found to be different from zero at the $2\sigma$ level in Ref.~\cite{Blanton:2019vdk},
is called into question in view of the change of sign in the new fit.

Our interpretation of the discrepancy is the following. First, a larger rebinning reduces 
the impact of the autocorrelation between samples, and leads to more reliable determinations
of the energy levels. 
Second, the increase in statistics enables the use of larger values of $t_\text{min}$,
which reduces the contamination from excited states. 
In summary, we now think that there may be systematic errors in
the spectrum of Ref.~\cite{Horz:2019rrn} that have not been accounted for.
Since the impact of three-particle interactions on the spectrum is small, such effects can
lead to significant systematic errors in the parameters in $\Kdf$. Note that in our preferred fit which includes $d$-wave interactions, the value of $M_\pi^2 \Kisozero$ is shifted by more than one standard deviation---compare \Cref{tab:compareD200} and fit 3 in \Cref{tab:fitpiD200}. This is another indication of the challenge in determining these parameters.

\begin{figure}[h!]
\begin{subfigure}{.5\textwidth}
  \centering
  \includegraphics[width=\linewidth]{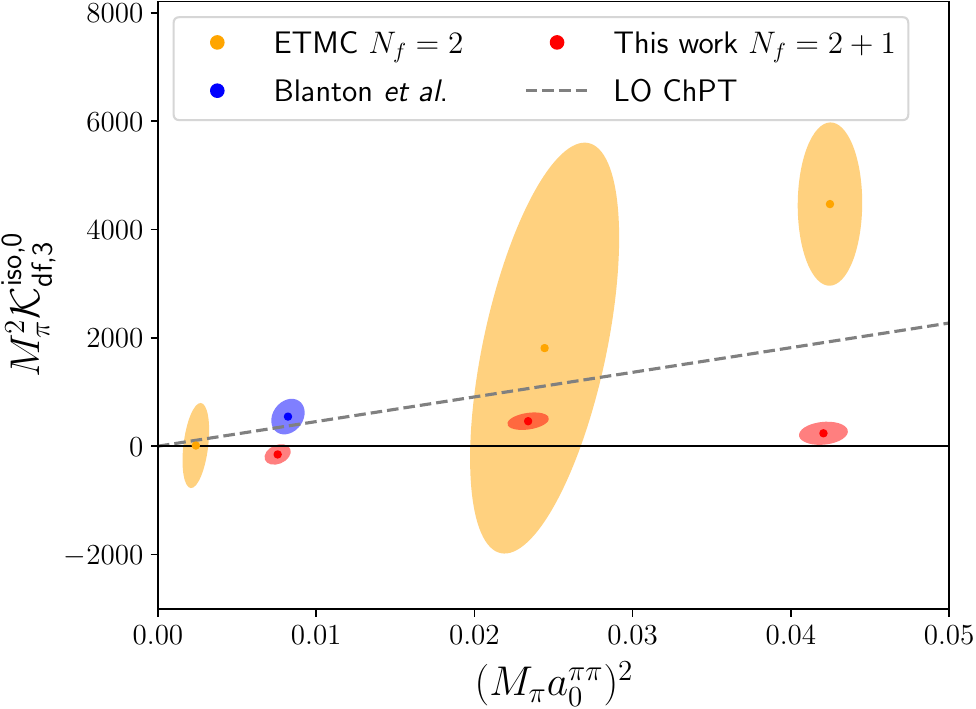}  
  \caption{$\kdf^\text{iso,0}$}
\end{subfigure}
\begin{subfigure}{.5\textwidth}
  \centering
  \includegraphics[width=\linewidth]{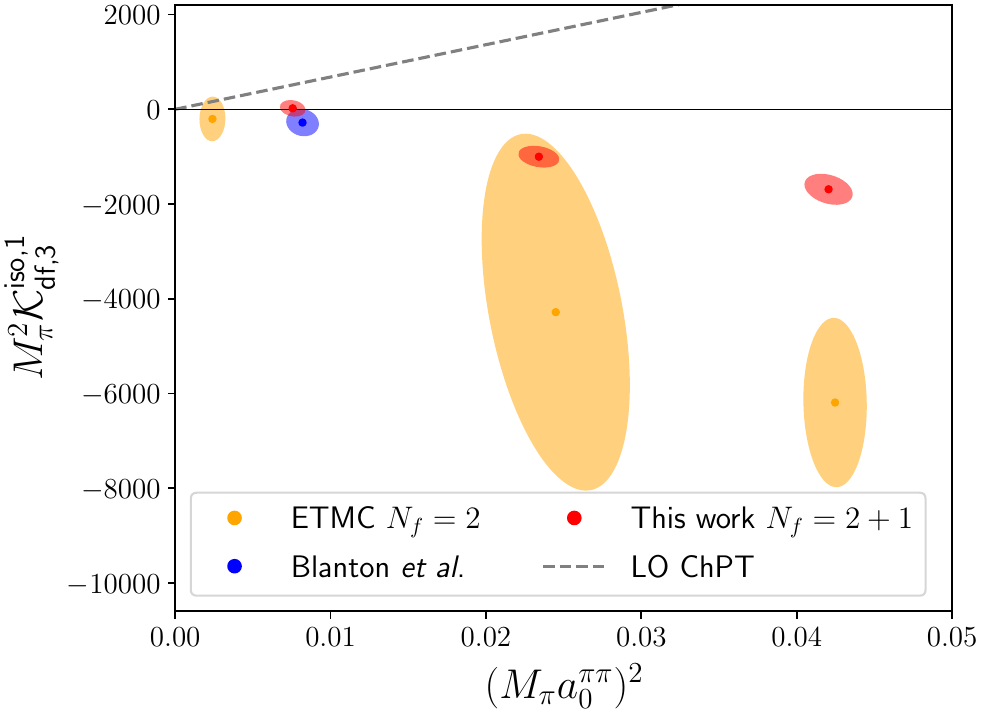}  
  \caption{$\kdf^\text{iso,1}$}
\end{subfigure}
\caption{Three-particle scattering quantities from fits using the RFT formalism, 
and the $s$-wave-only parametrization.
For our work this is fit 2 in \Cref{tab:fitpiN203,tab:fitpiN200,tab:fitpiD200}. 
Results for
Refs.~\cite{Blanton:2019vdk,Fischer:2020jzp} are also shown.
}
\label{fig:Kswave}
\end{figure}

We end this discussion with a plot comparing the determinations of 
the $s$-wave components of $\kdf$ from this work
with those from Ref.~\cite{Fischer:2020jzp}, as well as Ref.~\cite{Blanton:2019vdk}.
 Here we use the results of the full $s$-wave-only fits (fit 2), 
since these use the same fit functions as Refs.~\cite{Blanton:2019vdk,Fischer:2020jzp}.
 There are however some differences between our fit 2 and Refs.~\cite{Blanton:2019vdk,Fischer:2020jzp}: (i) fit 2 includes only trivial irreps, while Refs.~\cite{Blanton:2019vdk,Fischer:2020jzp} include both trivial and nontrivial irreps in the three-pion sector, and (ii) here we use more moving frames than in previous work.
The comparison is shown in \Cref{fig:Kswave}, which also shows the LO predictions from 
ChPT from \Cref{eq:Kdf_LO}. For this figure alone we use $M_\pi a_0^{\pi\pi}$ for the
$x$ axis, as this has been used in the prior works. As a consequence the error bars
now become ellipses.

We first note that, for the two higher-mass ensembles, most of our results (red ellipses) are statistically
different from zero, unlike for the D200 ensemble. The figure also shows that the just-discussed
tension with Ref.~\cite{Blanton:2019vdk} (blue ellipses) remains when we use a fit to
levels in the trivial irrep in  all frames.
What we see in addition is a significant tension at the heaviest pion mass 
with the results from the ETMC collaboration~\cite{Fischer:2020jzp} (orange ellipses).
Just as for the difference in the $d$-wave scattering length, it is difficult to understand this tension.
Possible sources of difference are that
the two calculations correspond to different lattice regularizations, 
and thus have different $O(a^2)$ discretization errors,
and the fact that the ETMC result is for $N_f=2$, 
while our N203 ensemble is close to the SU(3)-symmetric point. 
More investigation will be needed to understand how these effects enter in $\kdf$.

\subsubsection{Results for $\kdf$ from full fits including $d$ waves}

We conclude the discussion for pions by presenting our final results for the terms in $\kdf$,
which are obtained from fits to levels in all irreps including $d$-wave interactions.
We recall that such fits provide a much better description of
the spectrum than fit 2, the results from which are discussed
in the previous section for comparative purposes.
We have results for $\Kisozero, \Kisoone,$ and $\mathcal{K}_B$, and, as before,
use fit 3 for our central values.

\begin{figure}[h!]
\begin{subfigure}{.5\textwidth}
  \centering
  \includegraphics[width=\linewidth]{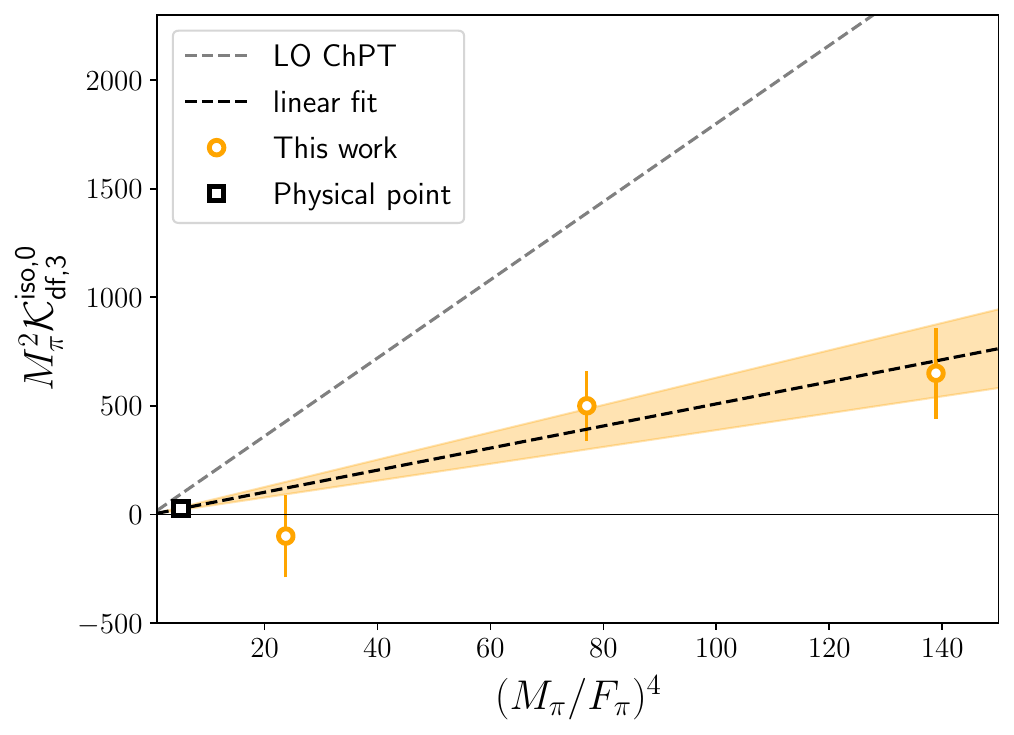}  
  \caption{$\kdf^\text{iso,0}$}
\end{subfigure}
\begin{subfigure}{.5\textwidth}
  \centering
  \includegraphics[width=\linewidth]{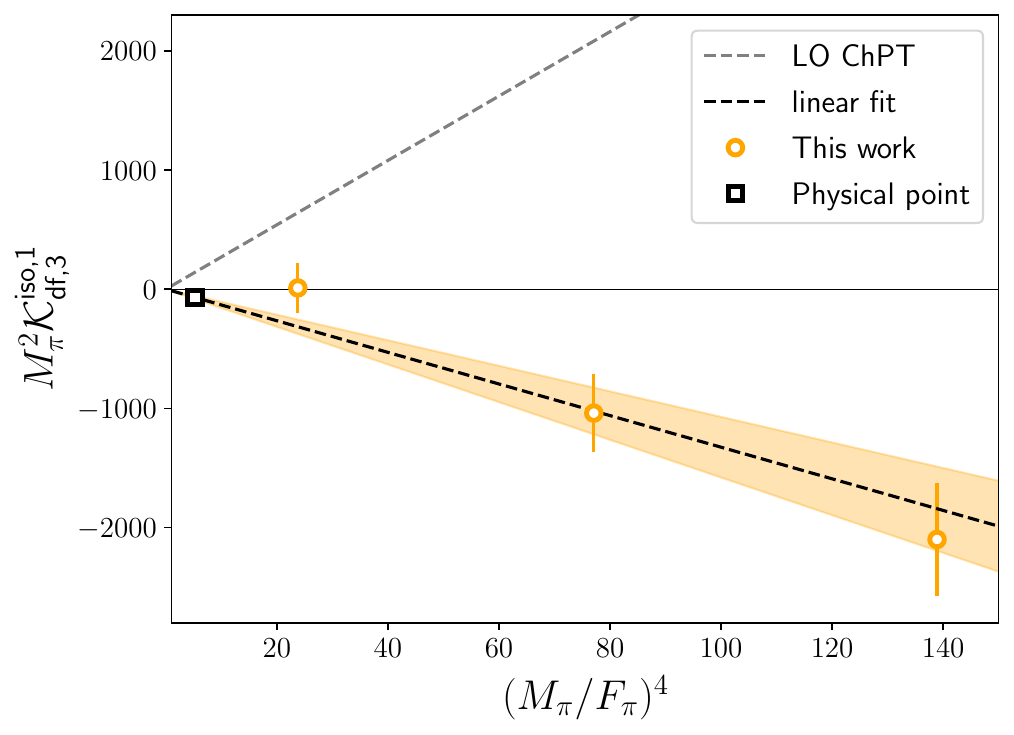}  
  \caption{$\kdf^\text{iso,1}$}
\end{subfigure}
\caption{Results for the isotropic parameters in $\kdf$ for pion scattering
plotted against the expected leading chiral behavior, $(M_\pi/F_\pi)^4$. 
Notation as in \Cref{fig:th2pi}.
We show the LO ChPT result for comparison.
}
\label{fig:Kisodwave}
\end{figure}

 The chiral dependence of the isotropic parameters is 
  shown in \Cref{fig:Kisodwave}.
We observe statistically significant deviations from zero
at heavier pion masses.
Our results are consistent with a linear chiral dependence on $(M_\pi/F_\pi)^4$ as expected from
ChPT, see \Cref{eq:Kdf_LO}. However, the respective slopes are in significant tension
with the ChPT predictions,
hinting at a possible breakdown in the convergence of ChPT.
Indeed, as noted already in Ref.~\cite{Blanton:2019vdk}, there are reasons to expect large
NLO corrections to $\Kdf$.
A final comment concerns the results extrapolated to the physical point: since $\kdf$ vanishes rapidly towards the chiral limit, it will be very difficult in practice to extract the three-pion interaction
directly at this point.

\begin{figure}[h!]
  \centering
  \includegraphics[width=0.5\linewidth]{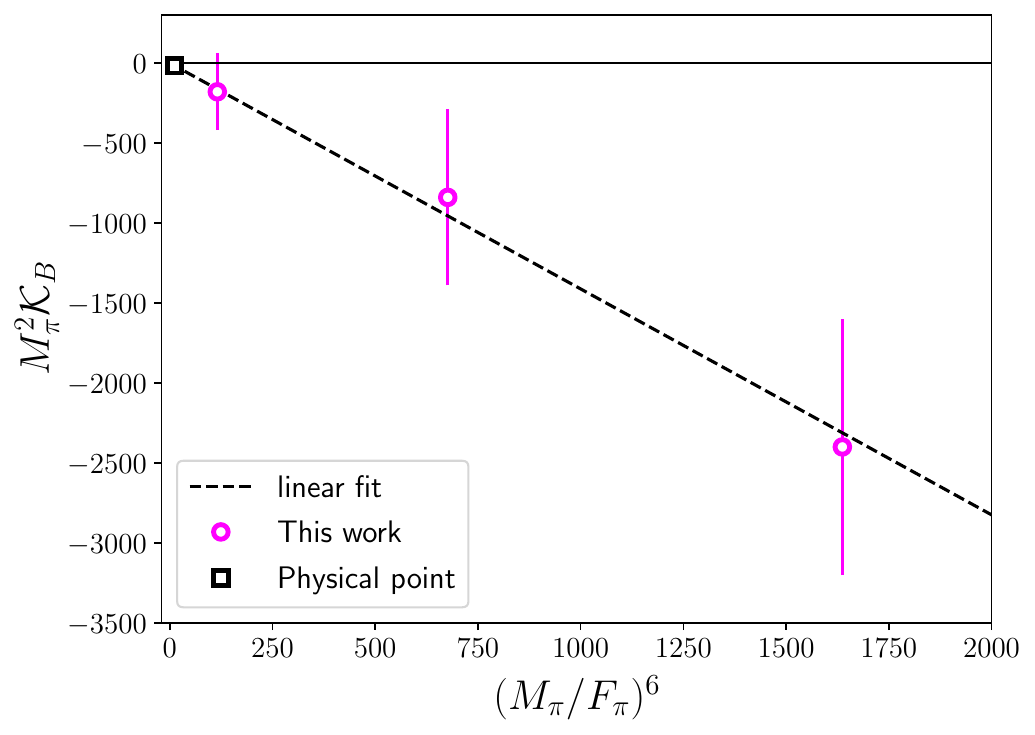}  
  \caption{Chiral scaling of $\cK_B$ for pions as a function of $(M_\pi/F_\pi)^6$,
  including a linear fit.  Notation as in \Cref{fig:th2pi}.}
\label{fig:KBpipi}
\end{figure}

We now turn to $\cK_B$.
As explained in \Cref{subsec:ChPT}, while no ChPT prediction is available, 
we expect the chiral scaling
\begin{equation}
M_\pi^2 \cK_B \propto \left( \frac{M_\pi}{F_\pi} \right)^6 \,,
\end{equation}
up to logarithms.
This expectation is borne out by our results shown in \Cref{fig:KBpipi}.
The extraction of $\cK_B$, despite being of quadratic order in the threshold expansion,
is aided by it being the only contribution from $\kdf$ to a set of energy levels in nontrivial irreps.
Similarly to the isotropic parameters, the extrapolation indicates a very small value for
$\cK_B$ at the physical point, which will presumably be difficult to extract directly from
simulations with close-to-physical quark masses.

\subsection{Results for multi-kaon systems}

Unlike for the isospin-2 $\pi\pi$ system, there are relatively few studies of two kaons at maximal isospin~\cite{Beane:2007uh,Sasaki:2013vxa,Helmes:2017smr,Alexandru:2020xqf}. 
Furthermore, the three-kaon K-matrix has yet to be explored, 
as the only other study of three kaons at maximal isospin set it to zero~\cite{Alexandru:2020xqf}. 
Thus we provide here the first exploration of this quantity.

The remainder of this section has a similar structure to that for pions. As discussed in \Cref{subsec:kaonfits},
we use fit 4, which includes the position of the Adler zero as a free parameter,
as our reference fit.

\subsubsection{Two-kaon threshold parameters }

As in the two-pion case, we start by looking at the two-kaon threshold parameters. 
The results are summarized in \Cref{tab:threshold}, 
along with the LO chiral predictions.
The central values are from fit 4 in \Cref{tab:fitN203K,tab:fitN200K,tab:fitD200K},
while the systematic uncertainty is given by the standard deviation of the results from
all the fit models, including the ERE fits on the D200 and N200 ensembles.
The deviation from LO ChPT is more pronounced for kaons than for pions (\Cref{tab:th3pi}), indicating poorer convergence of the chiral expansion.

\begin{table}[t!]
\centering
\begin{tabular}{c||c|c|c|c}
Ensemble & $M_K a^{KK}_0$& $\left(M_K a^{KK}_0\right)^\text{LO}$  & $M_K^2 r^{KK}_0 a^{KK}_0$  & $\left(M_K^2 r_0^{KK} a^{KK}_0\right)^\text{LO}$  \\ \hline \hline
N203 &$0.3012(44)_\text{st}(18)_\text{fit}$& 0.3431(12)  & $1.92(19)_\text{st}(14)_\text{fit}$ & 3   \\ \hline
N200 &$0.3358(50)_\text{st}(21)_\text{fit}$&  0.3761(19)& $1.36(23)_\text{st}(26)_\text{fit}$ & 3\\ \hline
D200 &$0.3648(59)_\text{st}(29)_\text{fit}$& 0.4052(16) & $0.77(24)_\text{st}(23)_\text{fit}$ & 3\\ 
\end{tabular}
\caption{Two-kaon threshold parameters from fit 4 in \Cref{tab:fitN203K,tab:fitN200K,tab:fitD200K},
with systematic errors due to choice of fit obtained as discussed in the text. } 
\label{tab:threshold}
\end{table}

\begin{figure}[t!]
\begin{subfigure}{.5\textwidth}
  \centering
  \includegraphics[width=\linewidth]{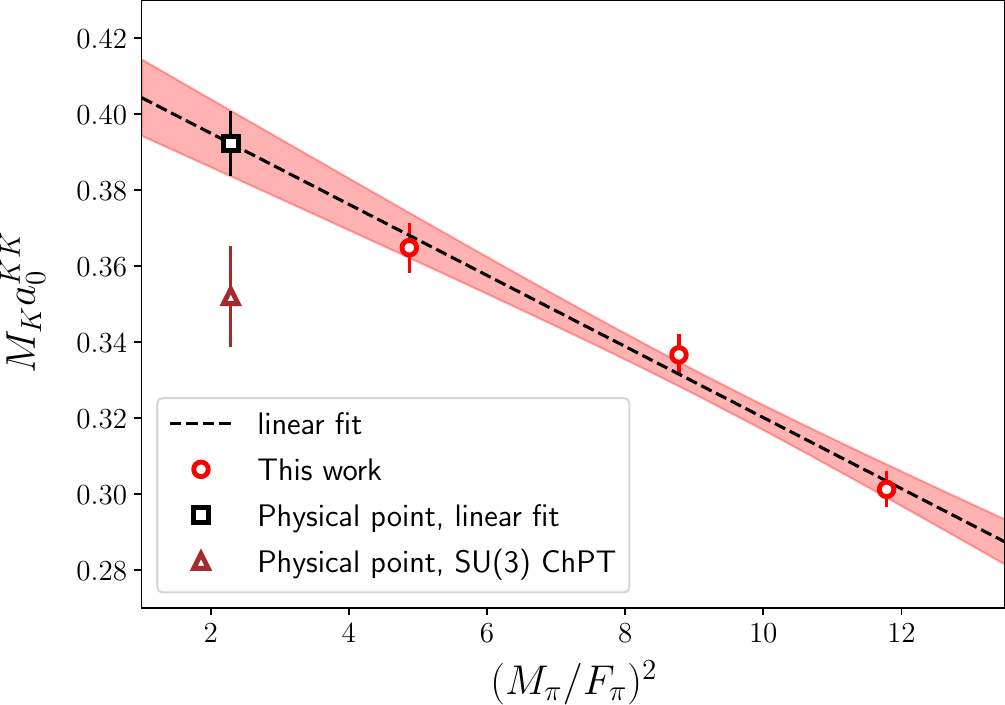}  
  \caption{$M_K a_0^{KK}$}
  \label{fig:scattlenK}
\end{subfigure}
\begin{subfigure}{.5\textwidth}
  \centering
  \includegraphics[width=\linewidth]{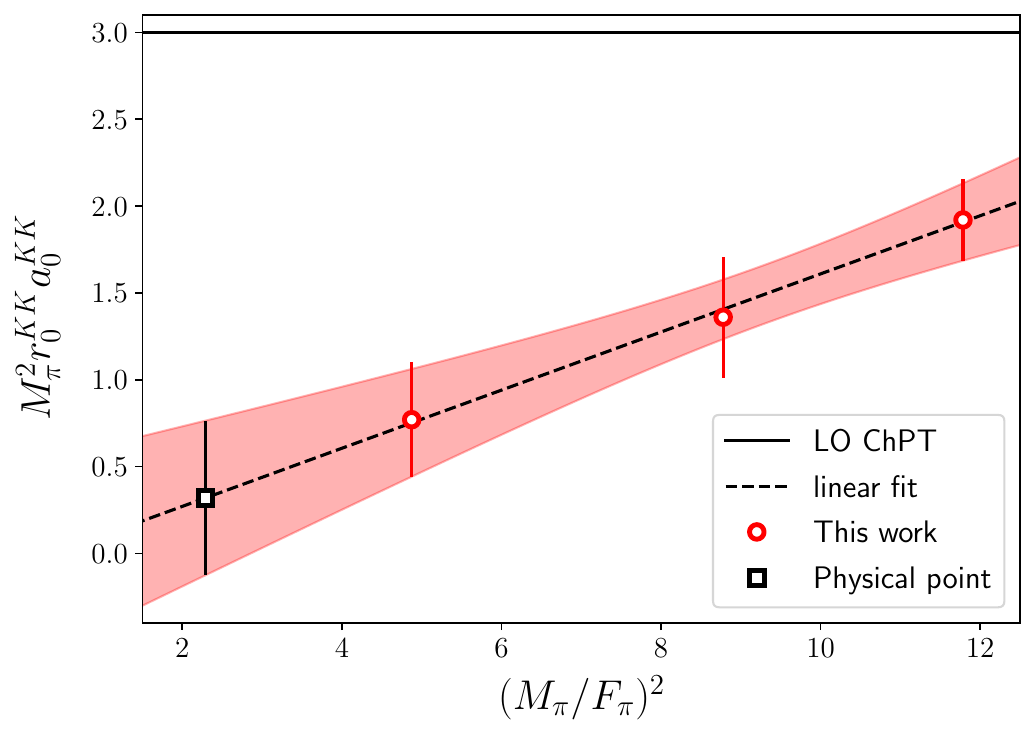}  
  \caption{$M_K^2 r_0^{KK} a_0^{KK}$}
  \label{fig:effrangeK}
\end{subfigure}
\caption{Results for the $s$-wave two-kaon scattering length and effective range.
Notation as in \Cref{fig:th2pi}, except that in the left panel we include, with an empty triangle,
the result for the chiral extrapolation from an SU(3) ChPT fit to both kaon and pion
scattering lengths.}
\label{fig:2Kth}
\end{figure}

The results for the scattering length and effective range are shown in \Cref{fig:2Kth}.
For both quantities we perform a linear fit in $(M_\pi/F_\pi)^2$, motivated by SU(2) ChPT as explained
in \Cref{subsec:ChPT},
leading to the following values at the physical point
\begin{equation}
\left(M_K a^{KK}_0\right)_\text{phys} = 0.390(9) \,, \qquad
\left(  M_K^2 r_0^{KK} a_0^{KK} \right)_\text{phys} = 0.3(4)\,,
\label{eq:a0KphysSU2}
\end{equation}
where the quoted uncertainties are only statistical.

We can also  perform a simultaneous SU(3) ChPT fit to the kaon and pion scattering lengths
using the NLO results \Cref{eq:a0pi_SU3,eq:a0K_SU3}.
We stress that a single, common LEC enters in both expressions, so that we have one
free parameter to describe six data points.
In the fit, we ignore the (small) uncertainties in $(M_\pi/F_\pi)^2$---as we do for all fits presented in
this section---as well as the correlations between $a^{KK}_0$ and $a_0^{\pi\pi}$
evaluated on the same ensemble.
The resulting percent-level determination of
\begin{equation}
L_{\pi\pi}(\mu = 4\pi F_\pi) = -1.13(3) \cdot 10^{-3}, \quad  \chi^2/\text{dof} = 0.98/(6-1),
\label{eq:Lpipires}
\end{equation}
does not reflect that there is
a family of NLO ChPT expressions, which differ only by higher-order terms,
but lead to slightly different results for $L_{\pi\pi}$.
Here we use the form in which both equations are written as functions of 
just $x_\pi$ and $x_K$, 
but it would be equally valid, for example, to replace
$F_K \to F_\pi$ in the NLO terms. 
Comparing the results from fitting with different allowed NLO forms,
we find that the resultant $L_{\pi\pi}$ varies at the $10\%$ level.
We thus assign a systematic error of this size, which we interpret as due to missing NNLO terms.
Including this uncertainty, the results of extrapolating to the physical point are
\begin{equation}
\left(M_\pi a^{\pi\pi}_0\right)_\text{phys} = 0.04291(4)_\text{st}(20)_\text{NNLO}\,,
\quad 
\left(M_K a^{KK}_0\right)_\text{phys} = 0.352(3)_\text{st}(13)_\text{NNLO}\,.
 \label{eq:a0physSU3}
\end{equation}
The kaon result is also shown in \Cref{fig:scattlenK}. 
We see that the statistical error is dominated by the systematic error from NNLO effects.

The result for the physical pion scattering length in \Cref{eq:a0physSU3} is in complete agreement
with that from the SU(2) ChPT extrapolation, given in \Cref{eq:a0physSU2}.
For the kaon scattering length, however, the result from the linear fit, \Cref{eq:a0KphysSU2},
disagrees significantly with that based on SU(3) ChPT, \Cref{eq:a0physSU3}.
This difference can be seen also in \Cref{fig:scattlenK}.  

We think that this roughly 10\% difference is mainly due to discretization effects.
In particular, we note that, 
along our chiral trajectory, the ratio $M_K/F_K$ is expected to increase
monotonically towards physical quark masses, yet its value on our most chiral
ensemble $4.513(9)$ (see \Cref{tab:decay_constants}) is larger than the physical value $4.472$.
Indeed, sizeable discretization effects of $3-4\%$ were observed in similar
ratios in Ref.~\cite{Bruno:2016plf}.
Since the expression for $M_K a_0^{KK}$, \Cref{eq:a0K_SU3}, is proportional to
$(M_K/F_K)^2$, the discretization errors in the SU(3) physical-point prediction
could be as big as $6-8\%$, and thus largely explain the discrepancy between the extrapolations based on SU(2) and SU(3) ChPT.

\subsubsection{Two-kaon $d$-wave interactions}

\begin{figure}[h!]
  \centering
  \includegraphics[width=0.5\linewidth]{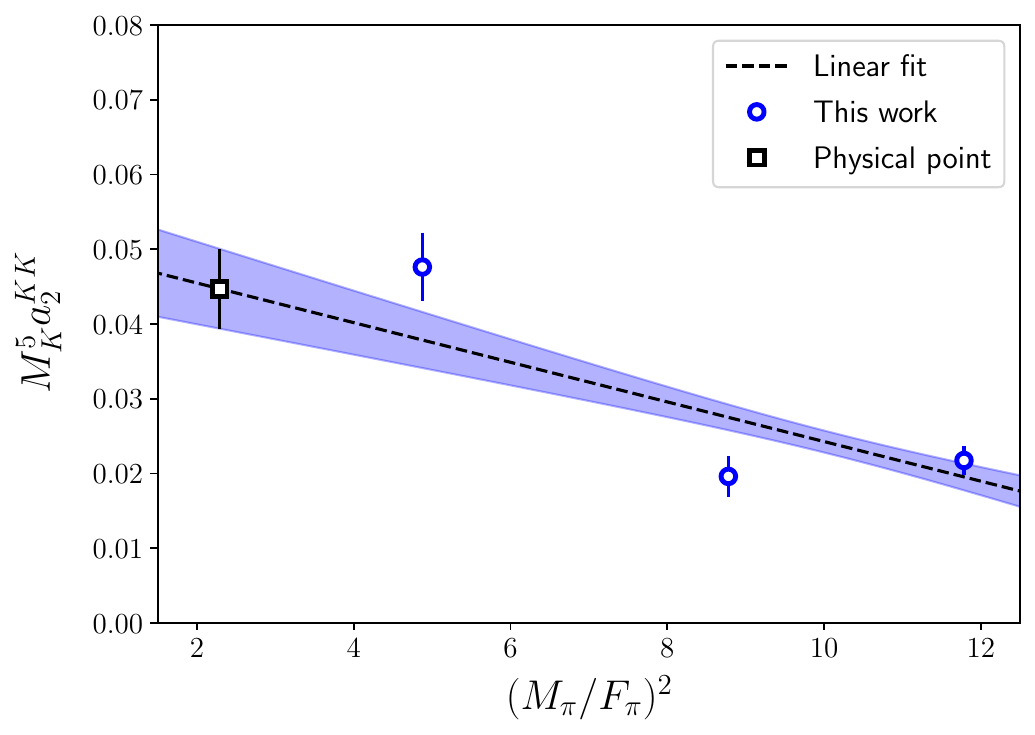}  
  \caption{Results for $M_K^5 a_2^{KK}$. Notation as in \Cref{fig:th2pi}.
  }
\label{fig:dwaveK}
\end{figure}

We now turn to the $d$-wave two-kaon interaction, 
which to our knowledge has not been previously studied in lattice calculations. 
In our fits, this is described by a single parameter, $D_0$, which is constrained quite well
by both two- and three-kaon levels in nontrivial irreps.
Our results for the $d$-wave scattering length, $M_K^5 a_2^{KK}$, are shown in \Cref{fig:dwaveK}.
Since we do not know of an SU(2) ChPT expression for this quantity, 
we perform a simple linear fit, which represents the data rather poorly.
The conclusion is that there is some evidence that the $d$-wave scattering length increases
as one approaches the physical point.

\subsubsection{Results for $\Kdf$ for three kaons}

Our final results are for the parameters of the three-kaon K-matrix: 
$\Kisozero$, $\Kisoone$, and $\cK_B$,
which have not previously been determined.
In order to simplify the notation, we use the same symbols for these parameters
as for pions, although they are different physical quantities.

\begin{figure}[t!]
\begin{subfigure}{.5\textwidth}
  \centering
  \includegraphics[width=\linewidth]{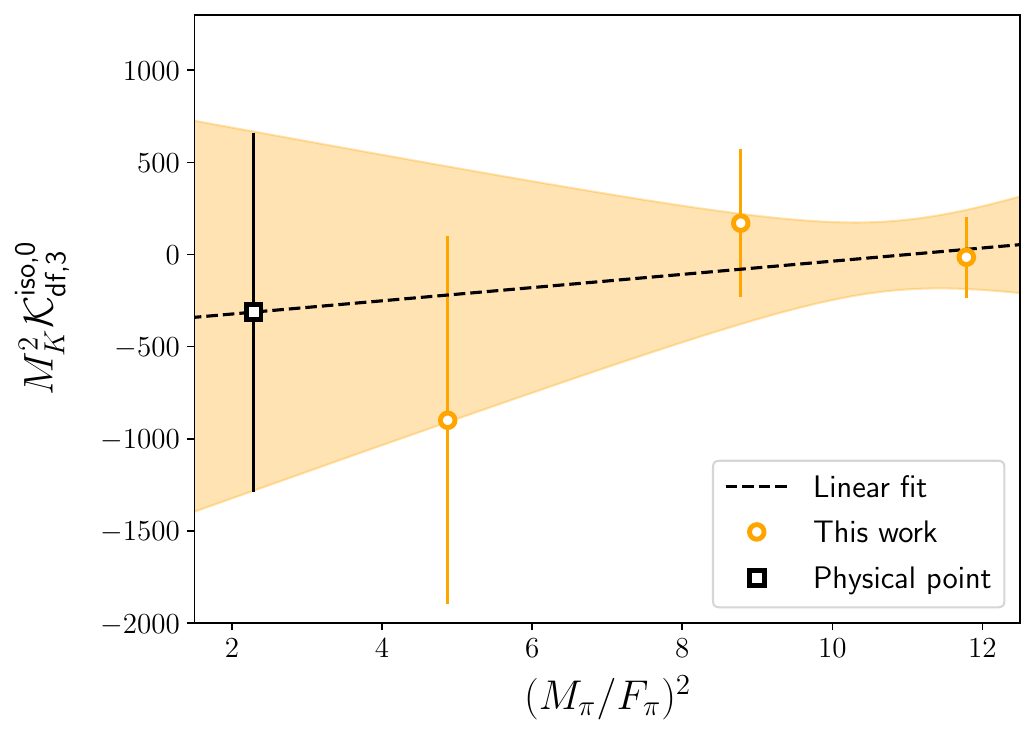}  
  \caption{$\Kisozero$}
  \label{fig:K0K}
\end{subfigure}
\begin{subfigure}{.5\textwidth}
  \centering
  \includegraphics[width=\linewidth]{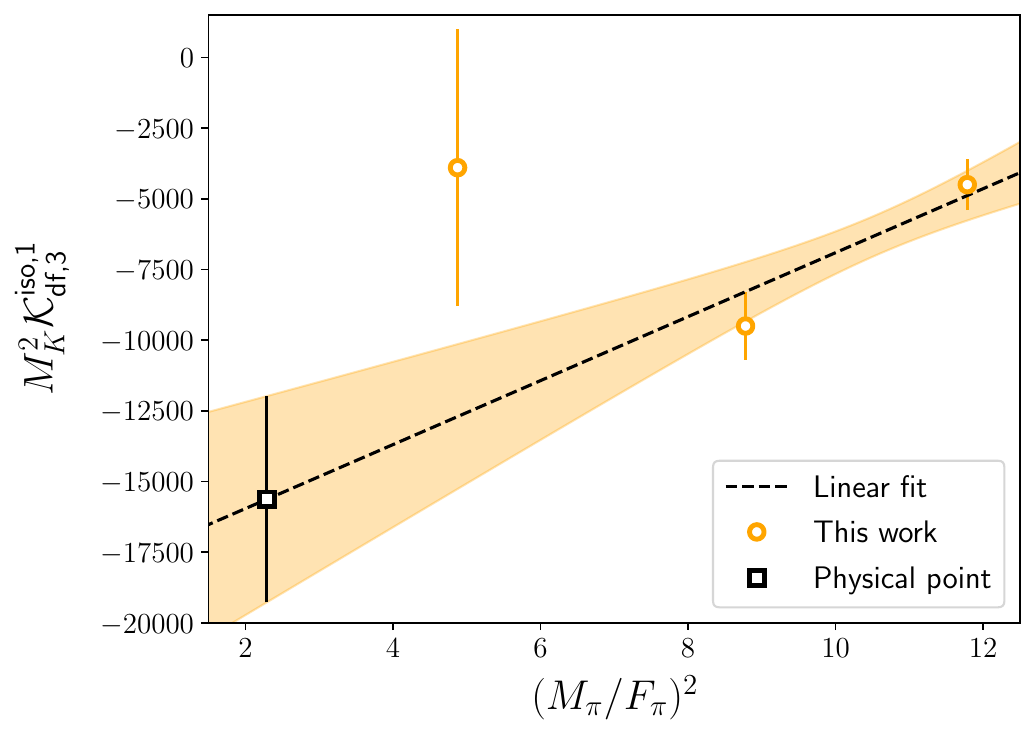}  
  \caption{$\Kisoone$}
  \label{fig:K1K}
\end{subfigure}
\caption{Results for the isotropic parameters in $\kdf$ for kaon scattering.
Notation as in \Cref{fig:th2pi}.
}
\label{fig:kisoK}
\end{figure}

The results for $\Kisozero$ and $\Kisoone$ are shown in \Cref{fig:kisoK}. 
From the left panel, we see that $\Kisozero$ is consistent with zero for all masses,
and also after linear extrapolation to the physical point.
For $\Kisoone$, by contrast, nonzero values are found for the two heaviest pion masses,
and a linear extrapolation to a nonzero value at the physical point is reasonable.
The only theoretical guidance we have is from the LO ChPT result of \Cref{eq:Kdf_LO},
which predicts proportionality to $M_K^4/F_K^4$.
Given our chiral trajectory,
this would lead to an increase as we move to smaller values of $M_\pi^2/F_\pi^2$.
However, the values predicted by LO ChPT are very far from those we find.
For instance, on the N203 ensemble the results are
\begin{equation}
M_K^2 \Kisozero \bigg \rvert_\text{LO} =  4100(100), \quad M_K^2 \Kisoone  \bigg \rvert_\text{LO} = 6200(200),
\end{equation}
which do not even appear in the plot ranges.

\begin{figure}[h!]
  \centering
  \includegraphics[width=0.5\linewidth]{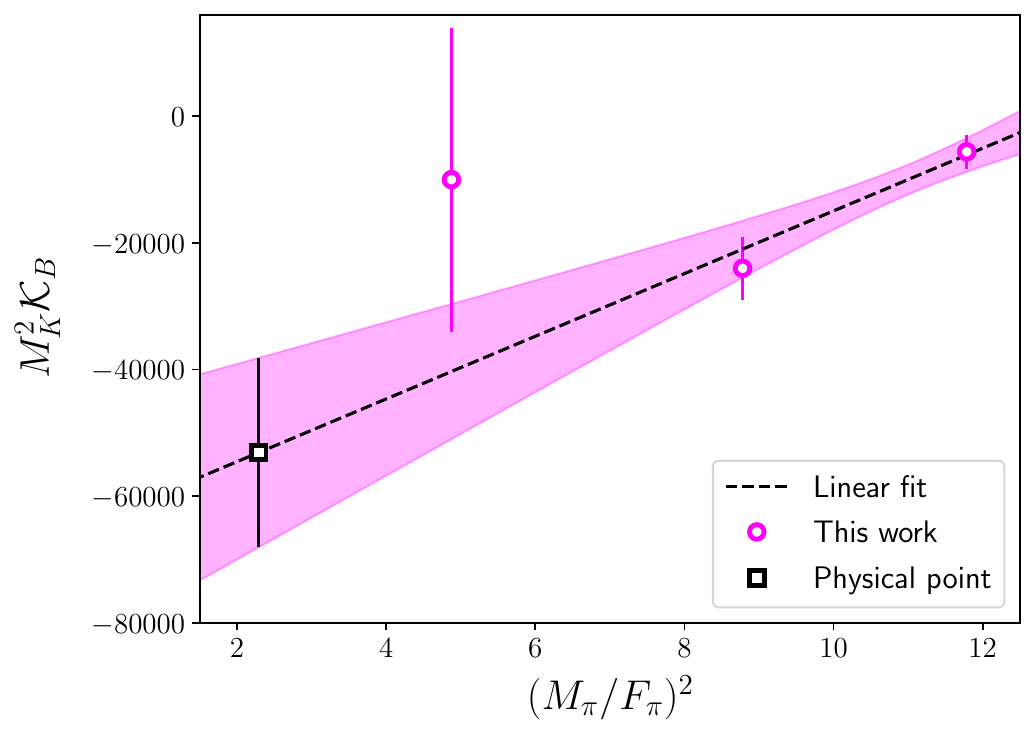}  
  \caption{Results for $\cK_B$ for kaon scattering. Notation as in \Cref{fig:th2pi}.}
\label{fig:KBK}
\end{figure}

Our results for $\cK_B$ are shown in \Cref{fig:KBK}. 
As can be seen, the result on the D200 ensemble has by far the largest statistical errors.
This is because the large value of $M_K L \sim 10$ suppresses the contribution 
to shifts in the spectrum from interactions that are of higher order in the threshold expansion.
Our results are consistent with an increase in $\cK_B$ as we approach the physical point.

\section{Conclusion}
 \label{sec:conc}
 


In this work we have extended the application of LQCD to multihadron systems by
utilizing state-of-the-art numerical methods to determine an order of magnitude more
two- and three-particle spectral levels than in previous work. 
 On each ensemble we have determined, for both pions and kaons, $50-80$ levels
 below the relevant inelastic thresholds,
roughly equally split between those for two and three particles.
The energies of the levels have been determined to a precision of $1-5\%$,
with the shifts from the corresponding free energies determined with errors of $5-15\%$.
This unprecedented number of levels with such high precision has been made possible with stochastic LapH and advanced contraction algorithms. 
The jackknife samples of all energies extracted in this work are provided in HDF5 format as ancillary files with the arXiv submission

We have found that these levels can be described well by the two- and three-particle
quantization conditions using a relatively small number of underlying parameters.
Since the major contribution to the energy shifts to three-particle levels
arises from two-particle interactions, 
our strategy has been to do a simultaneous fit to two- and three-particle levels, including
all correlations. 
We find that this leads to a better determination
of the two-particle scattering parameters than a fit to the two-particle levels alone.
The goodness of the fits is reasonable, with $\chi^2_{\rm red}=1.2-1.7$.

The $3\pi^+$ and $3K^+$ systems are nonresonant, with repulsive two-particle interactions that
are expected to have mild energy dependence. This allows the two-particle interactions to
be described by relatively few parameters in the relevant energy range. Our fits have been able to determine these
parameters accurately. For example, the pion and kaon scattering lengths are determined
with statistical errors of $1-2\%$.
These systems thus provide a good testbed for studying the significance of three-particle interactions, which themselves make only a small impact on the energy levels. Previous work has either found no statistical evidence
for such interactions, or a barely significant signal.

Our main conclusions are as follows. First, we find that a parametrization of the
two-pion K-matrix that includes the Adler zero expected from chiral symmetry is favored,
compared to an effective range expansion. For our lightest kaons, the presence of the Adler zero is also preferred.
In this regard, we stress that the three-particle quantization condition involves contributions
from subthreshold two-particle scattering, and thus is more sensitive to the Adler zero.
We also find that the position of the Adler zero is consistent with the expectations of
ChPT, albeit with relatively large errors.

Our second conclusion is that reasonable fits require the presence of $d$-wave parameters,
both for two- and three-particle interactions.
In the two-particle sector,
we determine the $d$-wave scattering length with $10-20\%$ statistical errors, 
and observe the expected chiral behavior. 
For three particles,
our results  provide the first determination of the $d$-wave three-particle parameter, 
$\cK_B$.
We note that, once $d$-wave parameters are included, there is no longer
a one-to-one connection between levels and scattering parameters, even in the two-particle
sector. Thus a global fit is required.

Our final conclusion is that the determination of three-particle interactions
requires the use of many energy levels, in a variety of irreps,
determined with sufficient accuracy.
In particular,  to extract $d$-wave contributions 
it is important to use levels in nontrivial irreps. 
We find nonzero results not only for $\cK_B$,
but also the $s$-wave parameters $\Kisozero$ and $\Kisoone$  in most of the cases.
The statistical significance of these results vary, but is greater than in previous studies,
exceeding $3\sigma$ in several cases.
For pions, the $s$-wave part of $\Kdf$ has the expected linear dependence on $M_\pi^4$, 
but the coefficient is in disagreement with leading order ChPT.
The $d$-wave part $\cK_B$ also has the expected $M_\pi^6$ dependence,
but in this case a ChPT prediction  for the coefficient is not available.
For both quantities our chiral extrapolations indicate that it will be very difficult to determine
the three-particle interaction for physical quark masses.
We also observe some tensions between our results for $\Kisozero$ and $\Kisoone$
and those of previous work~\cite{Fischer:2020jzp}.

A noteworthy feature of our fits using the quantization conditions is
that they continue to work quite well above
the energies that we have included in the fits (see \Cref{fig:n200_three_pion_spectrum}).
This shows that we have not forced the fits to work in a limited energy range only to
have them quickly fail outside that range, and thus supports the applicability of
the threshold expansion that we have used for the K-matrices.
In particular, for three kaons, these extra levels lie outside
the range that can be rigorously described by the quantization condition, 
indicating that, as expected, anomaly-induced transitions 
and overlaps onto mixed-flavor states are suppressed.

There are many exciting directions in which this work can be extended. Most straightforward is to
the maximal charge systems with hybrid flavor content, e.g. $\pi^+\pi^+ K^+$ and
$\pi^+ K^+ K^+$. 
The formalism for analyzing such ``$2+1$'' systems has very recently been developed~\cite{Blanton:2021mih}.
More challenging are systems of three pions or kaons with isospin less than the maximal value,
which involve quark annihilation diagrams and, in some cases, resonant behavior.
The relevant formalism has been derived in Ref.~\cite{\isospin}.
We hope to study such systems in the near future.

Another future direction is to extend the study of $3\pi^+$ and $3K^+$ interactions
to ensembles with different lattice spacings and volumes. This will provide a direct
check on the importance of discretization errors, and on whether our fits can accurately
describe the volume dependence. The former have been shown to be relevant in two-baryon systems~\cite{Green:2021qol}. In the latter regard, we note that the formalism that we use
drops corrections proportional to $e^{-m_\pi L}$, and there is no expectation in the short term
that such terms could be included in a rigorous way. Thus we must assume that they are small,
and the only consistency check that we have on this is to compare results at different volumes.

Finally, we note that, to obtain the three-particle S-matrix elements from the K-matrices determined 
here, one needs to solve the integral equations presented in Ref.~\cite{\HSQCb}.
To do so in the presence of $d$-wave interactions requires a generalization of the methods
developed in Refs.~\cite{Jackura:2020bsk,Hansen:2020otl}.

\acknowledgments

We thank G. Colangelo, J. Peláez, J. Ruiz de Elvira for useful discussions. We thank Marco C\`e for providing the pion and kaon decay constants computed in Ref.~\cite{decay_constants}.

The work of TDB and SRS is supported in part by the United States Department of Energy (USDOE)
grant No. DE-SC0011637, in part by the DOE.
The work of ADH is supported by: (i) The U.S. Department of Energy, Office of Science, Office of Nuclear Physics through the Contract No. DE-SC0012704 (S.M.);
(ii) The U.S. Department of Energy, Office of Science, Office of Nuclear Physics and Office of Advanced Scientific Computing Research, within the framework of Scientific Discovery through Advance Computing (SciDAC) award Computing the Properties of Matter with Leadership Computing Resources.
The work of BH is supported by the  U.S. Department of Energy, Office of Science, Office of Nuclear Physics under Award Number DE-AC02-05CH1123.
CJM acknowledges support from the U.S.~NSF under award PHY-1913158.
FRL acknowledges the support provided by the European project H2020-MSCA-ITN-2019//860881-HIDDeN, the Spanish project FPA2017-85985-P, and the Generalitat Valenciana grant PROMETEO/2019/083. The work of FRL also received funding from the European Union Horizon 2020 research and innovation program under the Marie Sk{\l}odowska-Curie grant agreement No. 713673 and ``La Caixa'' Foundation (ID 100010434, LCF/BQ/IN17/11620044). FRL also acknowledges financial support from Generalitat Valenciana through the plan GenT program (CIDEGENT/2019/040). 

Calculations for this project were performed on the HPC clusters ``HIMster II'' at the Helmholtz-Institut Mainz and ``Mogon II'' at JGU Mainz. 
We are grateful to our colleagues in the CLS initiative for sharing ensembles.

\appendix

\section{Little-group representation matrices}\label{app:irrepconventions}
\label{app:A}


Our approach to designing the single-hadron and multi-hadron operators used in this work
has been described in detail in Ref.~\cite{Morningstar:2013bda}.  However, 
single-hadron operators with momenta in directions such as $(0,1,2)$ and $(1,1,2)$,
where Cartesian components are used, were not treated in Ref.~\cite{Morningstar:2013bda}.  
Hence, we provide specific details concerning only these additional operators in this 
appendix.

For each class of momenta, we choose one representative reference momentum direction
$\bm{p}_{\rm ref}$.  We then construct operators that transform irreducibly under
the little group of $\bm{p}_{\rm ref}$.  Recall that the little group of $\bm{p}_{\rm ref}$
is the subset of the symmetry operations that leave the reference momentum $\bm{p}_{\rm ref}$
invariant.  For each momentum direction $\bm{p}$ in a class of momenta, we select one
reference rotation $R_{\rm ref}^{\bm{p}}$ that transforms $\bm{p}_{\rm ref}$ into $\bm{p}$.
As long as the selected rotation transforms $\bm{p}_{\rm ref}$ into $\bm{p}$, it does
not matter which group element is chosen, but a choice must be made in order to
specific all phases between the single-hadron operators of different momentum directions.
These phases must be known in order to properly construct the multi-hadron operators.
All single-hadron operators having a momentum in the direction of $\bm{p}$ are then
obtained by applying the reference rotation to the corresponding operators constructed
using the momentum in the direction of $\bm{p}_{\rm ref}$.  Our choices of reference
momenta directions and reference rotations for the additional operators used in this
work are listed in \Cref{tab:refrotates}.

\begin{table}[tp]
\centering
\begin{tabular}{c|cc|cc}
$\bm{p}_{\rm ref}$ direction & $\bm{p}$ direction & $R_{\rm ref}^{\bm{p}}$ &
   $\bm{p}$ direction & $R_{\rm ref}^{\bm{p}}$ \\
 \hline
$(0,1,2)$ & $(0, 1, 2)$  & $E$               & $(0, 1, -2)$  &$C_{2y}$            \\
          & $(0, 2, 1)$  & $C_{2e}$          & $(0, -2, 1)$  &$C_{4x}$            \\
          & $(1, 0, 2)$  & $C_{4z}^{-1}$     & $(1, 0, -2)$  &$C_{2a}$            \\
          & $(2, 0, 1)$  & $C_{3\delta}$     & $(-2, 0, 1)$  &$C_{3\gamma}$       \\
          & $(1, 2, 0)$  & $C_{3\delta}^{-1}$& $(1, -2, 0)$  &$C_{3\alpha}^{-1}$  \\
          & $(2, 1, 0)$  & $C_{4y}$          & $(-2, 1, 0)$  &$C_{4y}^{-1}$       \\
          & $(0, -1, 2)$ & $C_{2z}$          & $(0, -1, -2)$ &$C_{2x}$            \\
          & $(0, 2, -1)$ & $C_{4x}^{-1}$     & $(0, -2, -1)$ &$C_{2f}$            \\
          & $(-1, 0, 2)$ & $C_{4z}$          & $(-1, 0, -2)$ &$C_{2b}$            \\
          & $(2, 0, -1)$ & $C_{3\beta}$      & $(-2, 0, -1)$ &$C_{3\alpha}$       \\
          & $(-1, 2, 0)$ & $C_{3\gamma}^{-1}$& $(-1, -2, 0)$ &$C_{3\beta}^{-1}$   \\
          & $(2, -1, 0)$ & $C_{2c}$          & $(-2, -1, 0)$ &$C_{2d}$            \\
 \hline
$(1,1,2)$ & $(1, 1, 2)$    & $E$                & $(1, -2, 1)$   & $C_{4x}$          \\ 
          & $(-1, 1, 2)$   & $C_{4z}$           & $(-1, -2, 1)$  & $C_{3\beta}^{-1}$ \\ 
          & $(1, -1, 2)$   & $C_{4z}^{-1}$      & $(1, -2, -1)$  & $C_{3\alpha}^{-1}$\\ 
          & $(-1, -1, 2)$  & $C_{2z}$           & $(-1, -2, -1)$ & $C_{2f}$          \\ 
          & $(1, 1, -2)$   & $C_{2a}$           & $(2, 1, 1)$    & $C_{3\delta}$     \\ 
          & $(-1, 1, -2)$  & $C_{2y}$           & $(2, -1, 1)$   & $C_{2c}$          \\ 
          & $(1, -1, -2)$  & $C_{2x}$           & $(2, 1, -1)$   & $C_{4y}$          \\ 
          & $(-1, -1, -2)$ & $C_{2b}$           & $(2, -1, -1)$  & $C_{3\beta}$      \\ 
          & $(1, 2, 1)$    & $C_{3\delta}^{-1}$ & $(-2, 1, 1)$   & $C_{4y}^{-1}$     \\ 
          & $(-1, 2, 1)$   & $C_{2e}$           & $(-2, -1, 1)$  & $C_{3\gamma}$     \\ 
          & $(1, 2, -1)$   & $C_{4x}^{-1}$      & $(-2, 1, -1)$  & $C_{3\alpha}$     \\ 
          & $(-1, 2, -1)$  & $C_{3\gamma}^{-1}$ & $(-2, -1, -1)$ & $C_{2d}$        
 \end{tabular}
\caption{Our choices for the reference momentum $\bm{p}_{\rm ref}$ directions
and the reference rotations $R_{\rm ref}^{\bm{p}}$ for each of the additional
momentum $\bm{p}$ directions that we use.  See Table II of Ref.~\cite{Morningstar:2013bda} 
for all other momentum directions used.  See Ref.~\cite{Morningstar:2013bda} 
for definitions of the rotation operators $C_{nj}$ below. $E$ is the identity element.
 \label{tab:refrotates}}
\end{table}

The little groups associated with the additional momentum directions used
in this work are listed in \Cref{tab:Cs_classes}.  Although this work
involves only mesons, we wish to provide details concerning both the
single-valued and double-valued representations for possible future 
calculations with baryons.  The double-valued representations of a group $\mathcal{G}$
are constructed by extending the group elements to form the so-called ``double group''
$\mathcal{G}^D$.  The elements of the double groups associated with our
choices of additional reference momentum directions are explicitly presented
in \Cref{tab:Cs_classes}, grouped into their conjugacy classes.

\begin{table}[t]
\begin{center}
\begin{tabular}{c|l}
$\bm{p}_{\rm ref}$ & Conjugacy classes \\ \hline
$(0,1,2)$ & $\mathcal{C}_{1} =  \{E  \}$  \\
&$\mathcal{C}_{2} =  \{ I_{s} C_{2x}  \}$\\
&$\mathcal{C}_{3} =  \{ I_{s} \overline{C}_{2 x}  \}$ \\
&$\mathcal{C}_{4} =  \{ \overline{E}  \}$ \\ \hline
$(1,1,2)$ & $\mathcal{C}_{1} = \{E  \} $\\
&$\mathcal{C}_{2} = \{ I_{s} C_{2b}  \}$\\
&$\mathcal{C}_{3} = \{ I_{s} \overline{C}_{2 b}  \}$\\
&$\mathcal{C}_{4} = \{ \overline{E}  \} $
\end{tabular}
\end{center}
\caption{The little groups corresponding to reference momentum directions $(0,1,2)$
and $(1,1,2)$ are isomorphic to $C_s$.  The elements of the double groups $C_{s}^D$ 
for these momenta directions are listed above, grouped
into conjugacy classes.  $E$ is the identity element, $\overline{E}$ represents
a rotation by $2\pi$ about any axis, and $\overline{G}=\overline{E}G$ for each 
element $G$ in $C_s$. Spatial inversion is denoted by $I_s$.
\label{tab:Cs_classes}}
\end{table}

\begin{table}[t]
\begin{center}
\begin{tabular}{c|rrrr}
 & $\chi^\Lambda_{1}$ & $\chi^\Lambda_{2}$ & $\chi^\Lambda_{3}$ 
& $\chi^\Lambda_{4}$   \\
 \hline
 $A_{1}$  & 1 & 1 & 1 & 1 \\
 $A_{2}$  & 1 & $-1$ & $-1$ & $1$ \\
 $F_{1}$  & 1 & $i$ & $-i$ & $-1$ \\
 $F_{2}$  & 1 & $-i$ & $i$ & $-1$ \\
 \hline
\end{tabular}
\end{center}
\caption{Characters $\chi^\Lambda$ for the single-valued and double-valued 
irreducible representations $\Lambda$ of the group $C_{s}$.  $\chi^\Lambda_n$
denotes the character of $\Lambda$ for all group elements in class
$\mathcal{C}_n$.  See \Cref{tab:Cs_classes} for the definitions of 
the classes $C_n$.  Since all of
the representations are one dimensional, the characters are the
representation matrices.\label{tab:CsD_characters}}
\end{table}

The characters of the irreducible representations (irreps) of the little 
group $C_s$ for the additional momenta directions are presented in 
\Cref{tab:CsD_characters}.
The one-dimensional single-valued irreps are labeled by $A$, and the
one-dimensional double-valued irreps are denoted by $F$.  Since all of the
irreps are one dimensional, the characters are the representation matrices.

\section{Tables of interpolating operators}\label{app:interpolators}
\label{app:B}

The three- and two-hadron operators used in this work are listed in 
\Cref{tab:three-pion-ops,tab:three-pion-opsB,tab:three-pion-opsC,tab:three-kaon-ops,tab:three-kaon-opsB,tab:three-kaon-opsC,tab:two-pion-ops,tab:two-pion-opsB,tab:two-kaon-ops,tab:two-kaon-opsB,tab:two-kaon-opsC} below.
The notation for the irreps follows the conventions in Ref.~\cite{Morningstar:2013bda}.
The subscripts $g$/$u$ denote even/odd parity, and the superscripts $+$/$-$ denote even/odd $G$-parity.
The Clebsch-Gordan coefficients that fully define each operator are not given,
but are available upon request.

\begin{table}[!bp]
\centering
\begin{tabular}{c|c|ccc|c}
\multirow{2}{*}{$\vec{d}_{\rm ref}$}&\multirow{2}{*}{$[d_1^2,d_2^2,d_3^2]$}&\multicolumn{3}{c}{$E^{\rm free}/M_\pi$}&\multirow{2}{*}{operators}\\%
&&N203&N200&D200&\\%
\hline%
(0, 0, 0)&{[}0, 0, 0{]}&3.0&3.0&3.0&$A_{1u}^-$\\%
&{[}0, 1, 1{]}&4.0667&4.4761&4.5989&$A_{1u}^- \oplus E_u^-$\\%
&{[}0, 2, 2{]}&4.8483&&&$A_{1u}^- \oplus E_u^- \oplus T_{2u}^-$\\%
&{[}1, 1, 2{]}&4.9909&&&$A_{1u}^- \oplus E_u^- \oplus T_{1g}^- \oplus T_{2g}^- \oplus T_{2u}^-$\\%
\hline%
(0, 0, 1)&{[}0, 0, 1{]}&3.3367&3.4572&3.4925&$A_2^-$\\%
&{[}0, 1, 2{]}&4.3033&4.7763&4.917&$A_2^- \oplus B_2^- \oplus E^-$\\%
&{[}1, 1, 1{]}&4.4508&5.0166&5.1869&$2 A_2^- \oplus B_2^-$\\%
&{[}0, 1, 4{]}&4.9289&&&$A_2^-$\\%
&{[}0, 2, 3{]}&5.0399&&&$A_2^- \oplus B_1^- \oplus E^-$\\%
&{[}1, 1, 3{]}&5.1861&&&$A_2^- \oplus B_1^- \oplus E^-$\\%
&{[}1, 2, 2{]}&5.2547&&&$A_1^- \oplus 4 A_2^- \oplus 2 B_1^- \oplus 3 B_2^- \oplus 3 E^-$\\%
\hline%
(0, 1, 1)&{[}0, 0, 2{]}&3.5632&3.7392&3.7895&$A_2^-$\\%
&{[}0, 1, 1{]}&3.7197&3.9992&4.0834&$A_2^-$\\%
&{[}0, 1, 3{]}&4.4899&5.007&5.16&$A_2^- \oplus B_2^-$\\%
&{[}0, 2, 2{]}&4.5611&5.1094&&$A_1^- \oplus A_2^-$\\%
&{[}1, 1, 2{]}&4.7124&&&$A_1^- \oplus 4 A_2^- \oplus 3 B_1^- \oplus B_2^-$\\%
&{[}0, 1, 5{]}&5.0579&&&$A_2^- \oplus B_1^-$\\%
&{[}0, 2, 4{]}&5.2013&&&$A_2^- \oplus B_1^-$\\%
\hline%
(1, 1, 1)&{[}0, 0, 3{]}&3.7406&3.9535&4.0137&$A_2^-$\\%
&{[}0, 1, 2{]}&3.9769&4.3326&4.4386&$A_2^- \oplus E^-$\\%
&{[}1, 1, 1{]}&4.1361&4.5961&4.7358&$A_2^-$\\%
&{[}1, 1, 3{]}&4.9186&&&$2 A_2^- \oplus 2 E^-$\\%
&{[}1, 2, 2{]}&4.9909&&&$A_1^- \oplus 2 A_2^- \oplus 3 E^-$\\%
&{[}0, 1, 6{]}&5.1732&&&$A_2^- \oplus E^-$\\%
\end{tabular}
\caption{\label{tab:three-pion-ops} Three-pion operators with $\vec{d}_{\rm ref}^{\,2}\leq 3$ used in this study.
Each row specifies all linearly-independent operators corresponding to a particular free energy-level.
Ensembles with a missing value for the energy of the free level $E^{\rm free}$ did not use those operators.
For each set of momenta that are equivalent up to allowed rotations, one representative integer momentum specifying the total momentum of the operator $\vec{P} = (2\pi/L)\vec{d}_{\rm ref}$ is given.
The integer momentum squared of the individual single particles in each operator is given by $d_i^2$.
}
\end{table}

\begin{table}[tp]
\centering
\begin{tabular}{c|c|ccc|c}
\multirow{2}{*}{$\vec{d}_{\rm ref}$}&\multirow{2}{*}{$[d_1^2,d_2^2,d_3^2]$}&\multicolumn{3}{c}{$E^{\rm free}/M_\pi$}&\multirow{2}{*}{operators}\\%
&&N203&N200&D200&\\%
\hline%
(0, 0, 2)&{[}0, 1, 1{]}&3.3367&3.4572&3.4925&$A_2^-$\\%
&{[}0, 0, 4{]}&3.8888&4.1298&4.1975&$A_2^-$\\%
&{[}0, 2, 2{]}&4.2546&4.6973&4.8283&$A_2^- \oplus B_2^-$\\%
&{[}1, 1, 2{]}&4.4164&4.965&5.1302&$A_2^- \oplus B_2^- \oplus E^-$\\%
&{[}0, 1, 5{]}&4.7833&&&$A_2^- \oplus B_2^- \oplus E^-$\\%
&{[}0, 3, 3{]}&4.9801&&&$A_2^- \oplus B_1^-$\\%
&{[}1, 1, 4{]}&5.0919&&&$2 A_2^- \oplus B_2^-$\\%
&{[}1, 2, 3{]}&5.2104&&&$A_1^- \oplus 2 A_2^- \oplus 2 B_1^- \oplus B_2^- \oplus 3 E^-$\\%
&{[}2, 2, 2{]}&5.2836&&&$A_2^- \oplus B_1^- \oplus E^-$\\%
\hline%
(0, 1, 2)&{[}0, 1, 2{]}&3.6213&3.838&3.9019&$A_2^-$\\%
&{[}1, 1, 1{]}&3.7954&4.1332&4.237&$A_2^-$\\%
&{[}0, 0, 5{]}&4.0174&4.2812&4.355&$A_2^-$\\%
&{[}0, 1, 4{]}&4.3462&4.7937&4.9256&$A_2^-$\\%
&{[}0, 2, 3{]}&4.4716&4.9739&5.1219&$A_1^- \oplus A_2^-$\\%
&{[}1, 1, 3{]}&4.6358&5.2456&&$A_1^- \oplus A_2^-$\\%
&{[}1, 2, 2{]}&4.7124&&&$2 A_1^- \oplus 5 A_2^-$\\%
&{[}0, 1, 6{]}&4.9051&&&$A_1^- \oplus A_2^-$\\%
&{[}0, 2, 5{]}&5.0831&&&$A_1^- \oplus 2 A_2^-$\\%
&{[}1, 1, 5{]}&5.2427&&&$A_1^- \oplus 6 A_2^-$\\%
\hline%
(1, 1, 2)&{[}0, 1, 3{]}&3.8412&4.1215&4.204&$A_2^-$\\%
&{[}0, 2, 2{]}&3.9242&4.2453&4.3401&$A_2^-$\\%
&{[}1, 1, 2{]}&4.099&4.5398&4.6737&$A_1^- \oplus 2 A_2^-$\\%
&{[}0, 0, 6{]}&4.1317&4.4149&4.4937&$A_2^-$\\%
&{[}0, 1, 5{]}&4.4919&4.9723&5.1134&$A_1^- \oplus A_2^-$\\%
&{[}0, 2, 4{]}&4.6529&5.2012&&$A_2^-$\\%
&{[}1, 1, 4{]}&4.8192&&&$A_2^-$\\%
&{[}1, 2, 3{]}&4.9443&&&$3 A_1^- \oplus 3 A_2^-$\\%
&{[}2, 2, 2{]}&5.0214&&&$2 A_1^- \oplus 2 A_2^-$\\%
&{[}0, 2, 6{]}&5.2154&&&$A_1^- \oplus A_2^-$\\%
\hline%
(0, 2, 2)&{[}0, 2, 2{]}&3.5632&3.7392&3.7895&$A_2^-$\\%
&{[}1, 1, 2{]}&3.7549&4.0704&4.1674&$A_2^-$\\%
&{[}0, 1, 5{]}&4.1803&4.5478&4.6552&$A_2^- \oplus B_1^-$\\%
&{[}0, 0, 8{]}&4.3297&4.6447&4.732&$A_2^-$\\%
&{[}0, 3, 3{]}&4.4042&4.8696&5.0062&$A_2^-$\\%
&{[}1, 1, 4{]}&4.5302&5.0944&&$A_2^- \oplus B_1^-$\\%
&{[}1, 2, 3{]}&4.663&&&$A_1^- \oplus 2 A_2^- \oplus B_1^- \oplus 2 B_2^-$\\%
&{[}2, 2, 2{]}&4.7447&&&$A_1^- \oplus A_2^-$\\%
&{[}0, 2, 6{]}&4.9495&&&$A_1^- \oplus A_2^- \oplus B_1^- \oplus B_2^-$\\%
&{[}0, 4, 4{]}&5.0924&&&$A_2^-$\\%
&{[}1, 1, 6{]}&5.1198&&&$A_1^- \oplus A_2^- \oplus B_1^- \oplus B_2^-$\\%
\end{tabular}
\caption{\label{tab:three-pion-opsB} Same as \Cref{tab:three-pion-ops} except for
the three-pion operators with $4\leq \vec{d}_{\rm ref}^{\,2}\leq 8$.
}
\end{table}

\begin{table}[tp]
\centering
\begin{tabular}{c|c|ccc|c}
\multirow{2}{*}{$\vec{d}_{\rm ref}$}&\multirow{2}{*}{$[d_1^2,d_2^2,d_3^2]$}&\multicolumn{3}{c}{$E^{\rm free}/M_\pi$}&\multirow{2}{*}{operators}\\%
&&N203&N200&D200&\\%
\hline%
(0, 0, 3)&{[}1, 1, 1{]}&3.0&3.0&3.0&$A_2^-$\\%
&{[}0, 1, 4{]}&3.6721&3.8596&3.9127&$A_2^-$\\%
&{[}1, 2, 2{]}&4.099&4.5398&4.6737&$A_2^- \oplus B_2^-$\\%
&{[}0, 0, 9{]}&4.4171&4.7456&4.8365&$A_2^-$\\%
&{[}0, 2, 5{]}&4.5203&5.0078&5.1505&$A_2^- \oplus B_2^- \oplus E^-$\\%
&{[}1, 1, 5{]}&4.6991&&&$A_2^- \oplus B_2^- \oplus E^-$\\%
&{[}1, 2, 4{]}&4.8682&&&$A_2^- \oplus B_2^- \oplus E^-$\\%
&{[}1, 3, 3{]}&4.9186&&&$A_2^- \oplus B_1^-$\\%
&{[}2, 2, 3{]}&5.0005&&&$A_2^- \oplus B_1^- \oplus E^-$\\%
&{[}0, 3, 6{]}&5.2058&&&$A_2^- \oplus B_1^- \oplus E^-$\\%
\end{tabular}%
\caption{\label{tab:three-pion-opsC} Same as \Cref{tab:three-pion-ops} except for
the three-pion operators with $\vec{d}_{\rm ref}^{\,2}=9$.
}
\end{table}

\begin{table}[bp]
\centering
\begin{tabular}{c|c|ccc|c}
\multirow{2}{*}{$\vec{d}_{\rm ref}$}&\multirow{2}{*}{$[d_1^2,d_2^2,d_3^2]$}&\multicolumn{3}{c}{$E^{\rm free}/M_K$}&\multirow{2}{*}{operators}\\%
&&N203&N200&D200&\\%
\hline%
(0, 0, 0)&{[}0, 0, 0{]}&3.0&3.0&3.0&$A_{1u}$\\%
&{[}0, 1, 1{]}&3.7035&3.6505&3.3624&$A_{1u} \oplus E_u$\\%
&{[}0, 2, 2{]}&4.2585&4.1702&3.6762&$A_{1u} \oplus E_u \oplus T_{2u}$\\%
&{[}1, 1, 2{]}&4.3328&4.2356&3.7005&$A_{1u} \oplus E_u \oplus T_{1g} \oplus T_{2g} \oplus T_{2u}$\\%
\hline%
(0, 0, 1)&{[}0, 0, 1{]}&3.226&3.2095&3.1185&$A_2$\\%
&{[}0, 1, 2{]}&3.8757&3.8124&3.4627&$A_2 \oplus B_2 \oplus E$\\%
&{[}1, 1, 1{]}&3.9519&3.8795&3.4874&$2 A_2 \oplus B_2$\\%
&{[}0, 1, 4{]}&4.3331&4.2433&3.7352&$A_2$\\%
&{[}0, 2, 3{]}&&4.3062&3.7644&$A_2 \oplus B_1 \oplus E$\\%
&{[}1, 1, 3{]}&&4.3729&3.789&$A_2 \oplus B_1 \oplus E$\\%
&{[}1, 2, 2{]}&&4.4105&3.8058&$A_1 \oplus 4 A_2 \oplus 2 B_1 \oplus 3 B_2 \oplus 3 E$\\%
\hline%
(0, 1, 1)&{[}0, 0, 2{]}&3.3937&3.3676&3.2175&$A_2$\\%
&{[}0, 1, 1{]}&3.473&3.4371&3.2427&$A_2$\\%
&{[}0, 1, 3{]}&4.0168&3.946&3.55&$A_2 \oplus B_2$\\%
&{[}0, 2, 2{]}&4.0596&3.9847&3.567&$A_1 \oplus A_2$\\%
&{[}1, 1, 2{]}&4.1374&4.0531&3.5921&$A_1 \oplus 4 A_2 \oplus 3 B_1 \oplus B_2$\\%
&{[}0, 1, 5{]}&&4.341&3.8038&$A_2 \oplus B_1$\\%
&{[}0, 2, 4{]}&&4.4236&3.8431&$A_2 \oplus B_1$\\%
&{[}1, 1, 4{]}&&&3.8681&$A_2 \oplus B_1$\\%
\end{tabular}%
\caption{\label{tab:three-kaon-ops} Same as \Cref{tab:three-pion-ops} except for
the three-kaons operators with $\vec{d}_{\rm ref}^{\,2}\leq 2$.
}
\end{table}

\begin{table}[bp]
\centering
\begin{tabular}{c|c|ccc|c}
\multirow{2}{*}{$\vec{d}_{\rm ref}$}&\multirow{2}{*}{$[d_1^2,d_2^2,d_3^2]$}&\multicolumn{3}{c}{$E^{\rm free}/M_K$}&\multirow{2}{*}{operators}\\%
&&N203&N200&D200&\\%
\hline%
(1, 1, 1)&{[}0, 0, 3{]}&3.5304&3.4974&3.3036&$A_2$\\%
&{[}0, 1, 2{]}&3.6561&3.6086&3.3466&$A_2 \oplus E$\\%
&{[}1, 1, 1{]}&3.7368&3.6794&3.3722&$A_2$\\%
&{[}1, 1, 3{]}&4.2893&4.1963&3.6832&$2 A_2 \oplus 2 E$\\%
&{[}1, 2, 2{]}&4.3328&4.2356&3.7005&$A_1 \oplus 2 A_2 \oplus 3 E$\\%
&{[}0, 1, 6{]}&&4.4294&3.867&$A_2 \oplus E$\\%
\hline%
(0, 0, 2)&{[}0, 1, 1{]}&3.226&3.2095&3.1185&$A_2$\\%
&{[}0, 0, 4{]}&3.6474&3.609&3.3803&$A_2$\\%
&{[}0, 2, 2{]}&3.8505&3.7902&3.4545&$A_2 \oplus B_2$\\%
&{[}1, 1, 2{]}&3.9324&3.862&3.4803&$A_2 \oplus B_2 \oplus E$\\%
&{[}0, 1, 5{]}&4.2447&4.1632&3.6985&$A_2 \oplus B_2 \oplus E$\\%
&{[}0, 3, 3{]}&&4.2758&3.7517&$A_2 \oplus B_1$\\%
&{[}1, 1, 4{]}&&4.3198&3.7645&$2 A_2 \oplus B_2$\\%
&{[}1, 2, 3{]}&&4.3861&3.7948&$A_1 \oplus 2 A_2 \oplus 2 B_1 \oplus B_2 \oplus 3 E$\\%
&{[}2, 2, 2{]}&&4.4258&3.8123&$A_2 \oplus B_1 \oplus E$\\%
\hline%
(0, 1, 2)&{[}0, 1, 2{]}&3.4223&3.3926&3.2263&$A_2$\\%
&{[}1, 1, 1{]}&3.5084&3.4677&3.2528&$A_2$\\%
&{[}0, 0, 5{]}&3.7504&3.7076&3.4497&$A_2$\\%
&{[}0, 1, 4{]}&3.9328&3.8705&3.5172&$A_2$\\%
&{[}0, 2, 3{]}&4.0088&3.9393&3.5482&$A_1 \oplus A_2$\\%
&{[}1, 1, 3{]}&4.0919&4.0121&3.5743&$A_1 \oplus A_2$\\%
&{[}1, 2, 2{]}&4.1374&4.0531&3.5921&$2 A_1 \oplus 5 A_2$\\%
&{[}0, 1, 6{]}&&4.2552&3.7634&$A_1 \oplus A_2$\\%
&{[}0, 2, 5{]}&&4.3575&3.8123&$A_1 \oplus 2 A_2$\\%
&{[}1, 1, 5{]}&&4.4291&3.8382&$A_1 \oplus 6 A_2$\\%
&{[}1, 2, 4{]}&&&3.879&$A_1 \oplus 4 A_2$\\%
\hline%
(1, 1, 2)&{[}0, 1, 3{]}&3.5813&3.542&3.3198&$A_2$\\%
&{[}0, 2, 2{]}&3.6293&3.5851&3.3381&$A_2$\\%
&{[}1, 1, 2{]}&3.7161&3.661&3.3649&$A_1 \oplus 2 A_2$\\%
&{[}0, 0, 6{]}&3.8429&3.7964&3.5135&$A_2$\\%
&{[}0, 1, 5{]}&4.0451&3.9774&3.59&$A_1 \oplus A_2$\\%
&{[}0, 2, 4{]}&4.1442&4.0673&3.6316&$A_2$\\%
&{[}1, 1, 4{]}&4.2283&4.141&3.658&$A_2$\\%
&{[}1, 2, 3{]}&4.3046&4.2101&3.6892&$3 A_1 \oplus 3 A_2$\\%
&{[}2, 2, 2{]}&&4.2515&3.7071&$2 A_1 \oplus 2 A_2$\\%
\end{tabular}%
\caption{\label{tab:three-kaon-opsB} Same as \Cref{tab:three-pion-ops} except for
the three-kaons operators with $3\leq \vec{d}_{\rm ref}^{\,2}\leq 6$.
}
\end{table}

\begin{table}[tp]
\centering
\begin{tabular}{c|c|ccc|c}
\multirow{2}{*}{$\vec{d}_{\rm ref}$}&\multirow{2}{*}{$[d_1^2,d_2^2,d_3^2]$}&\multicolumn{3}{c}{$E^{\rm free}/M_K$}&\multirow{2}{*}{operators}\\%
&&N203&N200&D200&\\%
\hline%
(0, 2, 2)&{[}0, 2, 2{]}&3.3937&3.3676&3.2175&$A_2$\\%
&{[}1, 1, 2{]}&3.4864&3.4482&3.2453&$A_2$\\%
&{[}0, 1, 5{]}&3.8352&3.7824&3.4782&$A_2 \oplus B_1$\\%
&{[}0, 3, 3{]}&3.9715&3.906&3.5348&$A_2$\\%
&{[}0, 0, 8{]}&4.005&3.9523&3.6277&$A_2$\\%
&{[}1, 1, 4{]}&4.0279&3.9541&3.5483&$A_2 \oplus B_1$\\%
&{[}1, 2, 3{]}&4.1079&4.0265&3.5805&$A_1 \oplus 2 A_2 \oplus B_1 \oplus 2 B_2$\\%
&{[}2, 2, 2{]}&4.156&4.0697&3.5989&$A_1 \oplus A_2$\\%
&{[}0, 2, 6{]}&&4.2822&3.7766&$A_1 \oplus A_2 \oplus B_1 \oplus B_2$\\%
&{[}1, 1, 6{]}&&4.3574&3.8034&$A_1 \oplus A_2 \oplus B_1 \oplus B_2$\\%
&{[}0, 4, 4{]}&&4.3675&3.8194&$A_2$\\%
&{[}1, 2, 5{]}&&&3.854&$2 A_1 \oplus 4 A_2 \oplus 4 B_1 \oplus 2 B_2$\\%
\hline%
(0, 0, 3)&{[}1, 1, 1{]}&3.0&3.0&3.0&$A_2$\\%
&{[}0, 1, 4{]}&3.4868&3.4577&3.2848&$A_2$\\%
&{[}1, 2, 2{]}&3.7161&3.661&3.3649&$A_2 \oplus B_2$\\%
&{[}0, 2, 5{]}&4.0647&3.9954&3.599&$A_2 \oplus B_2 \oplus E$\\%
&{[}0, 0, 9{]}&4.0773&4.0219&3.6795&$A_2$\\%
&{[}1, 1, 5{]}&4.1538&4.0733&3.6264&$A_2 \oplus B_2 \oplus E$\\%
&{[}1, 2, 4{]}&4.2575&4.1673&3.6695&$A_2 \oplus B_2 \oplus E$\\%
&{[}1, 3, 3{]}&4.2893&4.1963&3.6832&$A_2 \oplus B_1$\\%
&{[}2, 2, 3{]}&&4.2398&3.7018&$B_1 \oplus E$\\%
\end{tabular}%
\caption{\label{tab:three-kaon-opsC} Same as \Cref{tab:three-pion-ops} except for
the three-kaons operators with $8\leq \vec{d}_{\rm ref}^{\,2}\leq 9$.
}
\end{table}

\begin{table}[bp]
\centering
\begin{tabular}{c|c|ccc|c}
\multirow{2}{*}{$\vec{d}_{\rm ref}$}&\multirow{2}{*}{$[d_1^2,d_2^2]$}&\multicolumn{3}{c}{$E^{\rm free}/M_\pi$}&\multirow{2}{*}{operators}\\%
&&N203&N200&D200&\\%
\hline%
(0, 0, 0)&{[}0, 0{]}&2.0&2.0&2.0&$A_{1g}^+$\\%
&{[}1, 1{]}&3.0667&3.4761&3.5989&$A_{1g}^+ \oplus E_g^+$\\%
&{[}2, 2{]}&3.8483&&&$A_{1g}^+ \oplus E_g^+ \oplus T_{2g}^+$\\%
\hline%
(0, 0, 1)&{[}0, 1{]}&2.2509&2.3401&2.3662&$A_1^+$\\%
&{[}1, 2{]}&3.2563&3.7211&3.8598&$A_1^+ \oplus B_1^+ \oplus E^+$\\%
&{[}1, 4{]}&3.8944&&&$A_1^+$\\%
\hline%
(0, 1, 1)&{[}0, 2{]}&2.4183&2.5477&2.5846&$A_1^+$\\%
&{[}1, 1{]}&2.5889&2.8358&2.9114&$A_1^+$\\%
&{[}1, 3{]}&3.4054&3.9089&4.0585&$A_1^+ \oplus B_1^+$\\%
&{[}2, 2{]}&3.4795&4.0155&&$A_1^+ \oplus A_2^+$\\%
\hline%
(1, 1, 1)&{[}0, 3{]}&2.5487&2.7046&2.7486&$A_1^+$\\%
&{[}1, 2{]}&2.8109&3.1313&3.2283&$A_1^+ \oplus E^+$\\%
\end{tabular}%
\caption{\label{tab:two-pion-ops} Same as \Cref{tab:three-pion-ops} except for
the two-pion operators with $\vec{d}_{\rm ref}^{\,2}\leq 3$.
}
\end{table}

\begin{table}[t]
\centering
\begin{tabular}{c|c|ccc|c}
\multirow{2}{*}{$\vec{d}_{\rm ref}$}&\multirow{2}{*}{$[d_1^2,d_2^2]$}&\multicolumn{3}{c}{$E^{\rm free}/M_\pi$}&\multirow{2}{*}{operators}\\%
&&N203&N200&D200&\\%
\hline%
(0, 0, 2)&{[}1, 1{]}&2.0&2.0&2.0&$A_1^+$\\%
&{[}0, 4{]}&2.6574&2.8333&2.8826&$A_1^+$\\%
&{[}2, 2{]}&3.0667&3.4761&3.5989&$A_1^+ \oplus B_1^+$\\%
&{[}1, 5{]}&3.6391&4.1982&&$A_1^+ \oplus B_1^+ \oplus E^+$\\%
&{[}3, 3{]}&3.8483&&&$A_1^+ \oplus B_2^+$\\%
\hline%
(0, 1, 2)&{[}1, 2{]}&2.28&2.4007&2.4384&$A_1^+$\\%
&{[}0, 5{]}&2.7513&2.9435&2.9971&$A_1^+$\\%
&{[}1, 4{]}&3.1243&3.5322&3.6542&$A_1^+$\\%
&{[}2, 3{]}&3.2636&3.7328&3.8728&$A_1^+ \oplus A_2^+$\\%
&{[}1, 6{]}&3.7359&&&$A_1^+ \oplus A_2^+$\\%
&{[}2, 5{]}&3.9268&&&$2 A_1^+ \oplus A_2^+$\\%
\hline%
(1, 1, 2)&{[}1, 3{]}&2.4883&2.6826&2.7422&$A_1^+$\\%
&{[}2, 2{]}&2.5889&2.8358&2.9114&$A_1^+$\\%
&{[}0, 6{]}&2.8347&3.0407&3.0979&$A_1^+$\\%
&{[}1, 5{]}&3.2466&3.6855&3.8164&$A_1^+ \oplus A_2^+$\\%
&{[}2, 4{]}&3.4263&3.9413&4.0944&$A_1^+$\\%
\hline%
(0, 2, 2)&{[}2, 2{]}&2.0&2.0&2.0&$A_1^+$\\%
&{[}1, 5{]}&2.7997&3.0889&3.1764&$A_1^+ \oplus B_2^+$\\%
&{[}0, 8{]}&2.9788&3.2072&3.2704&$A_1^+$\\%
&{[}3, 3{]}&3.0667&3.4761&3.5989&$A_1^+$\\%
&{[}2, 6{]}&3.6897&&&$A_1^+ \oplus A_2^+ \oplus B_1^+ \oplus B_2^+$\\%
&{[}4, 4{]}&3.8483&&&$A_1^+$\\%
\hline%
(0, 0, 3)&{[}1, 4{]}&2.0872&2.096&2.0979&$A_1^+$\\%
&{[}0, 9{]}&3.0423&3.2803&3.3461&$A_1^+$\\%
&{[}2, 5{]}&3.1647&3.5948&3.7235&$A_1^+ \oplus B_1^+ \oplus E^+$\\%
&{[}3, 6{]}&3.9457&&&$A_1^+ \oplus B_2^+ \oplus E^+$\\%
\end{tabular}%
\caption{\label{tab:two-pion-opsB} Same as \Cref{tab:three-pion-ops} except for
the two-pion operators with $4\leq \vec{d}_{\rm ref}^{\,2}\leq 9$.
}
\end{table}

\begin{table}[b]
\centering
\begin{tabular}{c|c|ccc|c}
\multirow{2}{*}{$\vec{d}_{\rm ref}$}&\multirow{2}{*}{$[d_1^2,d_2^2]$}&\multicolumn{3}{c}{$E^{\rm free}/M_K$}&\multirow{2}{*}{operators}\\%
&&N203&N200&D200&\\%
\hline%
(0, 0, 0)&{[}0, 0{]}&2.0&2.0&2.0&$A_{1g}$\\%
&{[}1, 1{]}&2.7035&2.6505&2.3624&$A_{1g} \oplus E_g$\\%
&{[}2, 2{]}&3.2585&3.1702&2.6762&$A_{1g} \oplus E_g \oplus T_{2g}$\\%
&{[}3, 3{]}&3.7319&&&$A_{1g} \oplus T_{2g}$\\%
\end{tabular}%
\caption{\label{tab:two-kaon-ops} Same as \Cref{tab:three-pion-ops} except for
the two-kaon operators with $\vec{d}_{\rm ref}^{\,2}=0$.
}
\end{table}

\begin{table}[bp]
\centering
\begin{tabular}{c|c|ccc|c}
\multirow{2}{*}{$\vec{d}_{\rm ref}$}&\multirow{2}{*}{$[d_1^2,d_2^2]$}&\multicolumn{3}{c}{$E^{\rm free}/M_K$}&\multirow{2}{*}{operators}\\%
&&N203&N200&D200&\\%
\hline%
(0, 0, 1)&{[}0, 1{]}&2.1688&2.1565&2.0886&$A_1$\\%
&{[}1, 2{]}&2.8389&2.7774&2.4396&$A_1 \oplus B_1 \oplus E$\\%
&{[}1, 4{]}&3.3047&3.216&2.7159&$A_1$\\%
&{[}2, 3{]}&3.3748&3.2798&2.7454&$A_1 \oplus B_2 \oplus E$\\%
&{[}2, 5{]}&3.7879&&&$A_1 \oplus B_1 \oplus E$\\%
\hline%
(0, 1, 1)&{[}0, 2{]}&2.2931&2.2738&2.1624&$A_1$\\%
&{[}1, 1{]}&2.3779&2.3479&2.1887&$A_1$\\%
&{[}1, 3{]}&2.9494&2.8818&2.5066&$A_1 \oplus B_1$\\%
&{[}2, 2{]}&2.9939&2.9219&2.5242&$A_1 \oplus A_2$\\%
&{[}1, 5{]}&3.3817&3.2895&2.767&$A_1 \oplus B_2$\\%
&{[}2, 4{]}&3.4746&3.3742&2.8072&$A_1 \oplus B_2$\\%
&{[}2, 6{]}&3.8627&&&$A_1 \oplus A_2 \oplus B_1 \oplus B_2$\\%
&{[}3, 5{]}&3.927&&&$A_1 \oplus A_2 \oplus B_1 \oplus B_2$\\%
\hline%
(1, 1, 1)&{[}0, 3{]}&2.3941&2.3698&2.2264&$A_1$\\%
&{[}1, 2{]}&2.5308&2.4902&2.2718&$A_1 \oplus E$\\%
&{[}1, 6{]}&3.4512&3.3559&2.814&$A_1 \oplus E$\\%
&{[}2, 5{]}&3.5628&3.4579&2.8632&$A_1 \oplus A_2 \oplus 2 E$\\%
&{[}3, 4{]}&3.6132&&2.8864&$A_1 \oplus E$\\%
\hline%
(0, 0, 2)&{[}1, 1{]}&2.0&2.0&2.0&$A_1$\\%
&{[}0, 4{]}&2.4802&2.452&2.2832&$A_1$\\%
&{[}2, 2{]}&2.7035&2.6505&2.3624&$A_1 \oplus B_1$\\%
&{[}1, 5{]}&3.1275&3.0509&2.6203&$A_1 \oplus B_1 \oplus E$\\%
&{[}3, 3{]}&3.2585&3.1702&2.6762&$A_1 \oplus B_2$\\%
&{[}2, 6{]}&3.6423&&&$A_1 \oplus B_2 \oplus E$\\%
\hline%
(0, 1, 2)&{[}1, 2{]}&2.1795&2.1654&2.0906&$A_1$\\%
&{[}0, 5{]}&2.5559&2.5245&2.3346&$A_1$\\%
&{[}1, 4{]}&2.759&2.7051&2.4073&$A_1$\\%
&{[}2, 3{]}&2.8426&2.7806&2.4406&$A_1 \oplus A_2$\\%
&{[}1, 6{]}&3.2026&3.1225&2.6699&$A_1 \oplus A_2$\\%
&{[}2, 5{]}&3.3225&3.2318&2.7216&$2 A_1 \oplus A_2$\\%
&{[}3, 6{]}&3.7977&&&$A_1 \oplus A_2$\\%
&{[}4, 5{]}&3.8364&&&$A_1$\\%
\hline%
(1, 1, 2)&{[}1, 3{]}&2.3217&2.2977&2.1684&$A_1$\\%
&{[}2, 2{]}&2.3779&2.3479&2.1887&$A_1$\\%
&{[}0, 6{]}&2.6237&2.5897&2.3817&$A_1$\\%
&{[}1, 5{]}&2.8508&2.7921&2.4649&$A_1 \oplus A_2$\\%
&{[}2, 4{]}&2.9604&2.8914&2.5099&$A_1$\\%
&{[}2, 6{]}&3.4076&3.3125&2.7757&$A_1 \oplus A_2$\\%
&{[}3, 5{]}&3.4802&3.3793&2.8092&$A_1 \oplus A_2$\\%
&{[}4, 6{]}&3.9303&&&$A_1 \oplus A_2$\\%
&{[}5, 5{]}&3.9473&&&$A_1 \oplus A_2$\\%
\end{tabular}%
\caption{\label{tab:two-kaon-opsB} Same as \Cref{tab:three-pion-ops} except for
the two-kaon operators with $1\leq \vec{d}_{\rm ref}^{\,2}\leq 6$.
}
\end{table}

\begin{table}[tp]
\centering
\begin{tabular}{c|c|ccc|c}
\multirow{2}{*}{$\vec{d}_{\rm ref}$}&\multirow{2}{*}{$[d_1^2,d_2^2]$}&\multicolumn{3}{c}{$E^{\rm free}/M_K$}&\multirow{2}{*}{operators}\\%
&&N203&N200&D200&\\%
\hline%
(0, 2, 2)&{[}2, 2{]}&2.0&2.0&2.0&$A_1$\\%
&{[}1, 5{]}&2.5441&2.5066&2.2989&$A_1 \oplus B_2$\\%
&{[}3, 3{]}&2.7035&2.6505&2.3624&$A_1$\\%
&{[}0, 8{]}&2.7423&2.7038&2.4658&$A_1$\\%
&{[}2, 6{]}&3.1555&3.0758&2.6294&$A_1 \oplus A_2 \oplus B_1 \oplus B_2$\\%
&{[}4, 4{]}&3.2585&3.1702&2.6762&$A_1$\\%
&{[}5, 5{]}&3.7319&&&$A_1 \oplus A_2$\\%
\hline%
(0, 0, 3)&{[}1, 4{]}&2.0744&2.0718&2.0529&$A_1$\\%
&{[}2, 5{]}&2.7803&2.7239&2.4138&$A_1 \oplus B_1 \oplus E$\\%
&{[}0, 9{]}&2.795&2.7546&2.5039&$A_1$\\%
&{[}3, 6{]}&3.3337&3.2418&2.7257&$A_1 \oplus B_2 \oplus E$\\%
\end{tabular}%
\caption{\label{tab:two-kaon-opsC} Same as \Cref{tab:three-pion-ops} except for
the two-kaon operators with $8\leq\vec{d}_{\rm ref}^{\,2}\leq 9$.
}
\end{table}

\clearpage
\section{Energy levels used in fits}\label{app:levels}
\label{app:C}

\begin{table}[bph!]
\centering
\begin{tabular}{c|c|cccc}
$\vec{d}_{\rm ref}$&{Type}&N203&N200&D200&\\%
\hline%
\multirow{4}{*}{(0,0,0)}&$2\pi$&$2A^+_{1g}+E^+_g$&$2A^+_{1g}+E^+_g$&$2A^+_{1g}+E^+_g$\\%
&$3\pi$&$2A^-_{1u} +E^-_u$&$2A^-_{1u} +E^-_u$&$A^-_{1u} +E^-_u$\\%
&$2K$&$2A_{1g} +E_g$&$2A_{1g} +E_g$&$2A_{1g} +E_g$\\%
&$3K$&$2A_{1u} +E_u$&$2A_{1u} +E_u$&$2A_{1u} +E_u$\\ \hline
\multirow{4}{*}{(0,0,1)}& $2\pi$ &$2A^+_1 + B_1^+ + E^+$ & $2A^+_1 + B_1^+ + E^+$&$A^+_1$\\%
&$3\pi$ &$A^-_2 + B_2^- + E^-$ & $2 A^-_2 + B_2^- + E^-$  &$A^-_2$\\%
&$2K$ & $A_1+B_1+ E$ &$ A_1$ &$2 A_1+B_1+ E$\\%
&$3K$&$A_2$ & $A_2$&$3 A_2+2B_2 + E$  \\ \hline
\multirow{4}{*}{(0,1,1)}&$2\pi$  & $3A_1^++ E^+$& $3A_1^++ B_1^+$& $2A_1^+$\\%
&$3\pi$ &$2A_2^-$ & $2A_2^-$& $2A_2^-$\\%
&$2K$ &$2A_1$ &$2A_1$ &$3A_1+A_2+B_1$\\%
&$3K$&$2A_2$ &$2A_2$ & $2A_2$ \\ \hline
\multirow{4}{*}{(1,1,1)}&$2\pi$  &$2A_1^+ + E^+$ &$2A_1^+ + E^+$ & $2A_1^+ + E^+$\\%
&$3\pi$ &$3A_2^- + E^-$ &$3A_2^- + E^-$ & $2A_2^- + E^-$\\%
&$2K$ &$2A_1+E$ &$2A_1+E$  & $2A_1+E$\\%
&$3K$&$3A_2+E$ &$2A_2+E$  & $3A_2+E$  \\ \hline
\multirow{4}{*}{(0,0,2)}&$2\pi$  &$3 A^+_1+B^+_1$ &$3 A^+_1+B^+_1$ & $3 A^+_1+B^+_1$\\%
&$3\pi$ &$3 A_2^- + B_2^- $ & $3 A_2^- + B_2^- $& $2 A_2^- $\\%
&$2K$ & $3 A_1+B_1$ &$3 A_1+B_1$ & $3 A_1+B_1$\\%
&$3K$&$2 A_2+B_2$ &$2 A_2$ & $3 A_2+2B_2 + E$  \\ \hline
\multirow{4}{*}{(0,1,2)}&$2\pi$  &$4A^+_1 + A^+_2$ &$4A^+_1 + A^+_2$ &$3A_1^+ $\\%
&$3\pi$ &$4 A_2^-$ &$4 A_2^-$  &$3 A_2^-$\\%
&$2K$ &$4A_1 + A_2$ &$3A_1$ & $4A_1 + A_2$\\%
&$3K$& $3A_2$ &$2A_2$ & $3A_2$ \\ \hline
\multirow{4}{*}{(1,1,2)}&$2\pi$  &$4A^+_1 + A^+_2$ &$5A^+_1 + A^+_2$  & $3A^+_1$\\%
&$3\pi$ &$5A^-_2 + A^-_1$ &$2A^-_2 + A^-_1$ & $4A^-_2 + A^-_1$\\%
&$2K$ & $4A_1 + A_2$&$3A_1$ & $5A_1 + A_2$\\%
&$3K$&$5A_2 + A_1$ &$3A_2 + A_1$ & $4A_2 + A_1$ \\ \hline
\multirow{4}{*}{(0,2,2)}&$2\pi$  &$4A^+_1 + B^+_2$ &$4A^+_1 + B^+_2$ & $4A^+_1 + B^+_2$ \\%
&$3\pi$ &$4A^-_2 + B^-_1$ &$3A^-_2 + B^-_1$ & $2A^-_2 + B^-_1$\\%
&$2K$ &$4A_1 + B_2$ & $4A_1 + B_2$& $4A_1 + B_2$\\%
&$3K$& $2A_2 + B_1$ &$2A_2$ & $3A_2 + B_1$ \\ \hline
\multirow{4}{*}{(0,0,3)}&$2\pi$  &$3A_1^+ + B^+_1 + E^+$ &$3A_1^+ + B^+_1 + E^+$ & $2A_1^+ + B^+_1 + E^+$\\%
&$3\pi$ &$3A^-_2 + B^-_2$ &$4A^-_2 + B^-_2$ & $2A^-_2 + B^-_2$\\%
&$2K$ & $A_1 + B_1 + E$&$2A_1 + B_1 + E$ & $3A_1 + B_1 + E$\\%
&$3K$&$3A_2 + B_2$ &$3A_2 + B_2$ & $3A_2 + B_2$ \\ \hline
\end{tabular}
\caption{ Energy levels used in the fits of this work. Notation is as follows: ``$2A^-_{1u} +E^-_u$'' means the lowest two levels in the $A^-_{1u}$ irrep, and the lowest in the $E^-_u$ irrep.
}
\end{table}

\clearpage
\bibliographystyle{JHEP}      
\bibliography{ref.bib}

\end{document}